\documentclass[fleqn,usenatbib]{mnras}

\usepackage{newtxtext,newtxmath}
\usepackage[T1]{fontenc}

\DeclareRobustCommand{\VAN}[3]{#2}
\let\VANthebibliography\thebibliography
\def\thebibliography{\DeclareRobustCommand{\VAN}[3]{##3}\VANthebibliography}

\usepackage{graphicx}	%
\usepackage{amsmath}	%
\usepackage[nolist]{acronym}
\usepackage{multirow}
\usepackage{soul, color} %

\begin{acronym}[TDMA]
	\acro{NEA}{near-Earth asteroid}
	\acro{PHA}{potentially hazardous asteroid}
	\acro{LCDB}{Asteroid Lightcurve Photometry Database}
	\acro{ALCDEF}{Asteroid Lightcurve Data Exchange Format}
	\acro{CCD}{Charge Coupled Device}
	
	\acro{ESO LP}{European Southern Observatory Large Programme}
	\acro{ORM}{Observatorio del Roque de los Muchachos}
	\acro{INT}{Isaac Newton Telescope}
	\acro{WFC}{Wide Field Camera}
	\acro{IDS}{Intermediate Dispersion Spectrograph}
	\acro{NTT}{New Technology Telescope}
	\acro{SofI}{Son of ISAAC}
	\acro{EFOSC2}{ESO Faint Spectrograph and Camera - version 2}
	\acro{VLT}{Very Large Telescope}
	\acro{UT}{Unit Telescope}
	\acro{VISIR}{VLT Imager and Spectrometer for mid-Infrared}
	\acro{VIMOS}{VIsible MultiObject Spectrograph}
	\acro{PDS}{Palmer Divide Station}
	\acro{NOT}{Nordic Optical Telescope}
	\acro{ALFOSC}{Alhambra Faint Object Spectrograph and Camera}
	\acro{LT}{Liverpool Telescope}
	
	\acro{DA}{(29075) 1950 DA}
	\acro{KZ66}{(68346) 2001 KZ66}

\end{acronym}

\title[Detection of YORP on NEA (68346) 2001 KZ66]{Detection of the YORP Effect on the contact-binary (68346) 2001 KZ66 from combined radar and optical observations}

\author[T. J. Zegmott et al.]{
	Tarik J. Zegmott,$^{1}$\thanks{E-mail: tzegmott@gmail.com}
	S. C. Lowry,$^{1}$
	A. Ro\.z{}ek,$^{1,2}$
	B. Rozitis,$^{3}$
	M. C. Nolan,$^{4}$
	E. S. Howell,$^{4}$
	S. F. Green,$^{3}$
	\newauthor
	C. Snodgrass,$^{2,3}$
	A. Fitzsimmons,$^{5}$
	and P. R. Weissman$^{6}$
	\\
	$^{1}$Centre for Astrophysics and Planetary Science, University of Kent, Canterbury, CT2 7NH, ~UK\\
	$^{2}$Institute for Astronomy, University of Edinburgh, Royal Observatory, Edinburgh, EH9 3HJ, UK\\
	$^{3}$Planetary and Space Sciences, School of Physical Sciences, The Open University, Milton Keynes, MK7 6AA, UK\\
	$^{4}$Lunar and Planetary Laboratory, University of Arizona, Tucson, AZ 85721, USA\\
	$^{5}$Astrophysics Research Centre, Queens University Belfast, Belfast, BT7 1NN, UK\\
	$^{6}$Planetary Sciences Institute, Tucson, AZ 85719, USA\\
}

\date{Accepted XXX. Received YYY; in original form ZZZ}

\pubyear{2021}

\begin{document}
	\label{firstpage}
	\pagerange{\pageref{firstpage}--\pageref{lastpage}}
	\maketitle

	\begin{abstract}
		The YORP effect is a small thermal-radiation torque experienced by small asteroids, and is considered to be crucial in their physical and dynamical evolution. It is important to understand this effect by providing measurements of YORP for a range of asteroid types to facilitate the development of a theoretical framework. We are conducting a long-term observational study on a selection of near-Earth asteroids to support this. We focus here on (68346) 2001 KZ66, for which we obtained both optical and radar observations spanning a decade. This allowed us to perform a comprehensive analysis of the asteroid’s rotational evolution. Furthermore, radar observations from the Arecibo Observatory enabled us to generate a detailed shape model. We determined that (68346) is a retrograde rotator with its pole near the southern ecliptic pole, within a $ 15\degr $ radius of longitude $ 170\degr $ and latitude $ -85\degr $. By combining our radar-derived shape model with the optical light curves we developed a refined solution to fit all available data, which required a YORP strength of $ (8.43\pm0.69)\times10^{-8} \rm~rad ~day^{-2} $. (68346) has a distinct bifurcated shape comprising a large ellipsoidal component joined by a sharp neckline to a smaller non-ellipsoidal component. This object likely formed from either the gentle merging of a binary system, or from the deformation of a rubble pile due to YORP spin-up. The shape exists in a stable configuration close to its minimum in topographic variation, where regolith is unlikely to migrate from areas of higher potential.
	\end{abstract}


	\section{Introduction}
	
	The Yarkovsky-O'Keefe-Radzievskii-Paddack (YORP) effect is a gentle torque that small asteroids can experience due to the reflection and thermal emission of sunlight from their surfaces \citep{Rubincam:2000fg}. 
	This torque causes a change in rotation rate and spin-axis obliquity.
	The YORP effect is a major driver in the spin-state evolution of small Solar System bodies, and can lead to substantial physical changes, including shape and structural changes, binary formation and even mass shedding \citep{Scheeres:2015hr}. 
	To date, the YORP effect has been detected on just seven objects: (54509) YORP, (1862) Apollo, (1620) Geographos, (3103) Eger, (25143) Itokawa, (161989) Cacus, and (101955) Bennu \citep{2007Sci...316..272L,2007Sci...316..274T,Kaasalainen:2007hq,Durech:2008di,Durech:2012bq,Lowry:2014cb,Durech:2018gg,Nolan:2019eib}. 
	Recently, there has also been the first indication of the YORP effect acting on an asteroid in an excited rotation state \citep{Lee:2021ei}. 
	Crucially, all of the detections have been rotational accelerations (i.e. in the spin-up sense). 
	For a population of asteroids with randomized shapes and spin-states, YORP should produce both spin-up and spin-down cases.
	While recent theoretical developments are being proposed to explain the apparent lack of spin-down cases \citep{Golubov:2012kt,2014ApJ...794...22G,Golubov:2017ky}, to fully understand this important process requires more observational detections of the YORP effect in action.

	We are therefore conducting a long-term monitoring campaign of a sample of near-Earth asteroids (NEAs) with the aim of detecting the signature of YORP through on-going minute changes in their rotation periods. 
	This programme began in April 2010 as part of an approved Large Programme at the European Southern Observatory (ESO LP).
	The ESO LP campaign focussed mainly on optical-imaging monitoring of our sample of NEAs using the \ac{NTT}, at La Silla Observatory (Chile). 
	Photometric light curves are extracted from the imaging data to monitor changes in the periodicity of the light curve caused by evolution in the rotational period, usually over at least three distinct epochs of observation. 
	In selected cases we are acquiring thermal-IR imaging data on our targets across wavelengths 9.8--12.4~$\mu$m, using the ESO VISIR instrument at the 8.2m VLT telescope.
	This is important for detailed thermophysical modelling to determine theoretical YORP values for comparison with observed strengths, when high quality shape and spin-state models are available \citep{Rozitis:2013de,Rozek:2019ij}.
	Many of our sample have been observed with planetary radar \citep{Rozek:2019kp,Rozek:2019ij}.
	This allows for a more detailed shape model to be obtained, which greatly improves the likelihood of detecting YORP from the optical light curve data and further improves the quality of the thermophysical modelling.

	Our sample asteroids were chosen to maximize their potential for detecting YORP.
	They tend to be small (sub-km effective radius), thus increasing their susceptibility to YORP. 
	All our targets are NEAs, and spend all of their time in close proximity to the Sun.
	Rotation rates are mainly around 2-3 hours, with some exceptions, as this makes them practical for light curve observations, as the NEA can be observed to make at least one full rotation in any given night.  
	This is also a very important regime, being close the rotational break-up limit for asteroid bodies.
	Extensive observational monitoring of such objects is important for understanding how NEAs disrupt due to YORP-induced rotational torques, leading to mass-shedding events for example \citep{2019EPSC...13.1561L}.
	
	Here we present the latest results on one of our target NEAs (68346) 2001 KZ66 (hereafter referred to as KZ66).
	This object is both a \ac{NEA} of the Apollo class and a \ac{PHA}, and was discovered on the 29 May 2001 at Haleakala by the Near-Earth Asteroid Tracking programme \citep{Pravdo:1999eh}. 
	It was observed by the NEOWISE survey which determined a geometric albedo of $0.291\pm0.110$. This was used to obtain a diameter of $0.736\pm0.208 \rm ~km$ \citep{Masiero:2017fo}.
	There have been several measurements of its synodic rotation period. 
	Optical observations from the Palmer Divide Station during May 2016 revealed a large amplitude, $\rm 0.63^{mag}$, and a rotation period of $\rm 4.987\pm0.005$ hours \citep{Warner:2016uq}. 
	Follow-up observations in July 2016 from the Palmer Divide Station displayed a lower amplitude of $\rm 0.35^{mag}$, but the rotation period was consistent with the earlier value \citep{Warner:2017vb}. 
	More recently, the object was observed at the Isaac Aznar Observatory. 
	The light curve had an amplitude of $\rm 0.77^{mag}$. However, the synodic rotation period was measured to be $\rm 5.633\pm0.002$ hours \citep{Aznar:2017ro}, much larger than earlier measurements.
	
	We have obtained nine optical light curves throughout the period April 2010 to January 2019 with the 3.6~m \ac{NTT}. 
	An additional light curve was obtained during a supporting programme on the 2.5~m \ac{INT}, Spain, in 2012. 
	Included in our dataset are 7 published light curves from the Palmer Divide Station taken between May 2016 and July 2016 \citep{Warner:2016uq,Warner:2017vb}. 
	Additionally, we use radar observations that were taken over two nights in October 2003 from Arecibo Observatory. 

	In this paper, we will present the results and analysis of a long-term photometric monitoring programme to model the asteroid and to detect changes in the object's rotation rate that could be due to YORP. 
	The format of this paper is the following: Sect. \ref{sec:observations} describes our observing campaign of KZ66. 
	In Sect. \ref{sec:modelling} and \ref{sec:spinstate} we present our analysis of the shape and spin-state modelling, and the approach to detect YORP-induced rotational accelerations. 
	Sect. \ref{sec:discussion} provides a general discussion of the results and their implications, and overall conclusions are drawn in Sect. \ref{sec:conclusions}.

	\section{Observations of (68346) 2001 KZ66}
	\label{sec:observations}
	
	\subsection{Optical light curves}
	\label{sec:lightcurves}
	
\begin{figure}
		\resizebox{\hsize}{!}{\includegraphics[width=\linewidth,trim=0 0 0 0,clip=true]{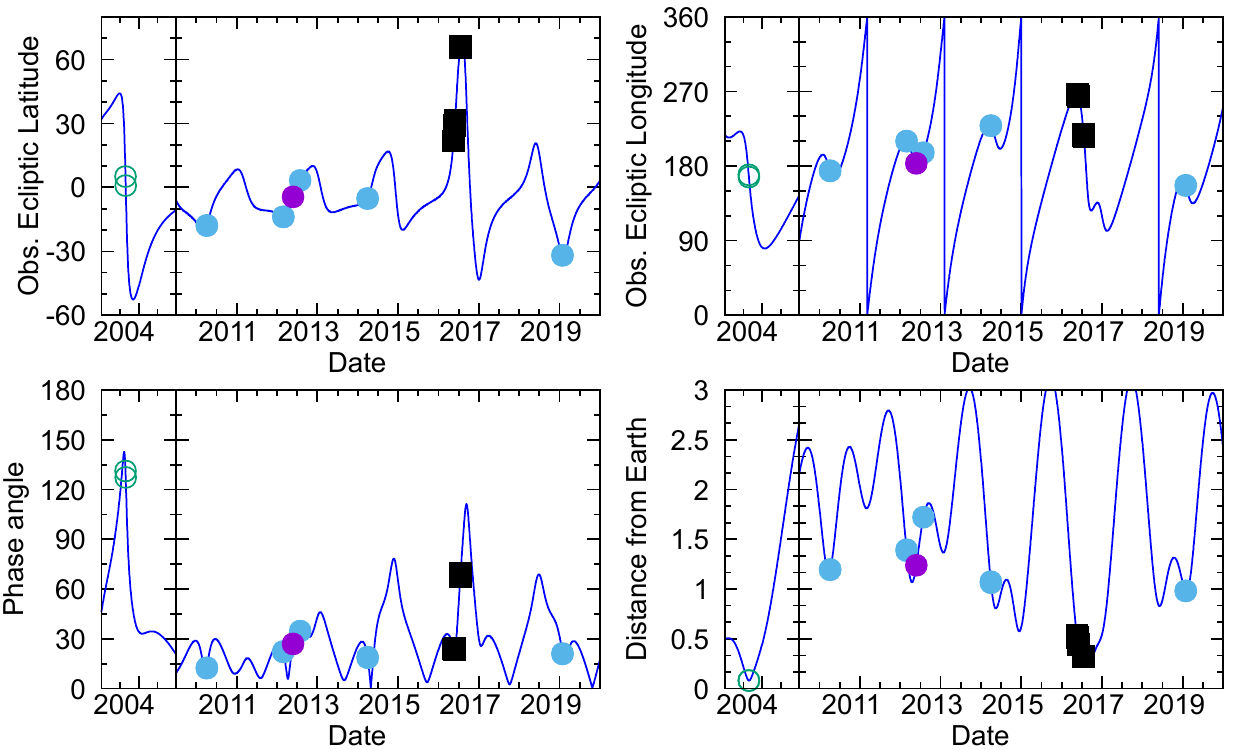}}
		\caption{
			Asteroid (68346) 2001 KZ66 observing geometries during the optical and radar observations over the period 2003 to 2019. The vertical line in all panels indicates the omission of the period from 2005 to 2010. 
			The top two panels show the position of the object in the ecliptic coordinate system, latitude and longitude, as observed from Earth. 
			The bottom two panels show the phase angle and geocentric distance of the asteroid. 
			Optical light curve data from the NTT are marked with filled blue circles, with light curve data from the INT marked with filled purple circles. Black squares represent the published light curve data. 
			The green circles mark when the Arecibo radar data were collected. The blue continuous line represents the object's observational ephemeris.
		}
		\label{fig:geom}
	\end{figure}

	The optical light curve dataset for KZ66 covers the period from April 2010 to January 2019, spanning a total of ten years. 
	A summary of all of the light curves used in this paper is reported in Table \ref{tab:obstable}, along with details of the observing conditions: observer-centred ecliptic longitude and latitude, heliocentric and geocentric distance, and orbital phase angle. 
	The light curves which were obtained as a part of our programme are those with IDs 1-8 and 16-17 in Table \ref{tab:obstable} and are presented with those IDs in Appendix Figs. \ref{fig:conv-lightcurvefit1}, \ref{fig:yorp-lightcurvefit1}, and \ref{fig:noyorp-lightcurvefit1}. 
	A graphical overview of the observing geometries for all data is given in Fig. \ref{fig:geom}.
	
	When observing the asteroid we used either sidereal or differential tracking, depending on the rate of motion of the object on any given night. 
	If the object was moving slowly enough, we opted for sidereal tracking and kept the exposure times short enough that the asteroid didn't move by more than the FWHM of the seeing during the exposure. 
	This ensured that the asteroid, and the background comparison stars, were not significantly trailed. 
	To optimize the light curve extracted we used circular photometric apertures which varied according to the varying seeing conditions from one exposure to the next. 
	Our chosen optimal aperture radius was set to 2~$\times$~FWHM of the profile of the asteroid. 
	This was not required for the comparison stars given their increased brightness so larger apertures were used. The apertures chosen were sufficiently large that they collected essentially all of the stars' light.
	The brightness of the asteroid was then compared with the average brightness of the background stars to produce relative light curves. 
	When this condition could not be achieved due to the higher rate of motion, we simply tracked at the projected rates of motion to maintain the stellar appearance of the asteroid. 
	Again, we chose an optimal aperture radius of 2~$\times$~FWHM of the profile of the asteroid.
	The background stars were now significantly trailed, but we limited exposure times such that the stars were never trailed by more than 5 arcsecs.  
	In cases where this would result in a low signal-to-noise ratio, multiple images were co-added. 
	However, this was only necessary for the July 2012 NTT dataset. 
	In the following sub-sections, we describe the instrumental set-up of each facility we used to observe KZ66. 

	\begin{table*}
		\centering          
		\begin{tabular}{ ccccccccccc}
			\hline \hline \noalign{\smallskip}
			{ID} &   UT Date    & {$r_\odot$} & {$\Delta_\oplus$} & {$\alpha$}  & $\lambda_O$ &  $\beta_O$  & Total  & Filter & Observing & Reference \\%& LC-only \\ %
			&    [\emph{dd/mm/yyyy}]    & {[AU]}  &   {[AU]}   & {[$\degr$]} & {[$\degr$]} & {[$\degr$]} & [hour] &        &          facility  &  \\%& model \\ %
			\noalign{\smallskip}
			1   & 04/04/2010 &  2.134  &   1.193    &    12.41     &    174.4    &    -18.0     &  2.0   &   R    &    NTT    &     \\% &\textbullet \\  %
			2   & 05/04/2010 &  2.133  &   1.197    &    12.84     &    174.0    &    -17.8     &  5.3   &   R    &    NTT    &     \\% &\textbullet \\   %
			3   & 26/02/2012 &  2.125  &   1.396    &    22.18     &    210.1    &    -13.8     &  3.3   &   R    &    NTT    &      \\%&\textbullet \\  %
			4   & 27/02/2012 &  2.124  &   1.385    &    21.90     &    210.0    &    -13.8     &  4.7   &   R    &    NTT    &      \\%&\textbullet \\  %
			5   & 24/05/2012 &  1.949  &   1.240    &    27.04     &    183.3    &     -4.4     &  3.1   &   R    &    INT    &      \\%&\textbullet \\  %
			6   & 28/07/2012 &  1.681  &   1.723    &    34.67     &    196.2    &      3.3     &  1.0   &   R    &   {NTT}   &      \\%&\textbullet \\  %
			7   & 30/03/2014 &  1.952  &   1.074    &    19.06     &    228.8    &     -5.3     &  3.5   &   V    &    NTT    &      \\%&\textbullet \\ %
			8   & 31/03/2014 &  1.949  &   1.063    &    18.63     &    228.6    &     -5.2     &  2.2   &   V    &   {NTT}   &      \\%&\textbullet \\ %
			9   & 27/05/2016 &  1.418  &   0.459    &    23.53     &    265.0    &     29.1     &  5.8   & clear  &    PDS    &  1   \\%&\textbullet \\ %
			10  & 28/05/2016 &  1.412  &   0.453    &    23.69     &    264.7    &     29.9     &  5.7   & clear  &    PDS    &  1  \\% &\textbullet \\ %
			11  & 29/05/2016 &  1.405  &   0.446    &    23.90     &    264.4    &     30.7     &  6.2   & clear  &    PDS    &  1  \\% &\textbullet \\ %
			12  & 30/05/2016 &  1.399  &   0.440    &    24.15     &    264.1    &     31.5     &  6.2   & clear  &    PDS    &  1  \\% &\textbullet \\  %
			13  & 17/07/2016 &  1.095  &   0.324    &    67.25     &    218.6    &     65.6     &  5.3   & clear  &    PDS    &  2  \\% &\textbullet \\  %
			14  & 18/07/2016 &  1.089  &   0.323    &    68.26     &    217.0    &     65.9     &  5.1   & clear  &    PDS    &  2  \\% &\textbullet \\ %
			15  & 19/07/2016 &  1.083  &   0.322    &    69.24     &    215.4    &     66.2     &  5.2   & clear  &    PDS    & {2} \\% &\textbullet \\ %
			16  & 27/01/2019 &  1.835  &   0.983    &    20.92     &    156.8    &    -31.8     &  6.1   &   V    &    NTT    &      \\%&\textbullet \\ %
			17  & 28/01/2019 &  1.839  &   0.981    &    20.54     &    156.3    &    -31.9     &  3.1   &   V    &    NTT    &      \\%&\textbullet \\ %
			\hline
		\end{tabular}

		\caption{
			A log of optical photometry datasets of asteroid (68346) 2001 KZ66 used in this study.
			Each light curve has a numerical ``ID'' listed,  
			then the Universal Time (UT) ``Date'' of the beginning of the night is given, 
			as well as the heliocentric ($r_\odot$) and geocentric ($\Delta_\oplus$) distances measured in AU, 
			the solar phase angle $(\alpha)$, 
			the observer-centred ecliptic longitude $(\lambda_O)$, 
			the observer-centred ecliptic latitude $(\beta_O)$,  
			and the ``Observing facility'' used to obtain the light curve. 
			Where relevant a ``Reference'' to published work is given: 
			(1)  \citet{Warner:2016uq}; (2) \citet{Warner:2017vb}
			Each line represents a single nightly light curve data set (for some of the nights listed, several light curve segments  have been obtained). 
			The light curves shown here were selected for the light-curve-only shape modelling. These light curves were also utilised for the initial radar observation modelling, but are omitted from the subsequent stages in order to make the model quasi-independant from them. 
			Observing facility key: 
			INT -- 2.5 m Isaac Newton Telescope (La Palma, Spain), 
			NTT -- European Southern Observatory 3.5 m New Technology Telescope (Chile), 
			PDS -- Palmer Divide Station (California, USA).
		}
		\label{tab:obstable} 
	\end{table*}

	\subsubsection{{New Technology Telescope} -- 2010, 2012, 2014, and 2019}
	
	The asteroid KZ66 was observed at the {ESO} $3.6\, \rm m$ \ac{NTT} telescope in La Silla (Chile), using the \ac{EFOSC2}. 
	The CCD detector of \ac{EFOSC2} has $2048 \times 2048$ pixels and a field of view of $4.1\arcmin \times 4.1\arcmin$. 
	The observations of KZ66 were performed in imaging mode using $2 \times 2$ binning on the detector, and with the Bessel R filter in 2010 and 2012, and Bessel V filter in 2014 and 2019. 
	The object was detected at the NTT on two different nights in 2010, three in 2012, two in 2014, and two in 2019 giving a total of nine light curves. 
	The data were reduced using the standard CCD reduction procedures. 
	The light curve with ID 6 from Table \ref{tab:obstable} required co-addition of the images to improve the signal-to-noise ratio.
	
	\subsubsection{{Isaac Newton Telescope}  -- 2012}
	
	KZ66 was also monitored with the  $2.5\, \rm m$ {INT} in La Palma (Spain), using the Wide-Field Camera (WFC). 
	The WFC is an array of four CCD chips, each with $2048\times 4100$ pixels, with a total field of view $34 \arcmin \times 34 \arcmin$. 
	However, for these observations, only CCD4 was used with a window of $10\arcmin \times 10\arcmin$ to reduce readout time between images. 
	The KZ66 observations were performed using the Harris R filter. 
	The target was observed over one night during 2012 on the 24 February for 3.1 hours. 
	The data were reduced using standard CCD reduction procedures.
	
	\subsubsection{Published optical light curves -- 2016}
	
	The previously published photometry data for KZ66 include ten light curves of which seven are used in this study. The other three were discarded due to poor signal-to-noise ratio. 
	These light curves have the IDs 9-15 (see Table \ref{tab:obstable}). 
	The observatory used to obtain these light curves is the Palmer Divide Station in California, USA, which hosts several small telescopes with diameters less than $0.5\rm ~m$. 
	The observations consist of four light curves taken in May 2016 \citep{Warner:2016uq} and a further three in July 2016 \citep{Warner:2017vb}, all of which were taken with a clear filter. 
	These processed light curves were obtained from the \ac{ALCDEF} database \citep{Warner:2011ty}.
	
	\subsection{Asteroidal radar observations}
	\label{sec:radarcamp}
	
	Radar observations were also used in this analysis, which included both delay-Doppler imaging and continuous-wave (cw) spectra. %
	Delay-Doppler images are obtained from a circularly polarized transmitted signal which is phase-modulated with a pseudo-random binary code \citep{Ostro:1993jv,Magri:2007io}. 
	This modulation pattern allows us to determine the distance between the observer and the parts of the object reflecting the signal. 
	The resolution of the delay is determined by the time-resolution of the modulated signal, the baud length. 
	The second axis in a delay-Doppler image is given by the Doppler shift measured in the returning signal. 
	The width of the shifted signal is dependant on a combination of the size of the object and its rotation rate. 
	Unlike the delay-Doppler images, the cw spectra contain no information on the delay of the radar signal. They solely record the Doppler shift of the emitted signal that returns from the object in both circular polarisation orientations.

	\subsubsection{Arecibo Observatory -- 2003}
	
	The William E. Gordon telescope in Arecibo, Puerto Rico, is a $305\,\rm m$ fixed-dish radio telescope equipped with an S-band  $2380\,\rm MHz$ Planetary Radar transmitter. 
	Observations of the asteroid KZ66 with Arecibo Observatory under the Planetary Radar programme (project number R1811) were performed on two consecutive nights: 28 and 29 October 2003. 
	The cw spectra were taken on each night, in addition to imaging with $0.1 \,\rm \mu s$ baud length code corresponding to ${\sim}15\,\rm m$ resolution in delay (see the detailed list of radar experiments in Table~\ref{tab:radartabA}).
	
	Modelling radar data is a computationally expensive process. 
	To minimise the computational time required we can either remove datasets with similar geometries or reduce the number of frames within a dataset by co-adding several frames at a time. 
	As only two nights of consecutive data were available with almost identical observing geometry, we opted for the latter. 
	Co-addition of pairs of frames was used in order to maintain maximal rotational coverage. 
	This also had the additional benefit of increasing the signal-to-noise ratio of the delay-Doppler images.

	\begin{table*}
		\centering          
		\begin{tabular}{cccccccccccc}
			\hline\hline
			\noalign{\smallskip}
			{UT Date}  & {RTT} & {Baud}	& Resolution	& {Start-Stop}           & {Runs}   & Radar    & SC/OC & Ranging\\
			{[\emph{yyyy-mm-dd}]} 	& {[s]} & {[$\mu\rm s$]} & [m]	& {[\emph{hh:mm:ss-hh:mm:ss}]} & {}  & model  & analysis &\\ \hline
			\noalign{\smallskip}
			2003-10-28 & 80 & cw  &	    & 12:21:46-12:36:34 & 6 & \textbullet & \textbullet & \\
			&    & cw  &	    & 12:39:38-12:40:51 & 1 & & & \textbullet \\
			&    & cw  &	    & 12:44:45-12:45:58 & 1 & & & \textbullet \\
			&    & 4   &	600 & 12:48:22-12:49:35 & 1 & & & \textbullet \\
			&    & 4   &	600 & 12:51:38-12:52:51 & 1 & & & \textbullet \\
			&    & 0.1 & 15  & 12:57:00-14:37:50 & 36 & \textbullet & & \\
			2003-10-29 & 79 & cw  &	    & 12:10:02-12:21:58 & 5 & \textbullet & \textbullet & \\
			&    & 0.1 &	15  & 12:27:18-13:57:37 & 33 & \textbullet & & \\
			&    & 4   &	600 & 14:01:20-14:02:32 & 1 & & & \textbullet \\
			&    & 4   &	600 & 14:04:20-14:05:32 & 1 & & & \textbullet \\
			&    & cw  &	    & 14:07:19-14:08:31 & 1 & & & \textbullet \\
			
		\end{tabular}
		\caption{
			Radar observations of asteroid  (68346) 2001 KZ66 obtained at Arecibo in October 2003. 
			``UT Date'' is the universal-time date on which the observation began.
			``RTT'' is the round-trip light time to the object.
			``Baud'' is the delay resolution of the pseudo-random code used for imaging; baud does not apply to cw data.
			The delay ``Resolution'' is dependant upon the baud and the number of samples taken per baud. 
			For a baud of $0.1 \rm ~\mu s$ and one sample taken per baud this corresponds to a delay resolution of $15 \rm ~m$.
			The timespan of the received data is listed by the UT \emph{start} and \emph{stop} times.
			``Runs'' is the number of completed transmit-receive cycles.
			``Radar model'' column indicates which radar observations were selected for the shape modelling.
			``SC/OC analysis'' column indicates which cw spectra were utilised to calculate the polarisation ratio.
			``Ranging'' column indicates which observations were taken to refine the ephemeris.
		}
		\label{tab:radartabA}
		
	\end{table*}
	
	\section{Modelling shape and spin-state}
	\label{sec:modelling}
	
	\subsection{Period and pole search with light curve data -- convex inversion results}
	\label{sec:convex}
	
\begin{figure}
		\resizebox{\hsize}{!}{\includegraphics[width=12cm, clip=true]{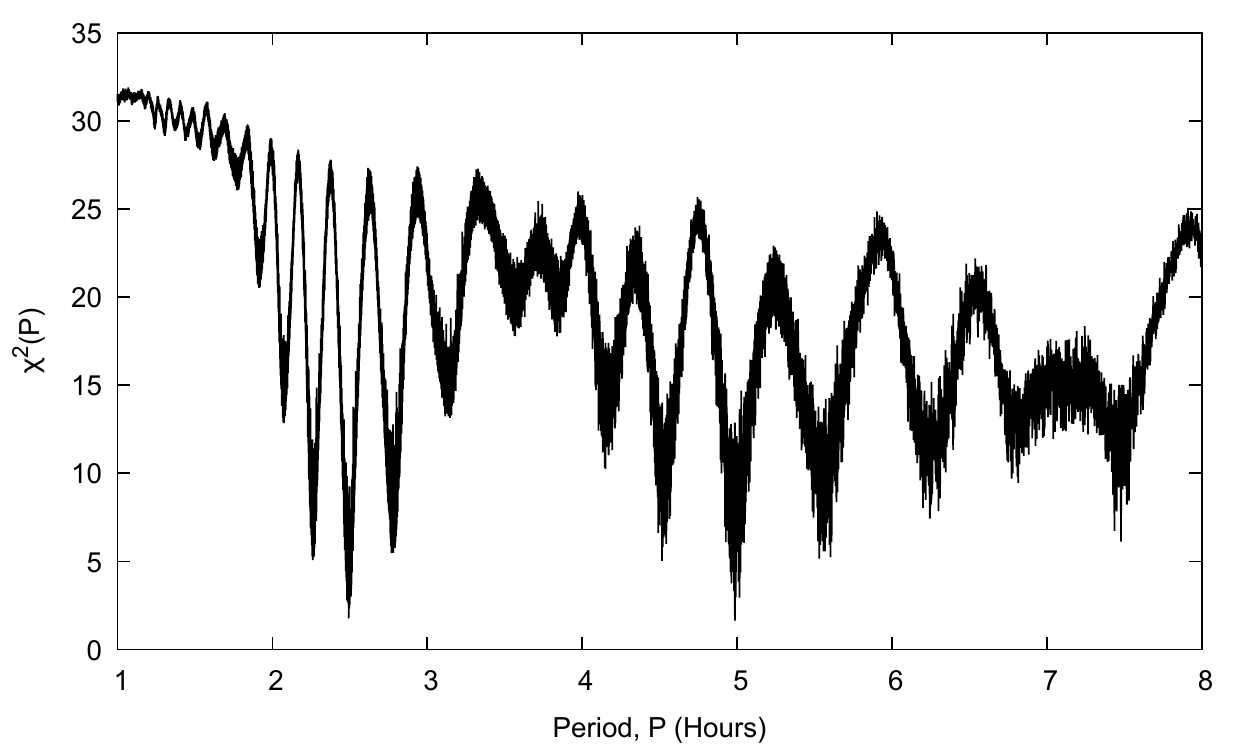}}
		\caption{Results of the sidereal rotation period scan for asteroid (68346) 2001 KZ66 described in Sect. \ref{sec:convex}. 
			The scan resulted in a rotational period of $\rm4.9860\pm0.0001$ hours,
			which was later refined to $4.985988 \pm 0.000020$ hours.}
		\label{fig:periodscan}
	\end{figure}

	The first step in the shape modelling procedure is to define an initial value for the sidereal rotation period of the asteroid, for which we use the method described in \cite{Kaasalainen:2001di}.
	With this approach, six pole orientations are initially spread evenly across the entire celestial sphere.
	We then set up a range of period values to scan across, and for each period we allow the shape to vary for each of the six selected poles, while each time performing a fit of the model to the observed light curve magnitudes. 
	When this is complete for a given period, we then record the lowest $\rm\chi^{2}$ value, and the remaining $\rm\chi^{2}$  values for the other five selected poles are discarded at this stage.
	The period values we scanned across range from 1--8 hours, which easily encompasses all of the previously reported periods for KZ66 \citep{Benner:2006du,Warner:2016uq,Warner:2017vb,Aznar:2017ro}.
	The result of the period search indicated two potential rotational periods, one at 2.493 hours and the other at 4.980 hours (Fig. \ref{fig:periodscan}).  
	However, the period of 2.493 h was subsequently eliminated during the pole orientation analysis as the shape models corresponding to the shorter period failed to reproduce the light curves well. 
	Moving forward we will only consider the rotational period of $\rm4.9860\pm0.0001$ h.
	
	To further constrain the asteroid's pole orientation and sidereal rotation period, and to determine a best-fit convex shape for the asteroid, we utilised the convex inversion techniques described by \cite{Kaasalainen:2001dx,Kaasalainen:2001di} in our customized procedures. 
	Therefore all shapes obtained from this section of the analysis are convex hulls, meaning that they approximate the real shape of the asteroid.
	
\begin{figure}
		\resizebox{\hsize}{!}{\includegraphics[width=12cm, clip=true]{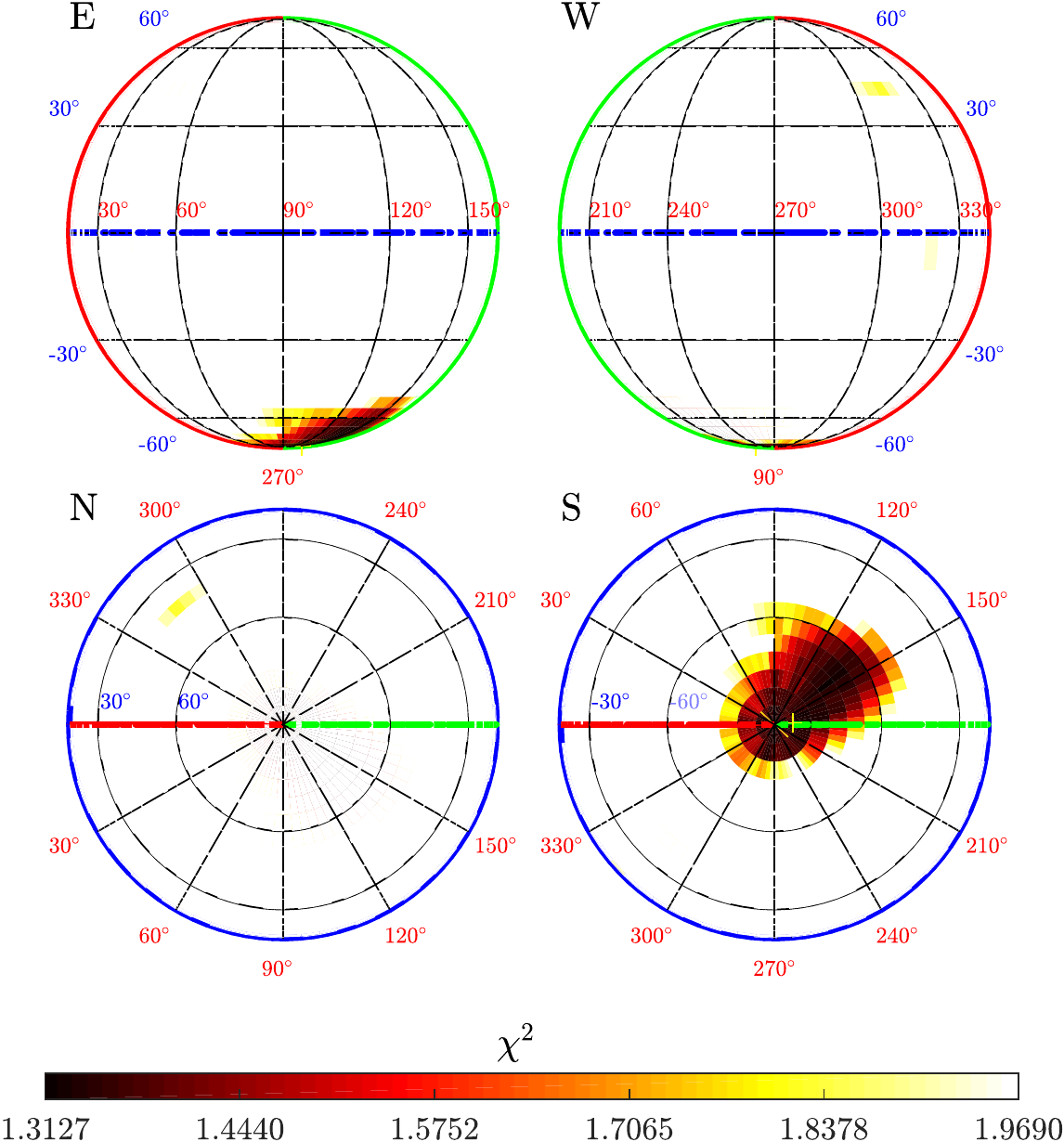}}
		
		\caption{Results of the convex inversion pole search for asteroid (68346) 2001 KZ66 projected on the surface of the celestial sphere described in ecliptic coordinates. The blue line marks the ecliptic plane with latitude $\beta=0\degr$, while additional circles of latitude are marked with black lines and labelled with blue numerals. The red line marks the longitude $\lambda=0\degr$ and the green line $\lambda=180\degr$, with selected meridians marked with black lines and labelled with red numerals. From top-left clockwise, the projections show the eastern (E), western (W), southern (S) and northern (N) hemispheres of the sky. The colour changes from black at the minimum $\chi^2$, with 1\% increments of the minimum $\chi^2$, and the white region representing all the solutions with $\chi^2$ more than 50\% above the minimum $\chi^2$. The best pole is marked by a yellow `+' which is found at $\lambda={170\degr}, \beta={-85\degr}$, with a 1$\sigma$ error of radius 15\degr.}
		\label{fig:convexsphere}
	\end{figure}

	Our approach first involves setting up a grid of pole positions covering the entire celestial sphere with a resolution of $5\degr\times5\degr$. 
	At each pole position, the rotation period and convex shape were optimised to fit the light curves. 
	The sidereal period determined previously is utilised in this step as an optimal starting point for the subsequent optimisation process.
	The initial epoch, $T_0$, and the initial rotation phase, $\varphi_{0}$, were held fixed during the optimisation. 
	The $T_{0}$ was set to 2455291.0, corresponding to the date of the first light curve (4 April 2010) and $\varphi_{0}$ set to $0\degr$. 
	The results of the pole search are shown in Fig. \ref{fig:convexsphere}.
	This model assumes a constant rotation period - a YORP factor is included later in Sect. \ref{sec:yorp-convinv}. 
	Due to the large range of observer-centred ecliptic latitude sampled by the light curve dataset, we were able to tightly constrain the pole which resides in the southern equatorial hemisphere.

\begin{figure}
		\resizebox{\hsize}{!}{
			\includegraphics[width=\textwidth,trim=2cm 8.5cm 2.1cm 8.5cm, clip=truef]{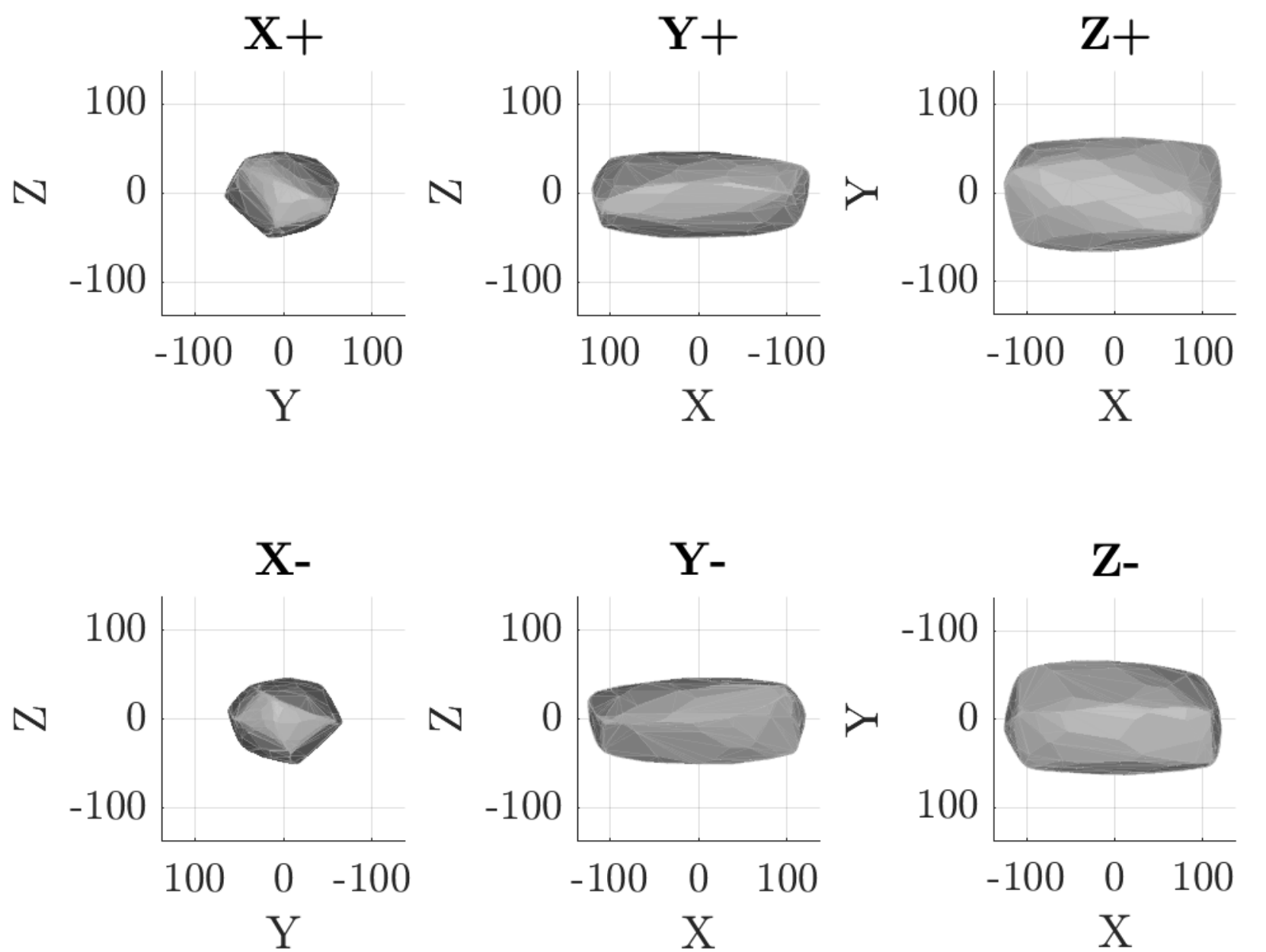}}
		\caption{The best-fit convex shape model of (68346) 2001 KZ66. The model was produced from a pole search using light curve data only, and assuming a constant period. The pole is located at $\lambda=170\degr$, $\beta=-85\degr$. \textit{Top row} (left to right): views along the X, Y and Z axes of the body-centric coordinate frame from the positive end of the axis. \textit{Bottom row} (left to right): views along the X, Y and Z axes from their negative ends. The model's Z axis is aligned with the rotation axis and axis of maximum inertia. The light curve convex inversion model is not scaled and the units shown are arbitrary.
			\label{fig:convexshape}}
	\end{figure}

	Our best model's pole is at an ecliptic longitude, $\lambda$, of $170\degr$ and an ecliptic latitude, $\beta$, of $-85\degr$ with a 1$\sigma$ error radius of 15\degr. 
	This pole is marked by a yellow cross in Fig. \ref{fig:convexsphere}. 
	We extracted the best-fit shape model and constant sidereal rotation period at this best-fit pole location.
	The latter was determined to be $4.985988 \pm 0.000020$ hours, and the best-fit shape model is shown in Fig. \ref{fig:convexshape}. 
	The best-fit convex shape can be described as a mix between an elongated ellipsoid and a cylinder. 
	The planar features are the result of the procedure attempting to match the large amplitude of the light curves. 
	
	\subsection{Determination of a shape model from radar observations -- \texttt{SHAPE} results} 
	\label{sec:SHAPE}
	
\begin{figure}
		\resizebox{\hsize}{!}{\includegraphics{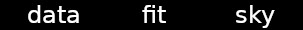}\includegraphics{pictures/radarlabel.jpg}}
		\resizebox{\hsize}{!}{\includegraphics{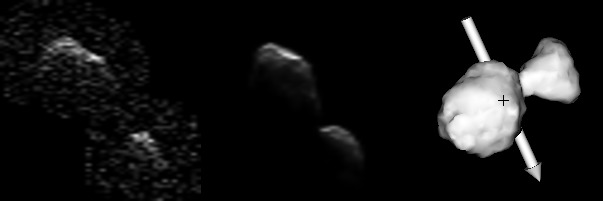} \includegraphics{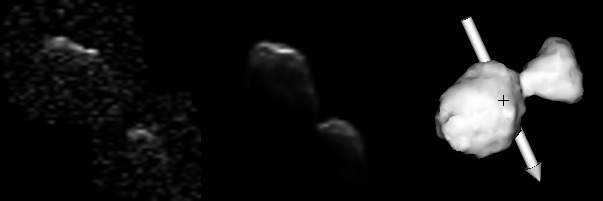}}
		\resizebox{\hsize}{!}{\includegraphics{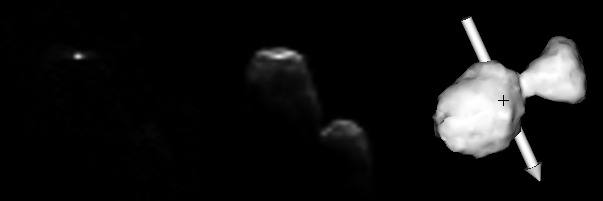} \includegraphics{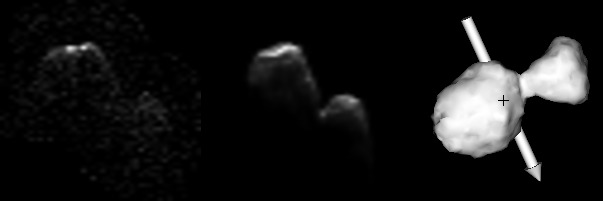}}
		\resizebox{\hsize}{!}{\includegraphics{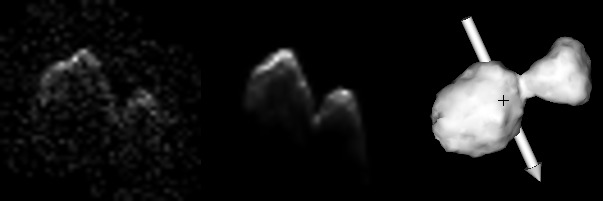} \includegraphics{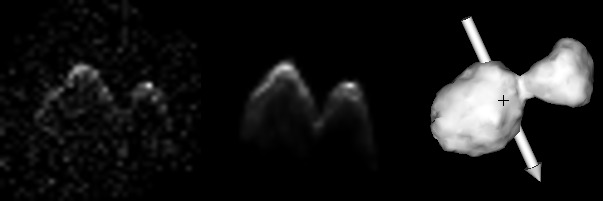}}
		\resizebox{\hsize}{!}{\includegraphics{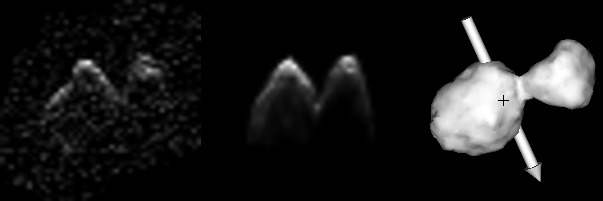} \includegraphics{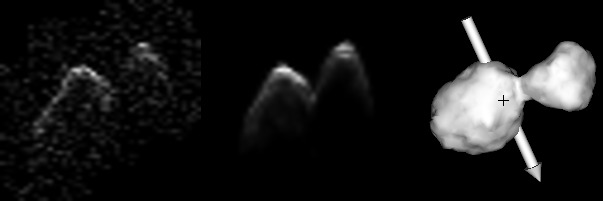}}
		\resizebox{\hsize}{!}{\includegraphics{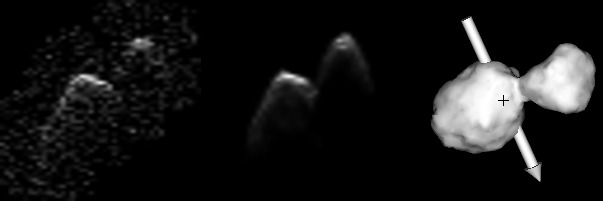} \includegraphics{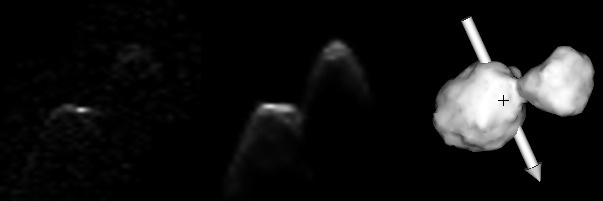}}
		\resizebox{\hsize}{!}{\includegraphics{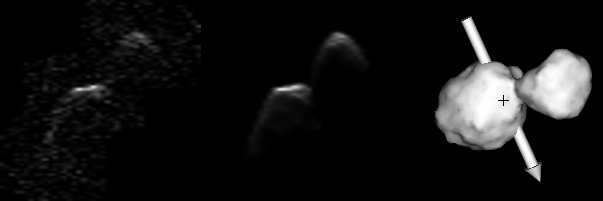} \includegraphics{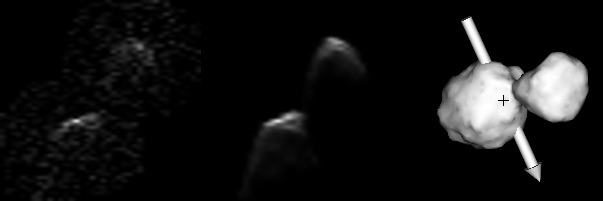}}
		\resizebox{\hsize}{!}{\includegraphics{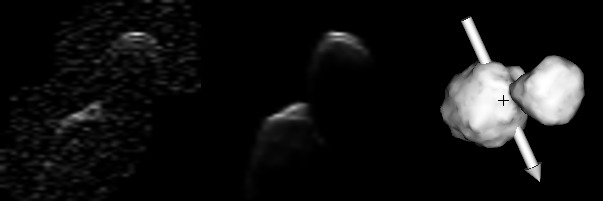} \includegraphics{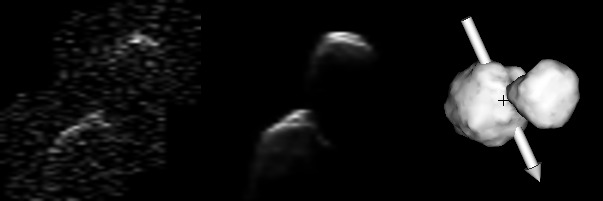}}
		\resizebox{\hsize}{!}{\includegraphics{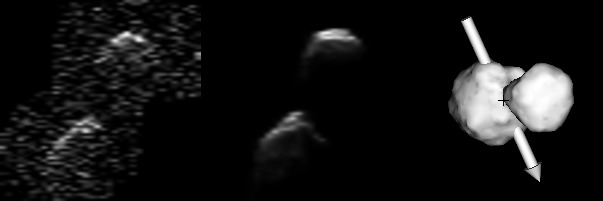} \includegraphics{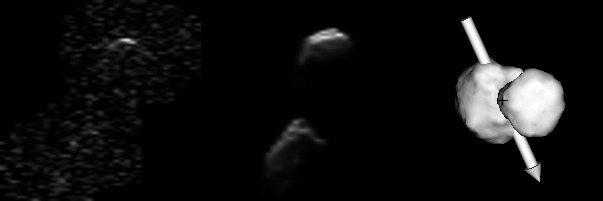}}
		\resizebox{\hsize}{!}{\includegraphics{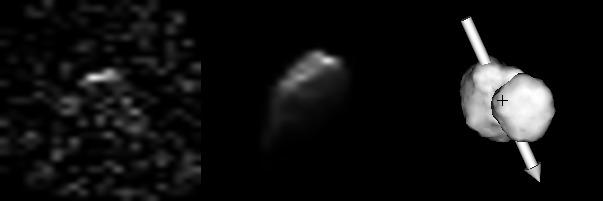} \includegraphics{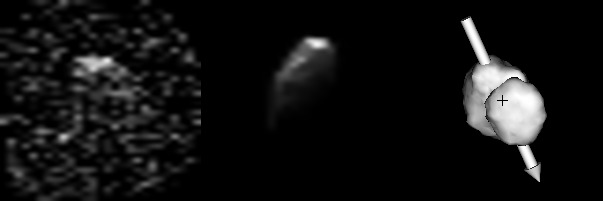}}
		\caption{Fit of the final radar-derived shape model of asteroid (68346) 2001~KZ66 to the radar data (model summary in Table~\ref{tab:models}). Each three-image sub-panel is comprised of: the observational data (left panel), synthetic echo (middle panel), and plane-of-sky projection of the best-fit model (right panel). On the data and synthetic-echo images the delay increases downwards and the frequency (Doppler) to the right. The plane-of-sky images are orientated with celestial north (in equatorial coordinates) to the top and east to the left. 
		The rotation vector (Z-axis of body-fixed coordinate system) is marked with a white arrow. This sequence of images corresponds to the Arecibo data collected on 28 October 2003.%
		\label{fig:radar:full}  }
	\end{figure}

\begin{figure}

		\resizebox{\hsize}{!}{\includegraphics{pictures/radarlabel.jpg}\includegraphics{pictures/radarlabel.jpg}}
		\resizebox{\hsize}{!}{\includegraphics{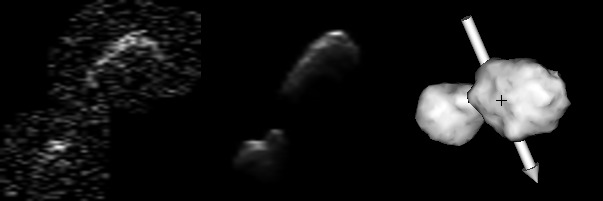} \includegraphics{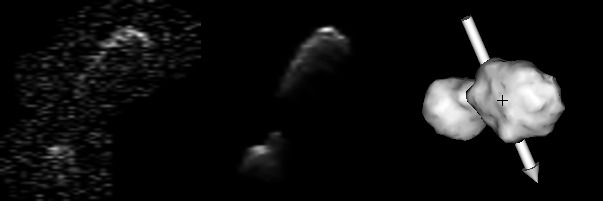}}
		\resizebox{\hsize}{!}{\includegraphics{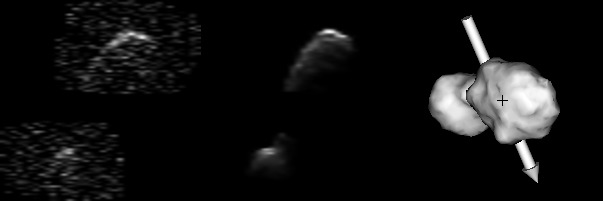} \includegraphics{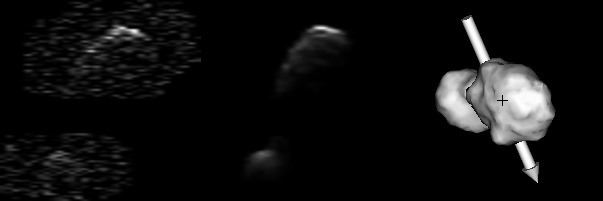}}
		\resizebox{\hsize}{!}{\includegraphics{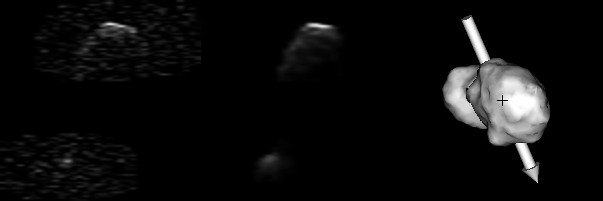} \includegraphics{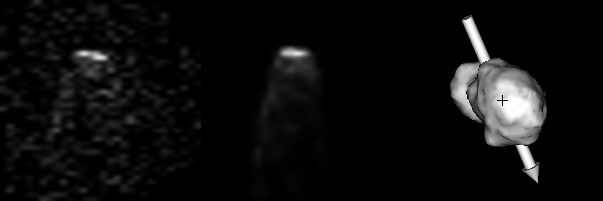}}
		\resizebox{\hsize}{!}{\includegraphics{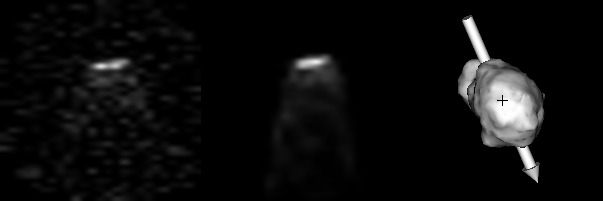} \includegraphics{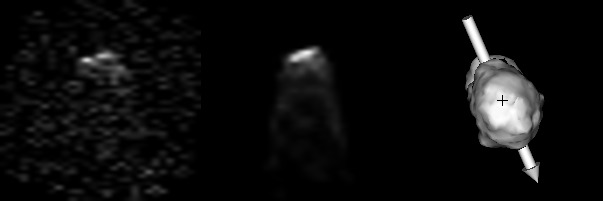}}
		\resizebox{\hsize}{!}{\includegraphics{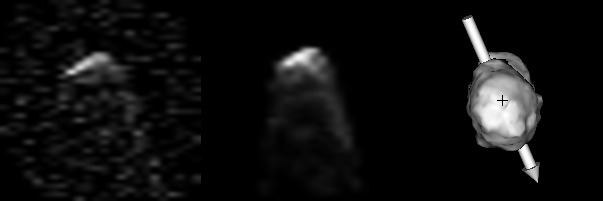} \includegraphics{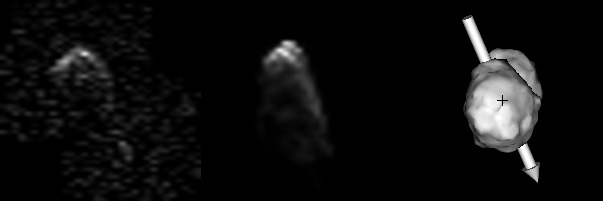}}
		\resizebox{\hsize}{!}{\includegraphics{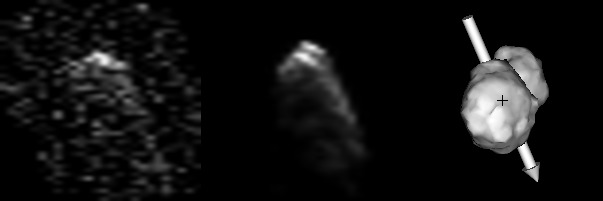} \includegraphics{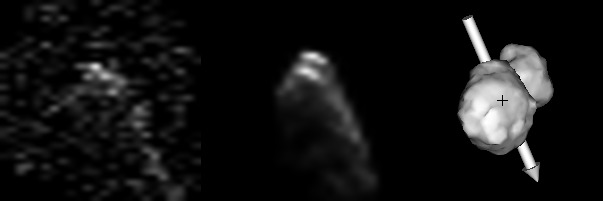}}
		\caption{Same as Fig. \ref{fig:radar:full},  but this sequence of images corresponds to the Arecibo data collected on 29 October 2003.%
			\label{fig:radar:full2}}
		
	\end{figure}

	The procedure described in Sect. \ref{sec:convex} can only produce convex models, and hence it will not produce the concavities of the neck region of the asteroid which can clearly be seen in the delay-Doppler images (Figs. \ref{fig:radar:full} and \ref{fig:radar:full2}) and the continuous-wave spectra (Fig. \ref{fig:cwspectra}).
	Shape modelling utilising radar data was performed using the SHAPE modelling software \citep{Hudson:1993uo,Magri:2007io}. 
	The efficiency of this process is greatly improved with good starting conditions, hence initial spin-state parameters were set to the values determined from our earlier analyses. 
	As for the starting point of the asteroid's shape, the delay-Doppler echoes indicate that this asteroid is bi-lobed. 
	Therefore we took the approach of constructing an initial two-component model comprised of two ellipsoids, with their radii estimated from the delay-Doppler images. 
	Each component is described by three axial lengths, three positional parameters, and three angular parameters.
	As the  asteroid is assumed to be a principal-axis rotator with a constant period, the rotational state of the model is described by five parameters: the ecliptic latitude and longitude which describe the model's pole orientation; the initial UT epoch, $T_0$; the initial rotation phase at $ T_0 $; and the period of rotation about the model's z-axis. 
	The initial parameters for the ellipsoid model were manually adjusted by visually matching the synthetic echoes output by SHAPE to a selection of the delay-Doppler images. 
	During this process the origin of the body-fixed coordinate system is overlapped with the model's centre-of-mass.
	All of the parameters above were optimised during the modelling, except for the pole orientation which is held fixed at the value determined from the convex inversion pole scan. 
	The resulting model consists of a large ellipsoid and a smaller spheroidal component. 
	The dimensions of the larger component along the body-centric coordinate axis are $0.847 \times 0.573 \times 0.575\rm ~km$. 
	The smaller component has axial lengths of $0.334 \times 0.341 \times 0.341\rm ~km$. Both components' centres are separated by a distance of $0.450\rm ~km$. 
	This initial ellipsoidal stage of modelling included both the light curve and radar observations. 
	However, all subsequent modelling relies solely on the radar observations in order to make the model quasi-independent from the light curves. 
	
\begin{figure}
		\resizebox{\hsize}{!}{\includegraphics[width=\textwidth,trim=2cm 8.5cm 2.1cm 8.5cm, clip=true]{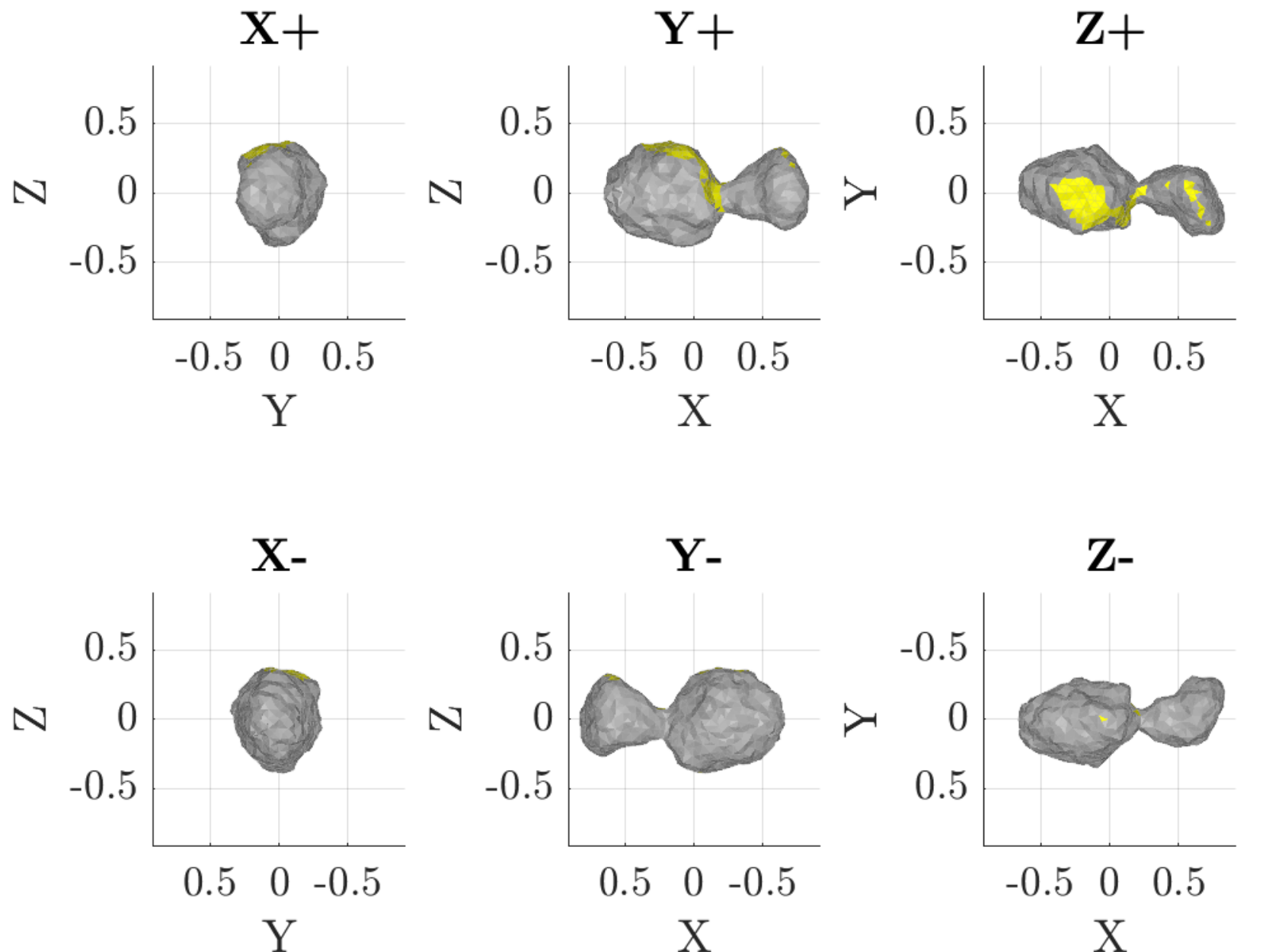}}
		\caption{Same as Fig. \ref{fig:convexshape}, but for the best-fit vertex shape model of (68346) 2001~KZ66 derived from the radar data. The model was derived from cw spectra and delay-Doppler images. The model was given a fixed pole orientation determined during the convex inversion pole search. Yellow facets indicate those not seen or seen at an incidence angle greater than $ 70\degr $ in the radar images. The axes scales are shown in kilometres. Details of the alignment between the  model's body-centric axes and the principal axes are detailed in Table \ref{table:Inertia&PAOrientation} of the appendix.
			\label{fig:vertexshape}}
	\end{figure}

	The complexity of the shape's description was gradually increased during the fitting procedure. 
	From the initial ellipsoid representation, the model was converted to spherical harmonic form during the intermediate stages before being converted to a vertex model. 
	The final model consists of 1000 vertices giving 1996 facets with a median facet edge length of 57 m. 
	The position of each vertex was optimised individually during the fitting procedure. In addition to the shape, we also fit for the rotation period and initial rotation phase. 
	During the fitting procedure, three penalty functions were applied to discourage unrealistic features and to improve the fitting procedure \citep{Magri:2007io}. These penalty functions increase the numerical value of the goodness-of-fit when the unphysical features are encountered. 
	Since SHAPE attempts to optimise the goodness-of-fit, it follows that the larger these penalties are the more strongly it discourages the features. 
	The first function penalised the deviation of the centre of mass away from the origin of body-fixed coordinates. 
	The second suppressed facet-scale topography, which discourages the appearance of unphysical spikes which can occur from fitting noise. 
	The final penalty function attempts to keep the third principal axis aligned with the model's z-axis.
	The resulting model is shown in Fig. \ref{fig:vertexshape} (Table \ref{tab:shapegeomparams} contains the geometric parameters). 
	The larger component has an ellipsoidal shape and it is joined to the smaller lobe by a tight neck region. 
	The smaller lobe is non-ellipsoidal with a curved body. 
	Inspection of the model's moments of inertia reveals that the largest axis of inertia is the y-axis rather than its spin (z) axis. 
	The moment of inertia of the y-axis is 5\% larger than that of the z-axis, although delay-Doppler images suffer from a north/south ambiguity, which causes aliasing in the plane-of-sky view. Generally, the images are resolved along the line-of-sight and perpendicular to the rotation axis, and the ambiguity occurs perpendicular to these. Although, with a near equatorial view, this ambiguity can lead to a worse constraint in the z-axis when compared to the x- and y-axis \citep{2002aste.book..151O}. This ambiguity can usually be broken with sufficient coverage; for KZ66 though, the images were obtained across only two days, about $ 5.3\degr $ of motion across the sky, and so the range of aspect angles covered is not large enough to break the ambiguity. 
	Due to this ambiguity, KZ66 is likely more compressed in the z-axis than demonstrated by our model, accounting for the difference in the moments of inertia. 
	For a spin-state analysis of this asteroid, this discrepancy is negligible, though this difference would be significant from a dynamical modelling point of view. 
	Table \ref{table:Inertia&PAOrientation} of the appendix contains a full description of the moments of inertia and the alignment of the principal axes to the body-centric axes for this model.
	The diameter of the model's equivalent-volume sphere has a value of $0.797 \rm ~ km$, which is in good agreement with the diameter of $0.736\pm0.208 \rm ~ km$ determined in the NEOWISE survey \citep{Masiero:2017fo}. 
	A comparison of the delay-Doppler images, a synthetic echo generated from the shape model, and a plane-of-sky image of the shape model are shown in Figs. \ref{fig:radar:full} and \ref{fig:radar:full2}. 
	Using this model we can accurately reproduce all of the data across both nights.
	
	\subsection{Surface structure of KZ66 from radar circular polarisation measurements}
	\label{sec:radarproperties}
\begin{figure}
		\centering
		\resizebox{0.75\hsize}{!}{
			\includegraphics[width=\textwidth]{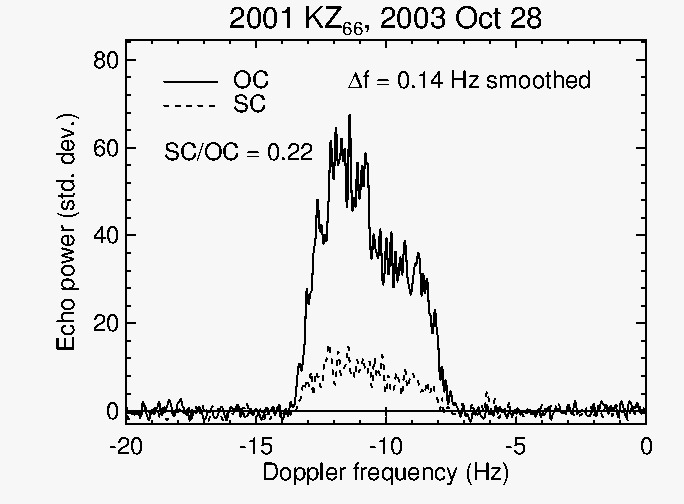}}
		\resizebox{0.75\hsize}{!}{
			\includegraphics[width=\textwidth]{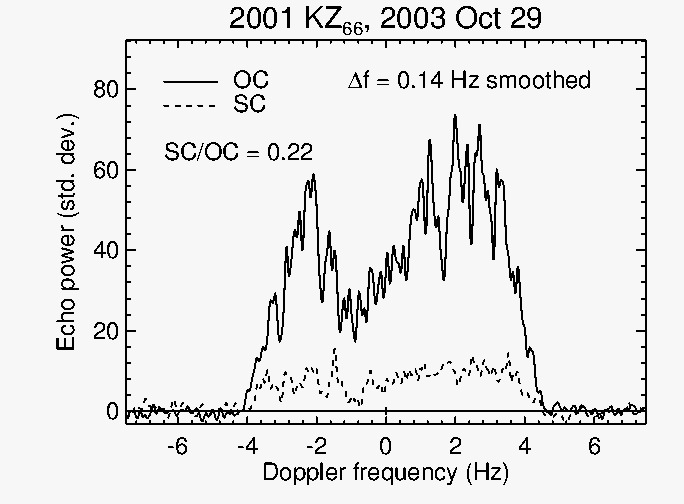}}
		\caption{Continuous wave (cw) spectra observations of asteroid (68346) 2001 KZ66 observed in October 2003 at Arecibo Observatory (detailed description of the observations given in Table \ref{tab:radartabA}). The received circularly polarised signal is recorded in both same circular (SC) polarisation as transmitted, shown by the dashed line, and the opposite circular (OC) polarisation, shown by the solid line. The bifurcation of the asteroid's shape is apparent in these cw spectra; it is particularly prominent in the right panel. 
			\label{fig:cwspectra}}
	\end{figure}

	One particularly useful product of radar observations is the circular polarisation ratio. The ratio, SC/OC, is determined from the detection of an asteroid's echo in a cw spectrum. The received signal is recorded in both same circular (SC) polarisation as transmitted and the opposite circular (OC) polarisation. For mirror-like backscattering, the SC component would be zero. These ratios have been used as a crude estimate of the near-surface complexity at scales near the wavelength of the observations \citep{2002aste.book..151O}, approximately $13\rm ~cm$ for the observations taken from Arecibo Observatory. However, recent observations from OSIRIS-Rex's OCAMS instrument, have shown that the relation between radar circular polarisation ratios and surface roughness are more complex than previously thought \citep{Lauretta:2019bn}. 
	The cw spectra obtained on 28 October 2003 recorded a ratio of $0.218 \pm 0.003$, and the subsequent night recorded $0.222 \pm 0.002$ (spectra are shown in Fig. \ref{fig:cwspectra}). This gives a mean polarisation ratio of $0.220\pm0.003$ for KZ66. This value places KZ66 within the mean value for \acp{NEA}, $0.34\pm0.25$ \citep{Benner:2008dd}. Compared to the polarisation ratios of other contact-binary asteroids with shape models, KZ66 has the lowest recorded value: Itokawa $0.27 \pm 0.04$ \citep{Ostro:2004ep}, 1996 HW1 $0.29 \pm 0.03$ \citep{Magri:2011fd}, 1999 JV6 $0.37 \pm 0.05$ \citep{Rozek:2019kp}. 
	
	\begin{table}
		\resizebox{\hsize}{!}{
			\begin{tabular}{cc}
				\hline \hline \noalign{\smallskip}
				Parameter & Value \\
				\hline \noalign{\smallskip}
				DEEVE dimensions (2a, 2b, 2c) & $1.570 \times 0.513 \times 0.629\rm ~km$\\
				Max. extent along (x, y, z) & $1.513 \times 0.635 \times 0.780\rm ~km$ \\		
				Surface area & $2.704\rm ~km^2$ \\
				Volume & $0.266\rm ~km^3$ \\
				$\rm D_{eq}$ & $0.797\rm ~km$ \\
				\noalign{\smallskip} \hline
			\end{tabular}
		}
		\caption{
			Summary of geometric parameters of (68346) 2001 KZ66. 
			The geometric parameters for the best fit radar-derived shape model of (68346) 2001 KZ66.
			DEEVE denotes the dynamically equivalent equal-volume ellipsoid.
			The maximum extents of the model are measured along the body-centric coordinate axis.
			The $D_{eq}$ is the diameter of a sphere with volume equal to that of the model.
		}
		\label{tab:shapegeomparams}
	\end{table}

	\begin{table}
		\centering
		\begin{tabular}[ht]{l  l l}
			\hline \hline \noalign{\smallskip}
			& Convex inversion & Radar inversion \\ 	\hline	\noalign{\smallskip} 
			\noalign{\smallskip}
			$T_0$ [JD]                                & $2455291.0$ & $2455290.98269$     \\
			\noalign{\smallskip}
			$P$ [h]                                 & $4.985988$ & $4.985997$       \\   %
			$\Delta P$ 								& $0.000020$ & $0.000042$ \\  %
			\noalign{\smallskip}
			$\nu \, [\times 10^{-8}\, \rm rad\, d^{-2}]$ &  $7.7^{+3.8}_{-13.2}$   & $8.43\pm0.69$ \\
			\noalign{\smallskip} \hline                                       & 
		\end{tabular}
		\caption{
			Summary of spin-state model parameters for asteroid (68346) 2001~KZ66. 
			Best-fit spin-state solutions from two approaches to shape modelling: convex light curve inversion, and modelling using the SHAPE software to invert radar data (which utilizes the pole position from light curve inversion). The pole orientation of both models have an ecliptic longitude of $ 170\degr $ and an ecliptic latitude of $ -85\degr $, the pole has a error radius of $ 15\degr $. The table lists: the model epoch ($T_0$), the sidereal rotation period ($P$), and  the YORP factor ($\nu$). }
		\label{tab:models}
	\end{table}

	\newpage
	
	\section{Direct detection of YORP}
	\label{sec:spinstate}
	
	The constant torque provided by the YORP effect produces a linear change in the rotation rate, which can be measured directly, as in \cite{2007Sci...316..272L}. However, the constant torque also manifests itself as a quadratic change in the rotation phase of an asteroid. To investigate the YORP effect in terms of rotation phase requires light curves with precise timing information and a good shape model and pole solution for the asteroid asteroid. If the YORP acceleration, $\nu$, is zero the change in rotation phase will be linear. This can clearly be seen in Eq. \ref{eq:rotation} below:
	\begin{equation}
	\varphi(t)=\varphi\left(T_0\right) + \omega\, \left(t-T_0\right) + \frac{1}{2}\nu\, \left(t-T_0\right)^2 \,,
	\label{eq:rotation}
	\end{equation}
	where:
	\[
	\begin{array}{lp{0.8\linewidth}}
	\varphi(t)  & rotation phase in radians,\\
	t          &  the time of observation (JD),\\
	\varphi(T_0)  & initial rotation phase in radians, \\
	T_0      & time (JD) at which the X-axis of the body crosses the plane-of-sky, also the epoch from which the model is propagated, \\ 
	\omega   & rotation rate in $\rm rad \, day^{-1}$; $\omega\equiv {2 \pi} / {P}$, $P$ is rotation period in days,\\
	\nu & the change of rotation rate in $\rm rad\, day^{-2}$; $\nu\equiv\dot{\omega}$ (the YORP strength). \\
	\end{array}
	\]
	
	\subsection{Convex inversion}
	\label{sec:yorp-convinv}
	
	Our first approach to detecting a YORP signature was based on the light curve only approach used in Sect. \ref{sec:convex} to determine the pole. This time the pole search was repeated while including a range of YORP strengths, $\nu$, between $-1.0 \times 10^{-6} \rm ~rad ~day^{-2}$ and $1.0 \times 10^{-6} \rm ~rad ~day^{-2}$. Performed in two stages, the first stage step-size in YORP strength was coarse with a resolution of $1.0 \times 10^{-7} \rm ~rad ~day^{-2}$. In the second stage, a finer scan between $-1.2 \times 10^{-7} \rm ~rad ~day^{-2}$ and $2.4 \times 10^{-7} \rm ~rad ~day^{-2}$ was performed with a resolution of $1.0 \times 10^{-9} \rm ~rad ~day^{-2}$. For each YORP strength, a grid of pole orientations covering the entire celestial sphere with a resolution of $5\degr \times 5\degr$ was sampled. The pole and YORP strength were held fixed, while period and convex shape were optimised to fit the light curves. 
	For each $\nu$ value we produced a $\chi^{2}$ map by projecting $\chi^{2}$ values for each grid point onto the celestial sphere (Fig. \ref{fig:convexsphere} is an example of such a $\chi^{2}$ map for $\nu=0$). These $\chi^{2}$ maps were examined for each value of $\nu$ with the minimum $\chi^{2}$ extracted from each. 
	The best fit to the light curve dataset is for a YORP value of $7.7 \times 10^{-8} \rm ~rad ~day^{-2}$, although plausible values of YORP range from $-5.20\times10^{-8} \rm ~rad ~day^{-2}$ to $1.15\times10^{-7} \rm ~rad ~day^{-2}$. It should be noted that the constant period convex inversion model, $\nu=0$, reproduces the light curves well (a full set of light curves are provided in Fig. \ref{fig:conv-lightcurvefit1}).
	
	\subsection{Phase-offset spin-state analysis}
	\label{sec:yorp-phoffs}
	
	Our second approach is to measure the rotational phase offsets, $\Delta\varphi$, between observed light curves and the synthetic light curves generated from the radar-derived model (described in Sect. \ref{sec:SHAPE}). 
	The final radar-derived model, shown in Fig. \ref{fig:vertexshape}, was generated independently of the light curves, with the exception of using the light curve-derived pole orientation. 
	
	We first ensure that the rotation phase of the synthetic light curves matches the observations for the first few optical light curves obtained on 4-5 April 2010. 
	This is where we set our $T_{0}$ value (Table \ref{tab:models}).
	To create a synthetic light curve, the model is propagated forward from $T_{0}$ to the epoch of each light curve, using the sidereal rotation period.
	We then determine which facets were illuminated and visible to the observer at the epochs of our light curves using asteroid-centred Sun and Earth vectors from JPL's Horizon service. 
	Our codes account for self-shadowing using ray-tracing. The scattering model employed to produce the synthetic light curves was a combination of the Lambertian and Lommel-Seelinger scattering models \citep{Kaasalainen:2001di}. At any given rotation phase, the relative flux contribution from each facet was then summed to produce the expected brightness of the asteroid, which was then converted to a relative magnitude. The synthetic light curve and observed light curve were then scaled so that they both oscillate about zero magnitude. 
	
	The synthetic and observed light curves may not be aligned at this stage, as we assume a zero YORP strength initially, and the initial rotation period used may be slightly inaccurate on the first iteration of the fitting procedure. 
	To quantify any phase offsets, we measure the phase offsets required in order to align the observed and synthetic light curves. 
	This is done by applying a range of phase offsets from 0\degr to 360\degr in steps of 0.5\degr to the synthetic light curves and recording the phase offset that minimises the $\chi^2$ fit between the light curves. 
	The associated error-bars are the formal 1-$ \sigma $ uncertainties from the  $\chi^2$ fitting process. 
	If the model can be described by a constant period, then a straight line should fit the phase offsets. A non-zero linear coefficient suggests an imprecise rotation period. 
	However, if KZ66 is undergoing a discernable YORP acceleration, then the phase offsets will be fit by a quadratic curve, like that found for (25143) Itokawa \citep{Lowry:2014cb} and (54509) YORP \citep{2007Sci...316..272L,2007Sci...316..274T}. During the initial fittings, the phase offsets may not be purely quadratic. They may also contain a linear component, caused by a small discrepancy in the rotation period, this component can be used to refine the initial rotation period. The process is iterated until the linear component becomes negligible, leaving only the quadratic change attributed to YORP to be fit. 
	
\begin{figure}
		\centering
		\resizebox{\hsize}{!}{\includegraphics[trim=0cm 0cm 0cm 0cm, clip=true]{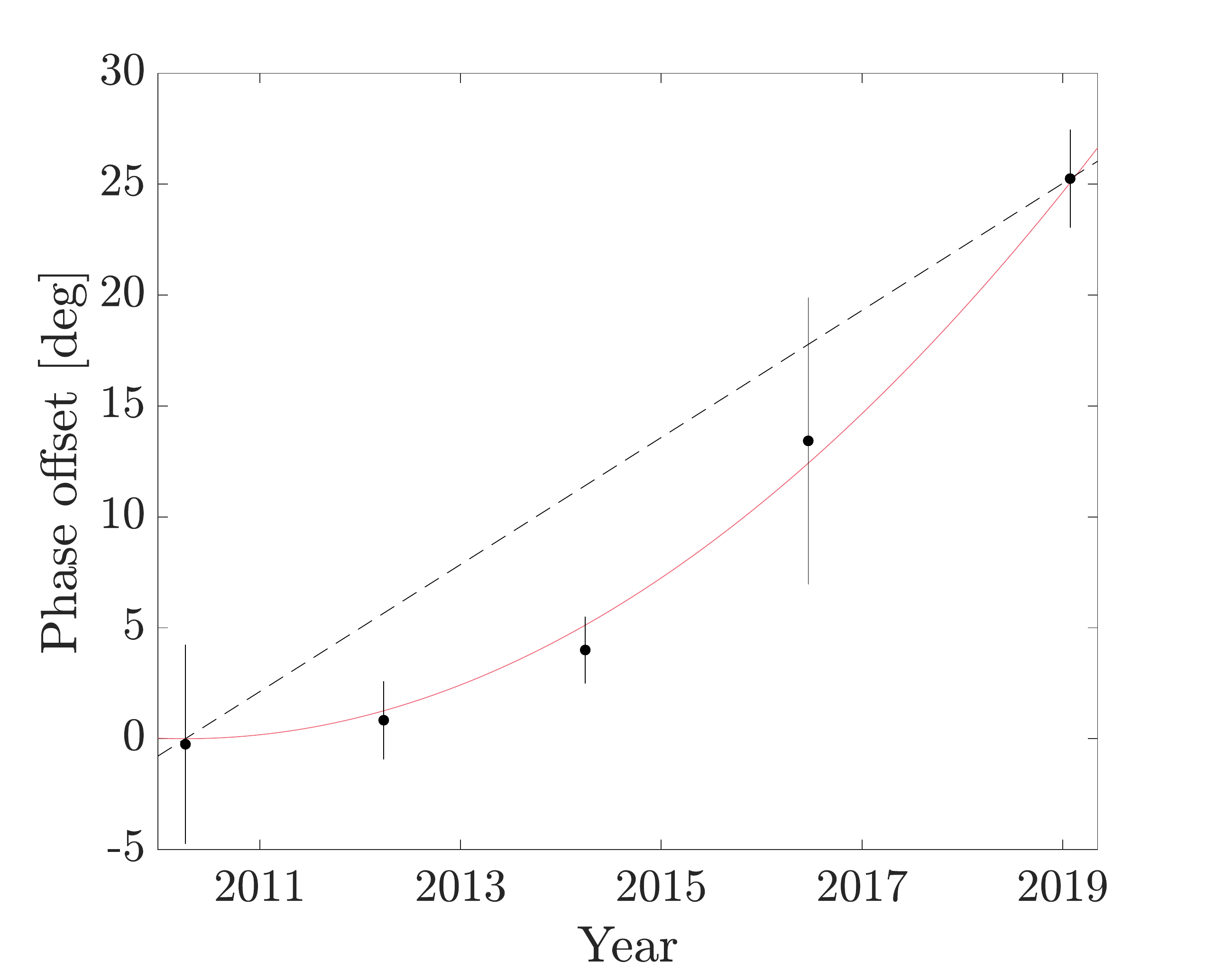}}
		\caption{Phase offset measurements for the non-convex radar-derived shape model of asteroid (68346) 2001~KZ66, with $\lambda={170}\degr$, $\beta={-85}\degr$, initial period $P=4.985997\pm0.000042 \, \rm hours$, and starting point $T_0=2455290.98269$ (April 2010). %
			The black circles represent averaged phase offset measurements for light curves grouped by year, and the associated uncertainties are given by the root-mean-square of the individual light curves' error within each year.
			The red solid line marks the best-fit YORP solution, $\nu=(8.43\pm0.69)\times10^{-8}\, \rm ~rad \, day^{-2}$. 
			The black dotted line is a straight line between the first and last points to highlight the deviation from a linear trend.}
		\label{fig:phoffs}
	\end{figure}

\begin{figure}
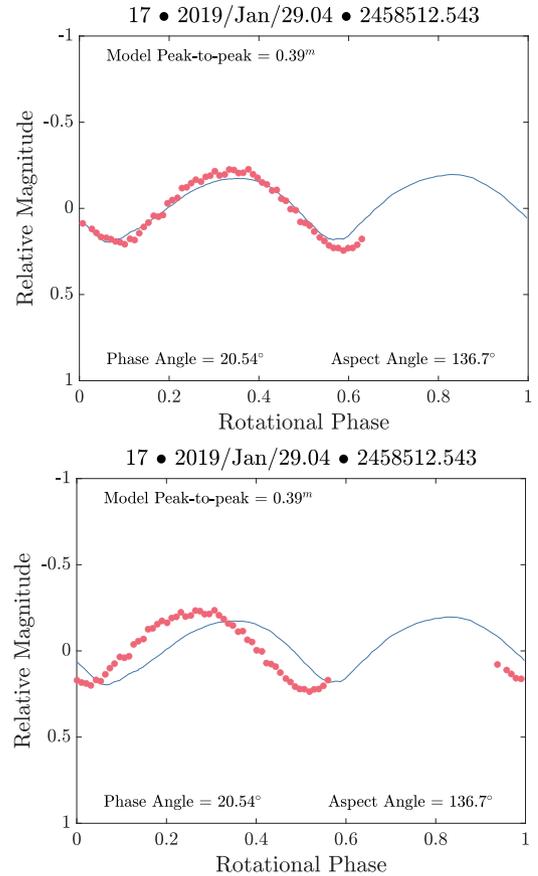

		\centering
		\resizebox{0.8\hsize}{!}{\includegraphics[width=12cm,trim=1.0cm 3.0cm 2.2cm 3cm, clip=true]{pictures/model_withYORP_190924_VertexModel__17}} \resizebox{0.8\hsize}{!}{\includegraphics[width=12cm,trim=1.0cm 3.0cm 2.2cm 3cm, clip=true]{pictures/model_withoutYORP_190924_VertexModel__17}}
		\caption{Example synthetic light curves generated using the radar-derived shape model of (68346) 2001 KZ66 (blue lines), with YORP (top) and without YORP (bottom), with the optical data over plotted (red dots). The complete dataset can be found in the Appendix.}
		\label{fig:yorpvsnoyorp}
	\end{figure}

	The results of the phase offset measurements are plotted in Fig. \ref{fig:phoffs}. In this figure the phase offsets have been grouped by similar epochs. The grouped phase offset is given by the mean of the individual light curve phase offsets. The uncertainties of grouped phase offsets are calculated as the root-mean-square of the phase offset errors of the light curves comprising the group. These grouped phase offset measurements result in a clear quadratic trend with a YORP strength of $\nu = (8.43\pm0.69) \times 10^{-8} \rm ~rad ~day^{-2}$, for a rotation period at $T_{0}$ of $4.985997 \pm0.000042\rm ~h$. 
	A figure of a sample light curve comparing the model with and without this YORP acceleration is plotted in Fig. \ref{fig:yorpvsnoyorp}.

	\section{Discussion}
	\label{sec:discussion}
	
	\subsection{Direct detection of YORP spin-up}
	By combining our detailed shape model with the optical light curves we measured a YORP acceleration of $(8.43\pm0.69) \times 10^{-8} \rm ~rad ~day^{-2}$. 
	This marks the eighth direct detection of YORP to date, and KZ66 is the fourth smallest asteroid of those with YORP detections; it is larger than asteroids YORP (2000 PH5), Itokawa, and Bennu (all detections are in Table \ref{tab:yorp:detections}).
	Intriguingly, all YORP detections to date have been positive accelerations (i.e. in the spin-up sense). 
	However, for a population of asteroids with randomized shapes and spin-states, YORP should produce both spin-up and spin-down cases.
	The probability of eight consecutive spin-up detections is therefore extremely small. This suggests that there is either a mechanism that favours YORP accelerations or a bias in the sample of YORP detections obtained to date. 
	One such mechanism for the preference of YORP spin-up is 'TYORP', which accounts for the thermal emission tangential to the surface from boulders \citep{Golubov:2012kt,2014ApJ...794...22G,Golubov:2017ky}. 
	However, while the TYORP process is certainly promising, further YORP detections are required to confirm this model.
	
	There may also be biases in the current sample.
	To date, the asteroids on which YORP has been detected have all been retrograde rotators; three out of eight of them have pole orientations within $10\degr$ of the southern ecliptic pole, and all are within $41\degr$. They all also have obliquities larger than $140\degr$ (Table \ref{tab:yorp:detections}). 
	With the exception of YORP and Bennu, all of the asteroids with YORP detections have elongated shapes \citep{Hudson:1999dt,Ostro:2004ep,2007Sci...316..274T,Kaasalainen:2007hq,Nolan:2013gj,Durech:2012bq,Durech:2018gg}. Under a rotational acceleration, like that of YORP, initially spherical rubble pile asteroids can be disrupted to form various shapes. The end state of this process ranges from ellipsoidal to bilobed shapes \citep{Sanchez:2018fg}.
	The magnitude and direction of the rotational torque induced by YORP is dependant on obliquity, but is also highly sensitive to the morphology. The asteroid shapes can roughly be classified into four types (I/II/III/IV) depending on their model response to YORP torque under zero-conductivity assumption \citep{Vokrouhlicky:2002cq}. The behaviour of both the spin and obliquity components of YORP for each type of asteroid vary with obliquity differently. In considering type I asteroids, the spin component of YORP is positive for obliquities of $0\degr$ to ${\sim}60\degr$ and ${\sim}120\degr$ to $180\degr$, with negative YORP falling in the region ${\sim}60\degr$ to ${\sim}120\degr$ \citep{Rubincam:2000fg,Vokrouhlicky:2002cq,Golubov:2019kv}. 
	
	Asteroids presenting large light curve amplitudes are favoured for direct detection of YORP as their rotation phases can be measured to a greater accuracy. This preference to obtain high-amplitude light curves limits the morphology and observation geometry of asteroids probed, which in turn restricts the type of YORP behaviour detected.

	\subsection{Gravitational slopes and topographic variation on (68346) 2001 KZ66}
\begin{figure}
		\centering
		\resizebox{\hsize}{!}{\includegraphics[trim=0cm 0cm 0cm 0cm, clip=true]{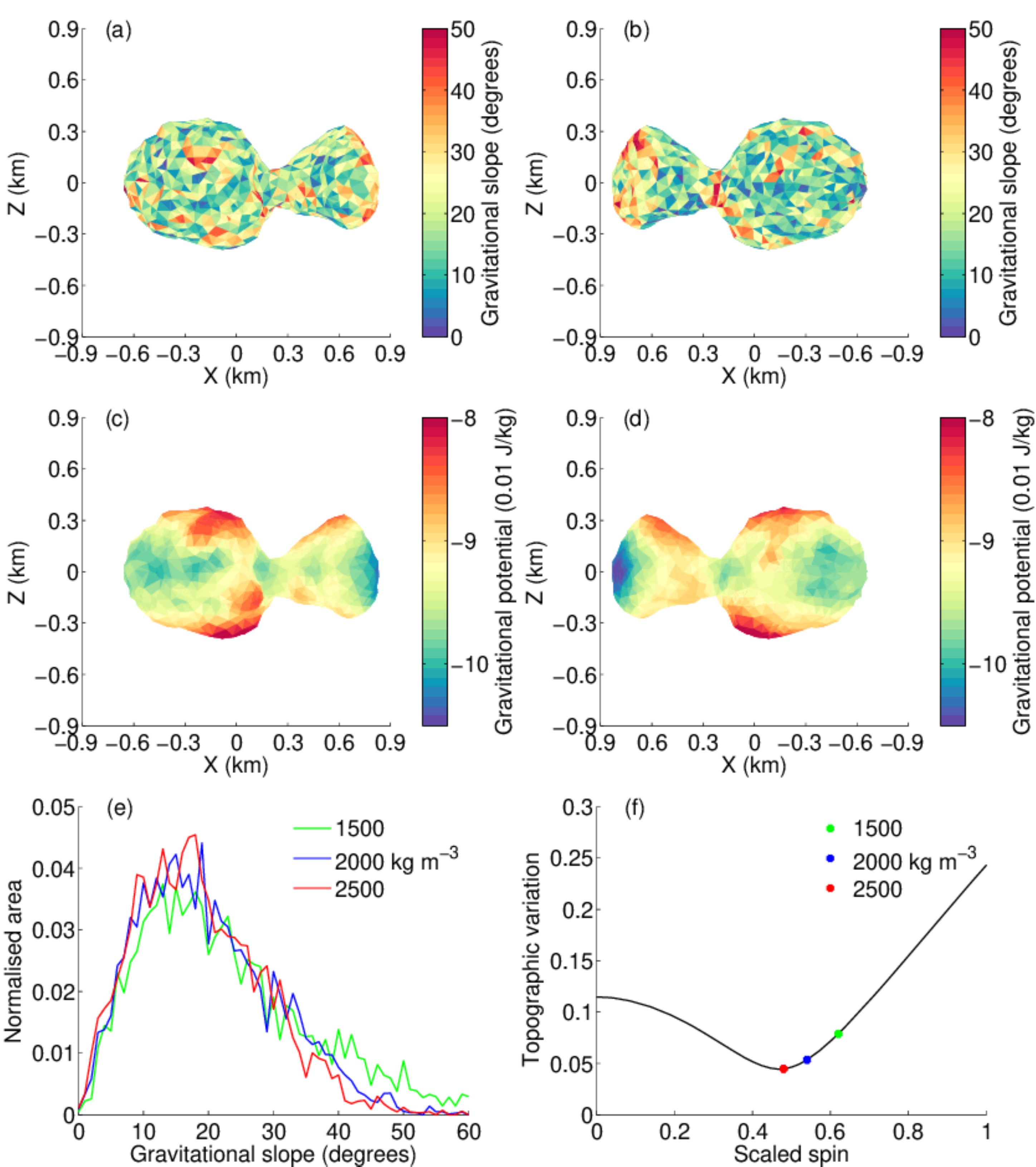}}
		\caption{Gravitational slopes and topographic variation on asteroid (68346) 2001 KZ66. (a) Gravitational slopes computed with the shape model assuming a bulk density of $ 2000 \rm ~kg ~m^{-3} $, view is along the positive y-axis. (b) Same as (a), but for view along the negative y-axis. (c) Gravitational potential computed with the shape model assuming a bulk density of $ 2000 \rm ~kg ~m^{-3} $, view is along the positive y-axis. (d) Same as (c), but view is along the negative y-axis. (e) Areal distribution of gravitational slope computed for three different values of bulk density. (f) Topographic variation in gravitational potential (i.e. the standard deviation of the gravitational potential variations normalised to the mean gravitational potential) as a function of scaled spin (black line). The asteroid's rotation period is fixed at the measured value, therefore, the scaled spin is solely a function of bulk density. The current topographic variation of KZ66 is identified for three different assumed values of bulk density (coloured data points).}
		\label{fig:topographic}
	\end{figure}

	The bifurcated shape of KZ66 with a small contact area between the two lobes raises questions of how stable its surface is against land sliding and other surface failure events, especially as YORP is very active on this body. To investigate this, we computed the gravitational potential and slopes of KZ66 using a polyhedron gravity model \citep{Werner:1996hv} that has been modified to account for rotational centrifugal forces \citep{Rozitis:2014bx}. The moderately high geometric albedo and average radar circular polarisation ratio of KZ66 suggests that it is likely to be an S-type asteroid \citep{Benner:2008dd}. Which agrees with the compositional characterisation of the asteroid obtained in the spectroscopic survey by \cite{deLeon:2010hl}. Therefore, we performed these calculations assuming uniform bulk density values of 1500, 2000, and 2500 $ \rm kg ~m^{-3} $ to cover the typical bulk density range for rubble-pile asteroids from this spectral class \citep{Carry:2012cw}. Fig. \ref{fig:topographic}a and \ref{fig:topographic}b show the gravitational slopes calculated from the shape model of KZ66, and Fig. \ref{fig:topographic}e shows their areal distribution. As shown, there are no large differences in the gravitational slopes between the neck region and the rest of the body. Furthermore, the majority of gravitational slopes are below $ 40\degr $, particularly for a bulk density of 2500 $ \rm kg ~m^{-3} $, which indicates that any land sliding occurring on the body would be rather limited in area, even if KZ66 lacked cohesion \citep{2015aste.book..767M}. 
	
	If land sliding did occur on KZ66 then it would cause mobilised material to migrate from areas of high gravitational potential to areas of low gravitational potential. The changes in shape and surface topography resulting from this material migration has the net effect of reducing the topographic variation in gravitational potential across the body. As such, deformable bodies prefer to exist in a state where this topographic variation is minimised \citep{Richardson:2014go,Richardson:2019jo}, and the YORP effect can induce migration of material when these bodies stray too far from this preferred state if a sufficient period of time has passed since the last migration \citep{Scheeres:2015hr}. To determine what topographic state KZ66 is currently in, we computed its topographic variation in gravitational potential as a function of scaled spin (i.e.  $ \omega / \sqrt{G\pi\rho} $) following the methodology outlined in \cite{Richardson:2019jo}. Fig. \ref{fig:topographic}c and \ref{fig:topographic}d show the spatial distribution of gravitational potential across the shape model of KZ66, and Fig. \ref{fig:topographic}f shows the functional dependence of KZ66’s topographic variation with scaled spin \citep{Holsapple:2004ff}. As shown, there are subtle variations in the gravitational potential across KZ66, particularly between its equator and poles, but intriguingly KZ66 currently exists at or near its preferred state where the topographic variation is minimised. The minima of the topographic variation is not a measurement of the object's bulk density, rather it represents an `erosional saddle-point', wherein the body is in its most eroded state for its current shape, topography, and spin-state \citep{Richardson:2014go}. However, as the detected YORP effect is causing KZ66 to spin-up, it will not exist in this state permanently. For instance, the scaled spin will be doubled in $ {\sim}1 \rm Myr $ at the current rate of YORP spin-up, which would lead to a factor of $ {\sim}5 $ increase in the amount of topographic variation. Therefore, whilst KZ66 seems to be stable in its current state, the YORP effect will eventually induce changes in its shape and surface topography. It is possible that the induced shape and topography changes would cause the YORP effect to switch from spin-up to spin-down \citep{CottoFigueroa:2015gp}, but if spin-up were to continue then KZ66 would ultimately fission to form an unbound asteroid pair \citep{Scheeres:2007io}.

	\subsection{Bifurcated shape of (68346) 2001 KZ66}
	We have observed the majority of KZ66's surface during the 2003 approach with Arecibo as shown in Fig \ref{fig:vertexshape}. With a median facet edge length of $57 \rm ~m$, the shape of large-scale topographical features are reported with confidence. KZ66 has a distinct bifurcated shape, which is dissimilar to most other contact binaries with radar shape models such as Kleopatra, and Itokawa as it has a much sharper concavity \citep{Shepard:2018eh,Ostro:2004ep}. In this regard, KZ66's shape bears more similarity with the \acp{NEA} 1996 HW1 \citep{Magri:2011fd} or 1999 JV6 \citep{Rozek:2019kp}. Other objects such as the Kuiper Belt Object 2014 MU69 \citep{Stern:2019fj} or comets such as 8P/Tuttle, 19P/Borrelly, 67P/Churyumov-Gerasimenko, and 103P/Hartley 2 \citep{2010Icar..207..499H,Britt:2004dq,Jorda:2016eb,Thomas:2013hh} also have bilobate shapes.
	The sharp neck line of KZ66 suggests that the formation of this object was not a highly energetic event, and both lobes must have merged gently.
	There are several mechanisms capable of forming a contact-binary asteroid like KZ66. We will briefly summarise the mechanisms capable of forming a contact-binary asteroid.

	One mechanism that can lead to the formation of a contact-binary is the collapse of a binary asteroid system. 
	If this collapse occurred at a low velocity the asteroid would preserve the bilobed shape and avoid the deformation that would occur with a catastrophic collapse. There are several possible processes that lead to the formation of binary asteroids. One is mutual capture which requires the components having relative speeds below their mutual escape velocities which are typically on the order of m/s. However, present-day relative speeds for the Main Belt and near-Earth asteroids are on the order of km/s. Hence such a scenario is extremely unlikely in  today's populations of asteroids \citep{Richardson:2006fo}. 
	Binaries can be formed from a single body rotationally fissioned as rotational acceleration leads the asteroid towards the spin-limit barrier for gravitational aggregates \citep{Pravec:2000dr}. Accelerated by YORP, an asteroid would reconfigure its shape before eventually fissioning to form a binary asteroid \citep{Jacobson:2011eq}. A good example of this is the \ac{NEA} 1994 KW4 with its rapidly rotating primary \citep{Ostro:2006dq}. Once the binary had been formed, binary-YORP or BYORP causes a decay in the orbital semi-major axis until the secondary gently collides with the primary and settles \citep{Cuk:2005hb}.

	Collisions between unbound pairs of asteroids are another mechanism that alter their shapes. Studies of catastrophic collisions show that they can form a large spectrum of shapes including contact-binaries \citep{Michel:2013kn,Sugiura:2018fy,Schwartz:2018jx}. Far more common, by at least an order of magnitude, are sub-catastrophic collisions in which at least 50\% of the impacted asteroid remains gravitationally bound \citep{Jutzi:2017ge}. \cite{Jutzi:2019ii} showed that these more frequent sub-catastrophic collisions between an ellipsoidal porous rubble-pile asteroid and a hyper-velocity impactor are able to change the overall structure and shape of the impacted asteroid. If the impactor strikes the centre of an ellipsoid asteroid, it can split into two separate components which during re-accumulation can form Itokawa-like contact-binary asteroids. Collisions between asteroids are also capable of forming binary systems, either by a collisionally induced rotational fission of the parent body due to a glancing impact, or gravitationally bound ejecta resulting from the collision between two asteroids \citep{2015aste.book..375W}. Formation via disruption is far more likely than collisionally induced rotational fission \citep{2002aste.book..289M}. 
	An additional mechanism for the formation of a contact-binary is the rotational evolution of a self-gravitating spherical aggregate with a weak core. \cite{Sanchez:2018fg} consider an inhomogeneous spherical asteroid with a concentric core that is weaker than its outer shell. The inclusion of a weak core means that by the time the shell starts to fail, the core will not provide any resistance. As the spherical asteroid is rotationally accelerated the core and shells start to deform asymmetrically, this is particularly prominent when the radius of the core is equal to half of the total radius of the asteroid. In this case, the shell develops a dent and the core becomes very deformed. When the simulations are continued the asteroid then starts to stretch to form a non-ellipsoidal shape with a distinct ``head'' and ``body''. The shape at this stage bears similarities to the asteroid Itokawa and the authors suggest this as a formation mechanism for Itokawa. When advanced further, the concavity between the lobes continues to deepen before finally fissioning to form a binary asteroid. The configurations at each stage are stable and only change when the asteroid is spun-up further. Hence, with the YORP-induced acceleration observed in KZ66 and a more pronounced ``neck'', it is possible that it has advanced further along this fission process than Itokawa.

	For comets it has been suggested that erosion due to out-gassing could play a role in their morphology, 
	and may have contributed to the deep neck region seen on comet 67P/Churyumov-Gerasimenko, by
	the Rosetta spacecraft \citep{Sierks:2015db}. 
	However, with a geometric albedo of $0.291\pm0.110$, it is unlikely that KZ66 is an extinct comet. 
	These objects have dark surfaces with geometric albedos generally less than 0.05 \citep{2004come.book..223L}.
	An issue with some of the suggested formation mechanisms is that they require a fast rotation rate, whereas KZ66 has a long rotation period close to five hours. 
	However, asteroids migrate through different spin-states over YORP-cycles caused by structural and small-scale topographical changes \citep{Statler:2009fwa}, presumably caused by YORP torques and perturbations \citep{Scheeres:2018dx}. Therefore it is possible that KZ66's shape as seen today was formed during a previous YORP cycle where it had a faster rotation period. 
	Recent work by \cite{Golubov:2019kv} on the dynamical evolution of asteroids showed that for an idealised system, ignoring thermal inertia and tangential YORP (TYORP), the YORP cycle drives the asteroid from the tumbling regime to disruption at high rotation rates, or back to the tumbling regime. Depending on the shape and rotation-state of the asteroid they migrate from obliquities of 0\degr or 180\degr to an obliquity of 90\degr, or from 90\degr to either 0\degr or 180\degr \citep{Golubov:2019kv}. The inclusion of TYORP allowed stable equilibria states to exist where asteroids would cease to follow these YORP-cycles, although until they encounter these equilibria they continue to migrate between tumbling states or disruption. With an obliquity of 158.5\degr \footnote{JPL solution number 206 from the Horizons ephemeris system (\url{https://ssd.jpl.nasa.gov/})}, it is possible that KZ66 has left the tumbling regime and is now in the process of migrating towards an obliquity of 90\degr.

	\begin{table*}
		\centering
		\begin{tabular}[tbp]{ccccccccc}
			\hline\hline \noalign{\smallskip}
			{Asteroid} & {Period} & {$\nu$} & {d}& Pole & Obliquity & Reference  \\
			& [h] & [ $\times10^{-8}~$rad day$^{-2}$] & [km] & [\degr] & [\degr] & \\
			\hline  \noalign{\smallskip}
			\multirow{2}{*}{YORP} &  \multirow{2}{*}{0.20283333(1)}  & \multirow{2}{*}{$349\pm10\%$} &  \multirow{2}{*}{0.113}  & \multirow{2}{*}{(180, -85)} & \multirow{2}{*}{174.3} & \cite{2007Sci...316..272L} \\
			& & & & & & \cite{2007Sci...316..274T} \\\noalign{\smallskip}
			\hline
			2001 KZ66 &4.985997(42) & $8.43\pm8\%$ & 0.797 & (170, -85) & 158.5 & This work \\\noalign{\smallskip}
			\hline
			\multirow{2}{*}{Apollo} & 3.0654476(30) & $5.3\pm25\%$ &  1.4 & (50, -71) & 159.6 & \cite{Kaasalainen:2007hq} \\
			& $3.065448(3)$ & $5.5\pm22\%$ & 1.45 & (48, -72) & 162.3 & \cite{Durech:2008bn}\\\noalign{\smallskip}
			\hline
			Bennu
			& 4.2960477(19)  & $4.61\pm40\%$  & 0.492 & (87, -65) & 159.6 & \cite{Nolan:2019eib} \\ \noalign{\smallskip}
			\hline
			Itokawa       & 12.132371(6) & $3.54\pm11\%$ & 0.33 & (128.5, -89.7) & 178.4 & \cite{Lowry:2014cb} \\\noalign{\smallskip}
			\hline
			Cacus  & 3.755067(2) & $1.9\pm16\%$ &  1.0 & (254, -62) & 143.2 & \cite{Durech:2018gg} \\ \noalign{\smallskip}
			\hline
			\multirow{2}{*}{Eger}   &  \multirow{2}{*}{5.710156(7)} & $1.4\pm14\%$  &  \multirow{2}{*}{{$\sim$}1.5} & \multirow{2}{*}{(226, -70)} & \multirow{2}{*}{155.6} & \cite{Durech:2012bq}  \\
			& & $1.1\pm15\%$ & & & & \cite{Durech:2018gg} \\
			\noalign{\smallskip}
			\hline
			Geographos     & 5.223336(2) & $1.15\pm4\%$ & 2.56 & (58, -49) & 149.9 & \cite{Durech:2008di}   \\
			\hline\noalign{\smallskip}
		\end{tabular}
		\caption{YORP detections to date. All detections of YORP as of May 2020 in order of YORP strength. The table lists: Asteroid's name, rotation period (with uncertainty given in parenthesis), YORP strength ($\nu$) and 1-$\sigma$ error, diameter of a sphere of equivalent volume, pole orientation ($\lambda$, $\beta$), orbital obliquity, and reference to published work. All obliquities were calculated using the pole orienations determined by the authors and the best oribtal solution from JPL Horizons as of Jan 2020. 
		}
		\label{tab:yorp:detections}
	\end{table*} 
	
	\section{Conclusions}
	\label{sec:conclusions}
	We have been monitoring the Apollo \ac{PHA} (68346) 2001 KZ66 for ten years during the period 2010 to 2019, obtaining ten optical light curves. 
	We also have two nights of radar observations from the Arecibo Observatory taken in 2003. 
	With these data and published optical light curves we have derived a robust shape model of KZ66. 
	KZ66 has a distinct bifurcated shape comprising a large ellipsoidal component joined by a sharp concavity to a smaller non-ellipsoidal curved component. 
	We have discussed four different formation mechanisms that could have played a role in the evolution of KZ66 - collapse of a binary system, rotational deformation, re-formation after collision, erosion. 
	We were able to rule out one of them by considering the geometric albedo of KZ66 and it is unlikely that out-gassing was responsible for the morphology that we see today. 
	The stability of KZ66's shape has also been discussed by calculating its gravitational slopes and investigating the topographic variation. KZ66 was found to currently exist at or near its preferred state with minimised topographic variation, where regolith is unlikely to migrate from areas of high potential energy to those of low potential.
	
	Using the radar-derived shape model we detected an acceleration of the asteroid's rotation rate which can be attributed to YORP. Using ten years of light curve data, the light-curve-only analysis resulted in a large range of possible YORP strengths with the best value at $(7.7^{+3.8}_{-13.2}) \times 10^{-8}\rm ~rad ~day^{-2}$. However, by combining the optical light-curves with our radar-derived shape model, we found that the model required a YORP strength of $(8.43\pm0.69) \times 10^{-8} \rm ~rad ~day^{-2}$ with an initial rotation period of $4.985997\pm0.000042 \rm ~h$ at epoch $ 2455290.98269 $ JD to fit all of our data. 
	This detection marks the eighth direct detection of YORP, all of which are positive accelerations.
	
	The SC/OC polarisation ratio of $0.220 \pm 0.003$ determined for KZ66 from the Arecibo cw spectra shows that it is a typical representative of the NEA population. Compared to other contact binaries with shape models, KZ66 has the lowest recorded value. A simplistic interpretation of this would indicate that its surface roughness is smoother than Itokawa's at the cm-to-m scale, but the relation between radar circular polarisation ratio and surface roughness has recently been shown to be more complex \citep{Lauretta:2019bn}. 
		
	KZ66 most likely formed as a result of either the gentle merging of both components due to the collapse of a binary system, or from the deformation of a rubble pile with a weak-tensile-strength core due to YORP spin-up. We rule out outgassing as a mechanism for producing its distinctive shape, as the geometric albedo of the object is inconsistent that of an extinct comet \citep{2004come.book..223L}. This shape currently exists in a stable configuration close to its minimum in topographic variation, where regolith is unlikely to migrate from areas of higher potential energy . However, this will eventually change as the asteroid is further accelerated by YORP. 

	Further insight into the formation of this object could be obtained by determining if the composition of both lobes are homogeneous; however, more data is required to search for compositional variation between the lobes. A more robust determination of the asteroid's bulk density could be determined via a measurement of Yarkovsky orbital drift \citep{Chesley:2003bk,Chesley:2014eg,Hanus:2018bm}. With a stronger constraint on the bulk density, we would be able to further narrow the region on which KZ66 resides on the topographic variation curve, allowing us to ascertain the stability of the asteroid's current state. In the future, we plan to perform a thermophysical analysis to determine the theoretical YORP strength, which could lead us to discover the need for heterogeneity to reconcile the theoretical and observed values - a method used by \cite{Lowry:2014cb} to determine the density inhomogeneity for the asteroid Itokawa. This will help deduce whether or not this asteroid was formed by one or more bodies. Another method to indicate differences between the lobes is to search for V-R colour variations over a complete rotation of the asteroid. With the addition of an observed standard star, these observations could also determine an accurate value of the asteroid's absolute magnitude, which has large variations reported in the literature \citep{deLeon:2010hl,Masiero:2017fo}. Alternatively, rotationally resolved spectral measurements of the asteroid could be obtained to determine the composition of each lobe. 
	We also plan to continue periodically monitoring KZ66 optically. 
	The orbit of KZ66 is such that the object is observable biennially from 2023, regularly reaching 19 magnitude or brighter. 
	These additional observations of KZ66 could be used to refine the YORP detection presented in this paper. 
	In the meantime, the shape model that we have developed can be used to further study the formation mechanisms of binary asteroids.
	
	\section*{Acknowledgements}
	
	We thank all the staff at the observatories involved in this study for their support.
	Based on observations collected at the European Organisation for Astronomical Research in the Southern Hemisphere under ESO programmes 185.C-1033(A, L, O, X) and 0102.C-0375(B).
	The Isaac Newton Telescope is operated on the island of La Palma by the Isaac Newton Group of Telescopes in the Spanish Observatorio del Roque de los Muchachos of the Instituto de Astrofísica de Canarias. The \ac{WFC} observations were obtained as part of I/2012A/08. 
	The authors are sincerely grateful for the Arecibo Observatory's telescope operations, electronics, and maintenance teams for making these observations possible. The Arecibo Planetary Radar Program is fully supported by NASA’s Near-Earth Object Observations Program in NASA’s Planetary Defense Coordination Office through grant no. 80NSSC19K0523 awarded to the University of Central Florida (UCF). The Arecibo Observatory is a facility of the US National Science Foundation operated under cooperative agreement by UCF, in alliance with Yang Enterprises, Inc., and Universidad Ana G. Mendéz
	We would also like to thank Sam Duddy, A. A. Hine, and V. Negrón for the observations that they performed which contribute to this work. 
	TJZ, SCL, AR, BR, SFG, and CS gratefully acknowledge support from the UK Science and Technology Facilities Council. 
	This work made use of the NASA/JPL HORIZONS ephemeris-generating programme. 
	All image reduction and processing were performed using the Image Reduction and Analysis Facility (IRAF) \citep{Tody:1986df,Tody:1993vm}. 
	IRAF is distributed by the National Optical Astronomy Observatories, which are operated by the Association of Universities for Research in Astronomy, Inc., under cooperative agreement with the National Science Foundation.
	The work makes use of the Asteroid Lightcurve Photometry Database repository \cite{Warner:2011ty}.
	We thank Chris Magri for providing the SHAPE software package.

	\section*{Data Availability}

	The light curve utilised in this article are available on VizieR at \url{https://vizier.u-strasbg.fr}. 
The shape model of the asteroid produced from this work is also available online at \url{https://3d-asteroids.space}, where it can be found and downloaded by searching for the asteroid by name.

	\bibliographystyle{mnras}
	\bibliography{KZ66_arxiv.bib} %

\begin{thebibliography}{}
\makeatletter
\relax
\def\mn@urlcharsother{\let\do\@makeother \do\$\do\&\do\#\do\^\do\_\do\%\do\~}
\def\mn@doi{\begingroup\mn@urlcharsother \@ifnextchar [ {\mn@doi@}
  {\mn@doi@[]}}
\def\mn@doi@[#1]#2{\def\@tempa{#1}\ifx\@tempa\@empty \href
  {http://dx.doi.org/#2} {doi:#2}\else \href {http://dx.doi.org/#2} {#1}\fi
  \endgroup}
\def\mn@eprint#1#2{\mn@eprint@#1:#2::\@nil}
\def\mn@eprint@arXiv#1{\href {http://arxiv.org/abs/#1} {{\tt arXiv:#1}}}
\def\mn@eprint@dblp#1{\href {http://dblp.uni-trier.de/rec/bibtex/#1.xml}
  {dblp:#1}}
\def\mn@eprint@#1:#2:#3:#4\@nil{\def\@tempa {#1}\def\@tempb {#2}\def\@tempc
  {#3}\ifx \@tempc \@empty \let \@tempc \@tempb \let \@tempb \@tempa \fi \ifx
  \@tempb \@empty \def\@tempb {arXiv}\fi \@ifundefined
  {mn@eprint@\@tempb}{\@tempb:\@tempc}{\expandafter \expandafter \csname
  mn@eprint@\@tempb\endcsname \expandafter{\@tempc}}}

\bibitem[\protect\citeauthoryear{Aznar~Macias, Predatu, Vaduvescu  \&
  Oey}{Aznar~Macias et~al.}{2017}]{Aznar:2017ro}
Aznar~Macias A.,  Predatu M.,  Vaduvescu O.,   Oey J.,  2017, Romanian Journal
  of Physics, 62, 904

\bibitem[\protect\citeauthoryear{Benner, Nolan, Ostro, Giorgini, Pray, Harris,
  Magri  \& Margot}{Benner et~al.}{2006}]{Benner:2006du}
Benner L.,  Nolan M.~C.,  Ostro S.~J.,  Giorgini J.~D.,  Pray D.~P.,  Harris
  A.~W.,  Magri C.,   Margot J.-L.,  2006, Icarus, 182, 474

\bibitem[\protect\citeauthoryear{Benner et~al.,}{Benner
  et~al.}{2008}]{Benner:2008dd}
Benner L.,  et~al., 2008, Icarus, 198, 294

\bibitem[\protect\citeauthoryear{Britt et~al.,}{Britt
  et~al.}{2004}]{Britt:2004dq}
Britt D.~T.,  et~al., 2004, Icarus, 167, 45

\bibitem[\protect\citeauthoryear{Carry}{Carry}{2012}]{Carry:2012cw}
Carry B.,  2012, Planetary and Space Science, 73, 98

\bibitem[\protect\citeauthoryear{Chesley et~al.,}{Chesley
  et~al.}{2003}]{Chesley:2003bk}
Chesley S.~R.,  et~al., 2003, Science, 302, 1739

\bibitem[\protect\citeauthoryear{Chesley et~al.,}{Chesley
  et~al.}{2014}]{Chesley:2014eg}
Chesley S.~R.,  et~al., 2014, Icarus, 235, 5

\bibitem[\protect\citeauthoryear{Cotto-Figueroa, Statler, Richardson  \&
  Tanga}{Cotto-Figueroa et~al.}{2015}]{CottoFigueroa:2015gp}
Cotto-Figueroa D.,  Statler T.~S.,  Richardson D.~C.,   Tanga P.,  2015, The
  Astrophysical Journal, 803, 25

\bibitem[\protect\citeauthoryear{{\'{C}}uk \& Burns}{{\'{C}}uk \&
  Burns}{2005}]{Cuk:2005hb}
{\'{C}}uk M.,  Burns J.~A.,  2005, Icarus, 176, 418

\bibitem[\protect\citeauthoryear{{\v{D}}urech et~al.,}{{\v{D}}urech
  et~al.}{2008a}]{Durech:2008bn}
{\v{D}}urech J.,  et~al., 2008a, Astronomy {\&} Astrophysics, 488, 345

\bibitem[\protect\citeauthoryear{{\v{D}}urech et~al.,}{{\v{D}}urech
  et~al.}{2008b}]{Durech:2008di}
{\v{D}}urech J.,  et~al., 2008b, Astronomy {\&} Astrophysics, 489, L25

\bibitem[\protect\citeauthoryear{{\v{D}}urech et~al.,}{{\v{D}}urech
  et~al.}{2012}]{Durech:2012bq}
{\v{D}}urech J.,  et~al., 2012, Astronomy {\&} Astrophysics, 547, A10

\bibitem[\protect\citeauthoryear{{\v{D}}urech et~al.,}{{\v{D}}urech
  et~al.}{2018}]{Durech:2018gg}
{\v{D}}urech J.,  et~al., 2018, Astronomy {\&} Astrophysics, 609, A86

\bibitem[\protect\citeauthoryear{Golubov}{Golubov}{2017}]{Golubov:2017ky}
Golubov O.,  2017, The Astronomical Journal, 154, 238

\bibitem[\protect\citeauthoryear{Golubov \& Krugly}{Golubov \&
  Krugly}{2012}]{Golubov:2012kt}
Golubov O.,  Krugly Y.~N.,  2012, The Astrophysical Journal, 752, L11

\bibitem[\protect\citeauthoryear{Golubov \& Scheeres}{Golubov \&
  Scheeres}{2019}]{Golubov:2019kv}
Golubov O.,  Scheeres D.~J.,  2019, The Astronomical Journal, 157, 105

\bibitem[\protect\citeauthoryear{Golubov, Scheeres  \& Krugly}{Golubov
  et~al.}{2014}]{2014ApJ...794...22G}
Golubov O.,  Scheeres D.~J.,   Krugly Y.~N.,  2014, The Astrophysical Journal,
  794, 22

\bibitem[\protect\citeauthoryear{Hanu{\v s} et~al.,}{Hanu{\v s}
  et~al.}{2018}]{Hanus:2018bm}
Hanu{\v s} J.,  et~al., 2018, Astronomy {\&} Astrophysics, 620, L8

\bibitem[\protect\citeauthoryear{{Harmon}, {Nolan}, {Giorgini}  \&
  {Howell}}{{Harmon} et~al.}{2010}]{2010Icar..207..499H}
{Harmon} J.~K.,  {Nolan} M.~C.,  {Giorgini} J.~D.,   {Howell} E.~S.,  2010,
  \mn@doi [\icarus] {10.1016/j.icarus.2009.12.026}, \href
  {https://ui.adsabs.harvard.edu/abs/2010Icar..207..499H} {207, 499}

\bibitem[\protect\citeauthoryear{Holsapple}{Holsapple}{2004}]{Holsapple:2004ff}
Holsapple K.~A.,  2004, Icarus, 172, 272

\bibitem[\protect\citeauthoryear{Hudson}{Hudson}{1993}]{Hudson:1993uo}
Hudson S.,  1993, Remote Sens. Rev., 8, 195

\bibitem[\protect\citeauthoryear{Hudson \& Ostro}{Hudson \&
  Ostro}{1999}]{Hudson:1999dt}
Hudson R.~S.,  Ostro S.~J.,  1999, Icarus, 140, 369

\bibitem[\protect\citeauthoryear{Jacobson \& Scheeres}{Jacobson \&
  Scheeres}{2011}]{Jacobson:2011eq}
Jacobson S.~A.,  Scheeres D.~J.,  2011, Icarus, 214, 161

\bibitem[\protect\citeauthoryear{Jorda et~al.,}{Jorda
  et~al.}{2016}]{Jorda:2016eb}
Jorda L.,  et~al., 2016, Icarus, 277, 257

\bibitem[\protect\citeauthoryear{Jutzi}{Jutzi}{2019}]{Jutzi:2019ii}
Jutzi M.,  2019, Planetary and Space Science, 177, 104695

\bibitem[\protect\citeauthoryear{Jutzi \& Benz}{Jutzi \&
  Benz}{2017}]{Jutzi:2017ge}
Jutzi M.,  Benz W.,  2017, Astronomy {\&} Astrophysics, 597, A62

\bibitem[\protect\citeauthoryear{Kaasalainen \& Torppa}{Kaasalainen \&
  Torppa}{2001}]{Kaasalainen:2001dx}
Kaasalainen M.,  Torppa J.,  2001, Icarus, 153, 24

\bibitem[\protect\citeauthoryear{Kaasalainen, Torppa  \& Muinonen}{Kaasalainen
  et~al.}{2001}]{Kaasalainen:2001di}
Kaasalainen M.,  Torppa J.,   Muinonen K.,  2001, Icarus, 153, 37

\bibitem[\protect\citeauthoryear{Kaasalainen, {\v{D}}urech, Warner, Krugly  \&
  Gaftonyuk}{Kaasalainen et~al.}{2007}]{Kaasalainen:2007hq}
Kaasalainen M.,  {\v{D}}urech J.,  Warner B.~D.,  Krugly Y.~N.,   Gaftonyuk
  N.~M.,  2007, Nature, 446, 420

\bibitem[\protect\citeauthoryear{Lamy, Toth, Fernandez  \& Weaver}{Lamy
  et~al.}{2004}]{2004come.book..223L}
Lamy P.~L.,  Toth I.,  Fernandez Y.~R.,   Weaver H.~A.,  2004, Comets II, pp
  223--264

\bibitem[\protect\citeauthoryear{Lauretta et~al.,}{Lauretta
  et~al.}{2019}]{Lauretta:2019bn}
Lauretta D.~S.,  et~al., 2019, \mn@doi [Nature] {10.1038/s41586-019-1033-6},
  568, 55

\bibitem[\protect\citeauthoryear{Lee et~al.,}{Lee et~al.}{2021}]{Lee:2021ei}
Lee H.-J.,  et~al., 2021, The Astronomical Journal, 161, 112

\bibitem[\protect\citeauthoryear{{\VAN{Leon}{De}{de}}~Le{\'o}n, Licandro,
  Serra-Ricart, Pinilla-Alonso  \& Campins}{{\VAN{Leon}{De}{de}}~Le{\'o}n
  et~al.}{2010}]{deLeon:2010hl}
{\VAN{Leon}{De}{de}}~Le{\'o}n J.,  Licandro J.,  Serra-Ricart M.,
  Pinilla-Alonso N.,   Campins H.,  2010, Astronomy {\&} Astrophysics, 517, A23

\bibitem[\protect\citeauthoryear{Lowry et~al.,}{Lowry
  et~al.}{2007}]{2007Sci...316..272L}
Lowry S.~C.,  et~al., 2007, Science, 316, 272

\bibitem[\protect\citeauthoryear{Lowry et~al.,}{Lowry
  et~al.}{2014}]{Lowry:2014cb}
Lowry S.~C.,  et~al., 2014, Astronomy {\&} Astrophysics, 562, A48

\bibitem[\protect\citeauthoryear{{Lowry} et~al.,}{{Lowry}
  et~al.}{2019}]{2019EPSC...13.1561L}
{Lowry} S.,  et~al., 2019, in EPSC-DPS Joint Meeting 2019. pp
  EPSC--DPS2019--1561

\bibitem[\protect\citeauthoryear{Magri, Ostro, Scheeres, Nolan, Giorgini,
  Benner  \& Margot}{Magri et~al.}{2007}]{Magri:2007io}
Magri C.,  Ostro S.~J.,  Scheeres D.~J.,  Nolan M.~C.,  Giorgini J.~D.,  Benner
  L.,   Margot J.-L.,  2007, Icarus, 186, 152

\bibitem[\protect\citeauthoryear{Magri et~al.,}{Magri
  et~al.}{2011}]{Magri:2011fd}
Magri C.,  et~al., 2011, Icarus, 214, 210

\bibitem[\protect\citeauthoryear{Masiero et~al.,}{Masiero
  et~al.}{2017}]{Masiero:2017fo}
Masiero J.~R.,  et~al., 2017, The Astronomical Journal, 154, 168

\bibitem[\protect\citeauthoryear{Merline, Weidenschilling, Durda, Margot,
  Pravec  \& Storrs}{Merline et~al.}{2002}]{2002aste.book..289M}
Merline W.~J.,  Weidenschilling S.~J.,  Durda D.~D.,  Margot J.~L.,  Pravec P.,
    Storrs A.~D.,  2002, Asteroids III, pp 289--312

\bibitem[\protect\citeauthoryear{Michel \& Richardson}{Michel \&
  Richardson}{2013}]{Michel:2013kn}
Michel P.,  Richardson D.~C.,  2013, Astronomy {\&} Astrophysics, 554, L1

\bibitem[\protect\citeauthoryear{Murdoch, S{\'a}nchez, Schwartz  \&
  Miyamoto}{Murdoch et~al.}{2015}]{2015aste.book..767M}
Murdoch N.,  S{\'a}nchez P.,  Schwartz S.~R.,   Miyamoto H.,  2015, Asteroids
  IV, pp 767--792

\bibitem[\protect\citeauthoryear{Nolan et~al.,}{Nolan
  et~al.}{2013}]{Nolan:2013gj}
Nolan M.~C.,  et~al., 2013, Icarus, 226, 629

\bibitem[\protect\citeauthoryear{Nolan et~al.,}{Nolan
  et~al.}{2019}]{Nolan:2019eib}
Nolan M.~C.,  et~al., 2019, Geophysical Research Letters, 46, 1956

\bibitem[\protect\citeauthoryear{Ostro}{Ostro}{1993}]{Ostro:1993jv}
Ostro S.~J.,  1993, Reviews of Modern Physics, 65, 1235

\bibitem[\protect\citeauthoryear{Ostro, Hudson, Benner, Giorgini, Magri, Margot
   \& Nolan}{Ostro et~al.}{2002}]{2002aste.book..151O}
Ostro S.~J.,  Hudson R.~S.,  Benner L. A.~M.,  Giorgini J.~D.,  Magri C.,
  Margot J.~L.,   Nolan M.~C.,  2002, Asteroids III, pp 151--168

\bibitem[\protect\citeauthoryear{Ostro et~al.,}{Ostro
  et~al.}{2004}]{Ostro:2004ep}
Ostro S.~J.,  et~al., 2004, Meteoritics {\&} Planetary Science, 39, 407

\bibitem[\protect\citeauthoryear{Ostro et~al.,}{Ostro
  et~al.}{2006}]{Ostro:2006dq}
Ostro S.~J.,  et~al., 2006, Science, 314, 1276

\bibitem[\protect\citeauthoryear{Pravdo et~al.,}{Pravdo
  et~al.}{1999}]{Pravdo:1999eh}
Pravdo S.~H.,  et~al., 1999, The Astronomical Journal, 117, 1616

\bibitem[\protect\citeauthoryear{Pravec \& Harris}{Pravec \&
  Harris}{2000}]{Pravec:2000dr}
Pravec P.,  Harris A.~W.,  2000, Icarus, 148, 12

\bibitem[\protect\citeauthoryear{Richardson \& Bowling}{Richardson \&
  Bowling}{2014}]{Richardson:2014go}
Richardson J.~E.,  Bowling T.~J.,  2014, Icarus, 234, 53

\bibitem[\protect\citeauthoryear{Richardson \& Walsh}{Richardson \&
  Walsh}{2006}]{Richardson:2006fo}
Richardson D.~C.,  Walsh K.~J.,  2006, Annual Review of Earth and Planetary
  Sciences, 34, 47

\bibitem[\protect\citeauthoryear{Richardson, Graves, Harris  \&
  Bowling}{Richardson et~al.}{2019}]{Richardson:2019jo}
Richardson J.~E.,  Graves K.~J.,  Harris A.~W.,   Bowling T.~J.,  2019, Icarus,
  329, 207

\bibitem[\protect\citeauthoryear{Ro{\.{z}}ek et~al.,}{Ro{\.{z}}ek
  et~al.}{2019a}]{Rozek:2019ij}
Ro{\.{z}}ek A.,  et~al., 2019a, Astronomy {\&} Astrophysics, 627, A172

\bibitem[\protect\citeauthoryear{Ro{\.{z}}ek et~al.,}{Ro{\.{z}}ek
  et~al.}{2019b}]{Rozek:2019kp}
Ro{\.{z}}ek A.,  et~al., 2019b, Astronomy {\&} Astrophysics, 631, A149

\bibitem[\protect\citeauthoryear{Rozitis, Duddy, Green  \& Lowry}{Rozitis
  et~al.}{2013}]{Rozitis:2013de}
Rozitis B.,  Duddy S.~R.,  Green S.~F.,   Lowry S.~C.,  2013, Astronomy {\&}
  Astrophysics, 555, A20

\bibitem[\protect\citeauthoryear{Rozitis, MacLennan  \& Emery}{Rozitis
  et~al.}{2014}]{Rozitis:2014bx}
Rozitis B.,  MacLennan E.,   Emery J.~P.,  2014, Nature, 512, 174

\bibitem[\protect\citeauthoryear{Rubincam}{Rubincam}{2000}]{Rubincam:2000fg}
Rubincam D.,  2000, Icarus, 148, 2

\bibitem[\protect\citeauthoryear{S{\'a}nchez \& Scheeres}{S{\'a}nchez \&
  Scheeres}{2018}]{Sanchez:2018fg}
S{\'a}nchez P.,  Scheeres D.~J.,  2018, Planetary and Space Science, 157, 39

\bibitem[\protect\citeauthoryear{Scheeres}{Scheeres}{2007}]{Scheeres:2007io}
Scheeres D.~J.,  2007, Icarus, 189, 370

\bibitem[\protect\citeauthoryear{Scheeres}{Scheeres}{2015}]{Scheeres:2015hr}
Scheeres D.~J.,  2015, Icarus, 247, 1

\bibitem[\protect\citeauthoryear{Scheeres}{Scheeres}{2018}]{Scheeres:2018dx}
Scheeres D.~J.,  2018, Icarus, 304, 183

\bibitem[\protect\citeauthoryear{Schwartz, Michel, Jutzi, Marchi, Zhang  \&
  Richardson}{Schwartz et~al.}{2018}]{Schwartz:2018jx}
Schwartz S.~R.,  Michel P.,  Jutzi M.,  Marchi S.,  Zhang Y.,   Richardson
  D.~C.,  2018, Nature Astronomy, 2, 379

\bibitem[\protect\citeauthoryear{Shepard et~al.,}{Shepard
  et~al.}{2018}]{Shepard:2018eh}
Shepard M.~K.,  et~al., 2018, Icarus, 311, 197

\bibitem[\protect\citeauthoryear{Sierks et~al.,}{Sierks
  et~al.}{2015}]{Sierks:2015db}
Sierks H.,  et~al., 2015, Science, 347, aaa1044

\bibitem[\protect\citeauthoryear{Statler}{Statler}{2009}]{Statler:2009fwa}
Statler T.~S.,  2009, Icarus, 202, 502

\bibitem[\protect\citeauthoryear{Stern et~al.,}{Stern
  et~al.}{2019}]{Stern:2019fj}
Stern S.~A.,  et~al., 2019, Science, 364, aaw9771

\bibitem[\protect\citeauthoryear{Sugiura, Kobayashi  \& Inutsuka}{Sugiura
  et~al.}{2018}]{Sugiura:2018fy}
Sugiura K.,  Kobayashi H.,   Inutsuka S.,  2018, Astronomy {\&} Astrophysics,
  620, A167

\bibitem[\protect\citeauthoryear{Taylor et~al.,}{Taylor
  et~al.}{2007}]{2007Sci...316..274T}
Taylor P.~A.,  et~al., 2007, Science, 316, 274

\bibitem[\protect\citeauthoryear{Thomas et~al.,}{Thomas
  et~al.}{2013}]{Thomas:2013hh}
Thomas P.~C.,  et~al., 2013, Icarus, 222, 550

\bibitem[\protect\citeauthoryear{Tody}{Tody}{1986}]{Tody:1986df}
Tody D.,  1986, in Crawford D.~L.,  ed., 1986 Astronomy Conferences. SPIE, pp
  733--17

\bibitem[\protect\citeauthoryear{Tody}{Tody}{1993}]{Tody:1993vm}
Tody D.,  1993, Astronomical Data Analysis Software and Systems II, 52, 173

\bibitem[\protect\citeauthoryear{Vokrouhlick{\'{y}} \& {\v
  C}apek}{Vokrouhlick{\'{y}} \& {\v C}apek}{2002}]{Vokrouhlicky:2002cq}
Vokrouhlick{\'{y}} D.,  {\v C}apek D.,  2002, Icarus, 159, 449

\bibitem[\protect\citeauthoryear{Walsh \& Jacobson}{Walsh \&
  Jacobson}{2015}]{2015aste.book..375W}
Walsh K.~J.,  Jacobson S.~A.,  2015, Asteroids IV, pp 375--393

\bibitem[\protect\citeauthoryear{Warner}{Warner}{2016}]{Warner:2016uq}
Warner B.~D.,  2016, Minor Planet Bulletin, 43, 311

\bibitem[\protect\citeauthoryear{Warner}{Warner}{2017}]{Warner:2017vb}
Warner B.~D.,  2017, Minor Planet Bulletin, 44, 22

\bibitem[\protect\citeauthoryear{Warner, Stephens  \& Harris}{Warner
  et~al.}{2011}]{Warner:2011ty}
Warner B.~D.,  Stephens R.~D.,   Harris A.~W.,  2011, The Minor Planet
  Bulletin, 38, 172

\bibitem[\protect\citeauthoryear{Werner \& Scheeres}{Werner \&
  Scheeres}{1997}]{Werner:1996hv}
Werner R.~A.,  Scheeres D.~J.,  1997, Celestial Mechanics and Dynamical
  Astronomy, 65, 313

\makeatother
\end{thebibliography}

	\appendix
	
	\section{Additional figures and tables}

	\begin{figure*}
		
		\resizebox{\hsize}{!}{	
			\includegraphics[width=.33\textwidth, trim=1cm 3.0cm 2.2cm 1.5cm, clip=true]{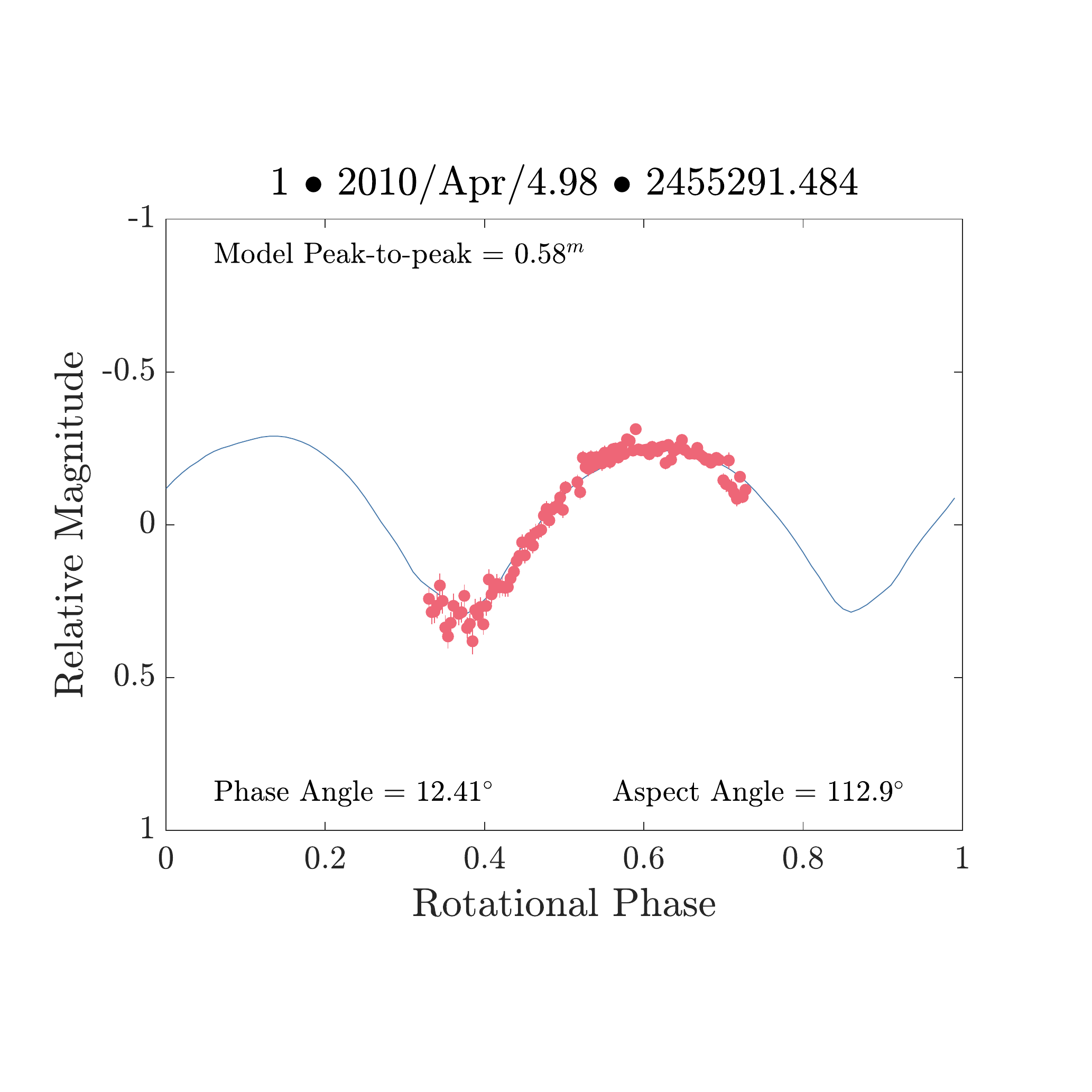}
			\includegraphics[width=.33\textwidth, trim=1cm 3.0cm 2.2cm 1.5cm, clip=true]{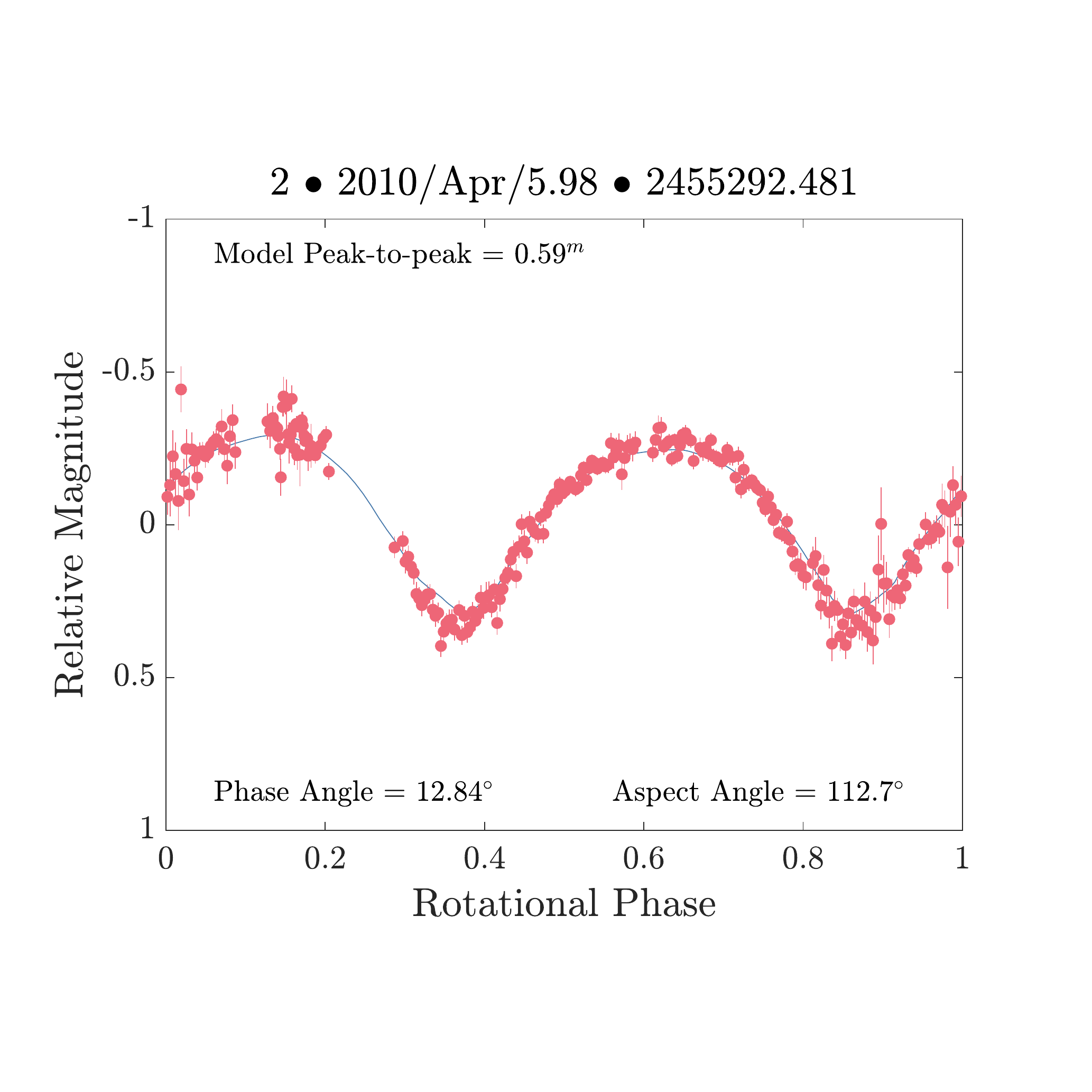}
			\includegraphics[width=.33\textwidth, trim=1cm 3.0cm 2.2cm 1.5cm, clip=true]{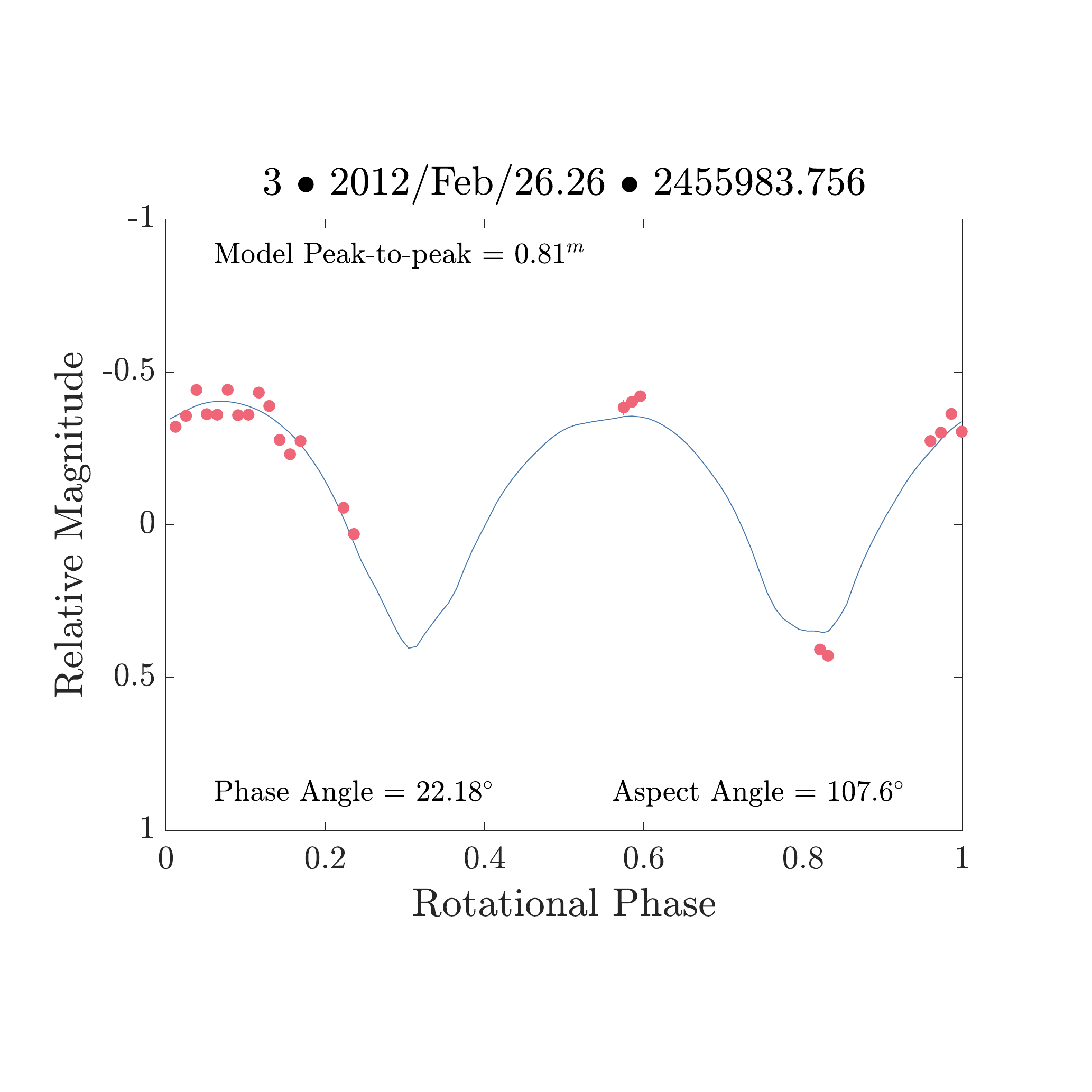}	
		}
		
		\resizebox{\hsize}{!}{	
			\includegraphics[width=.33\textwidth, trim=1cm 3.0cm 2.2cm 1.5cm, clip=true]{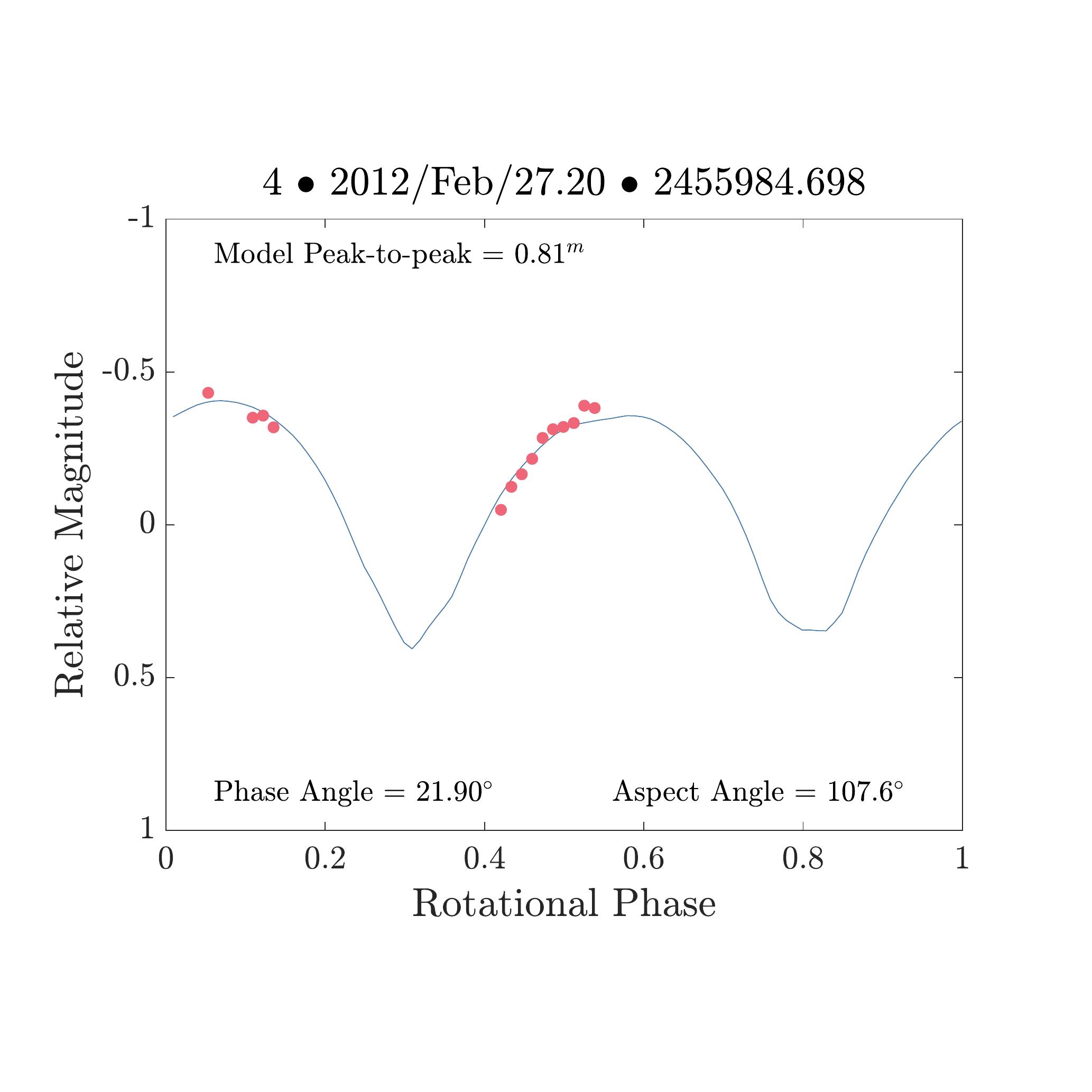}
			\includegraphics[width=.33\textwidth, trim=1cm 3.0cm 2.2cm 1.5cm, clip=true]{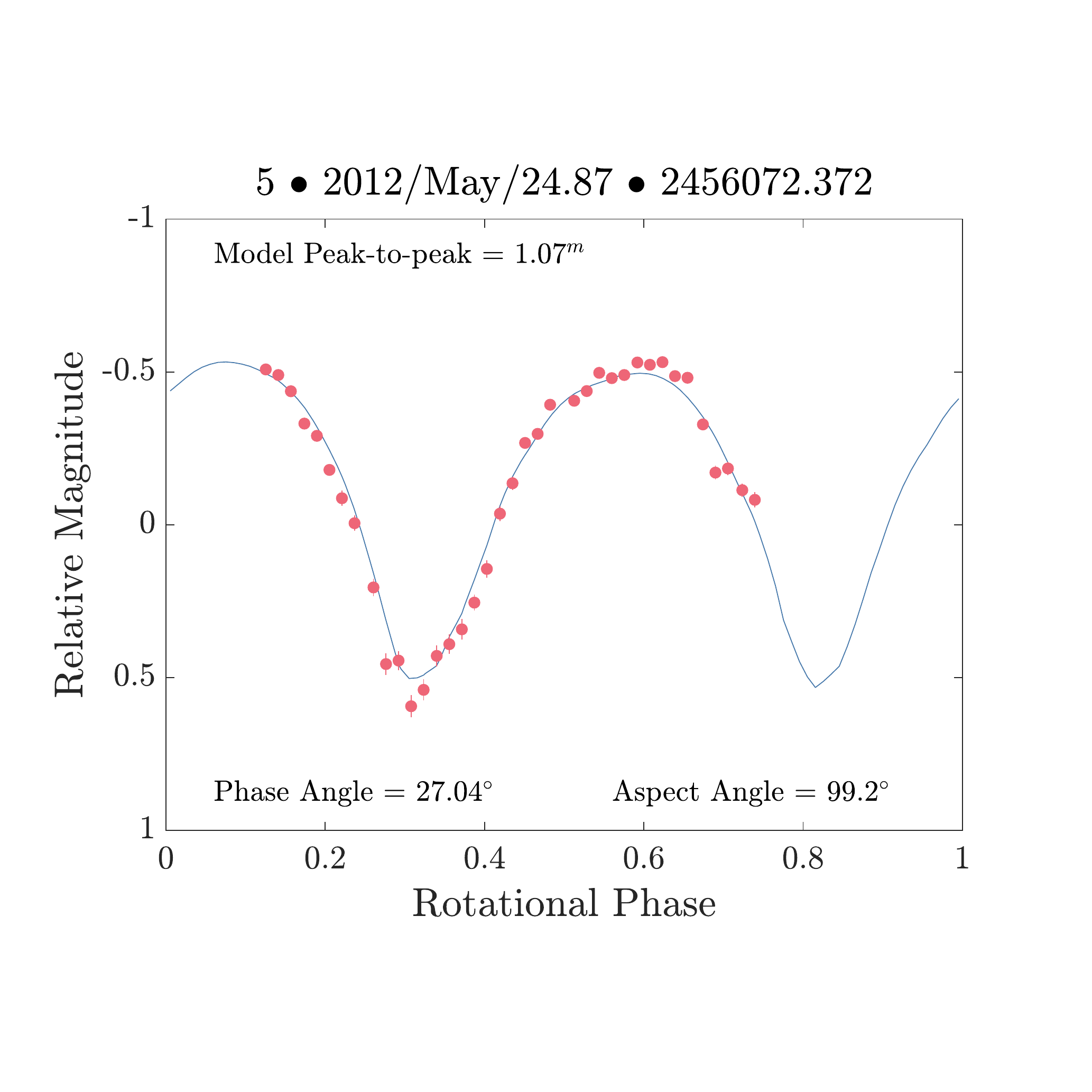}
			\includegraphics[width=.33\textwidth, trim=1cm 3.0cm 2.2cm 1.5cm, clip=true]{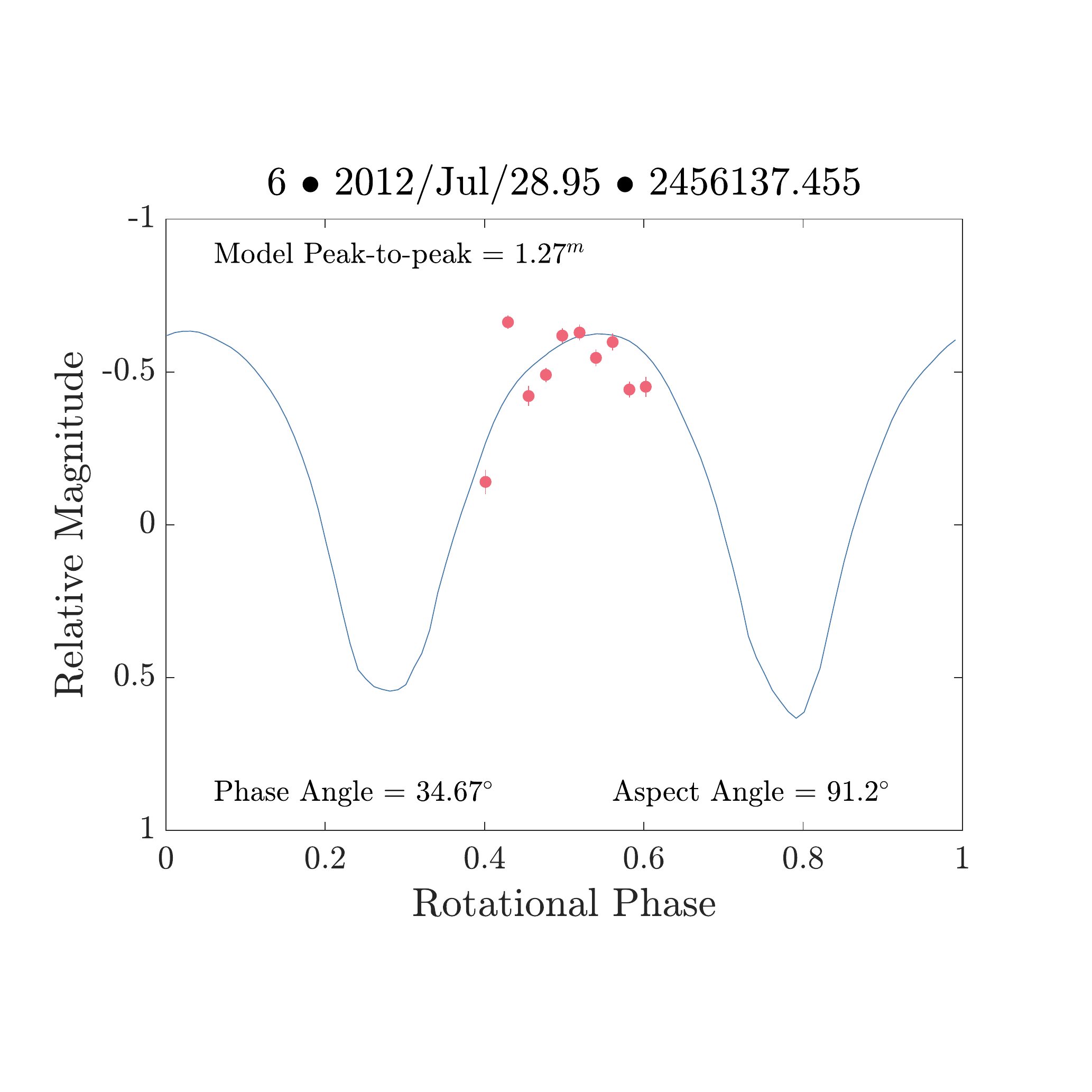}	
		}
		
		\resizebox{\hsize}{!}{	
			\includegraphics[width=.33\textwidth, trim=1cm 3.0cm 2.2cm 1.5cm, clip=true]{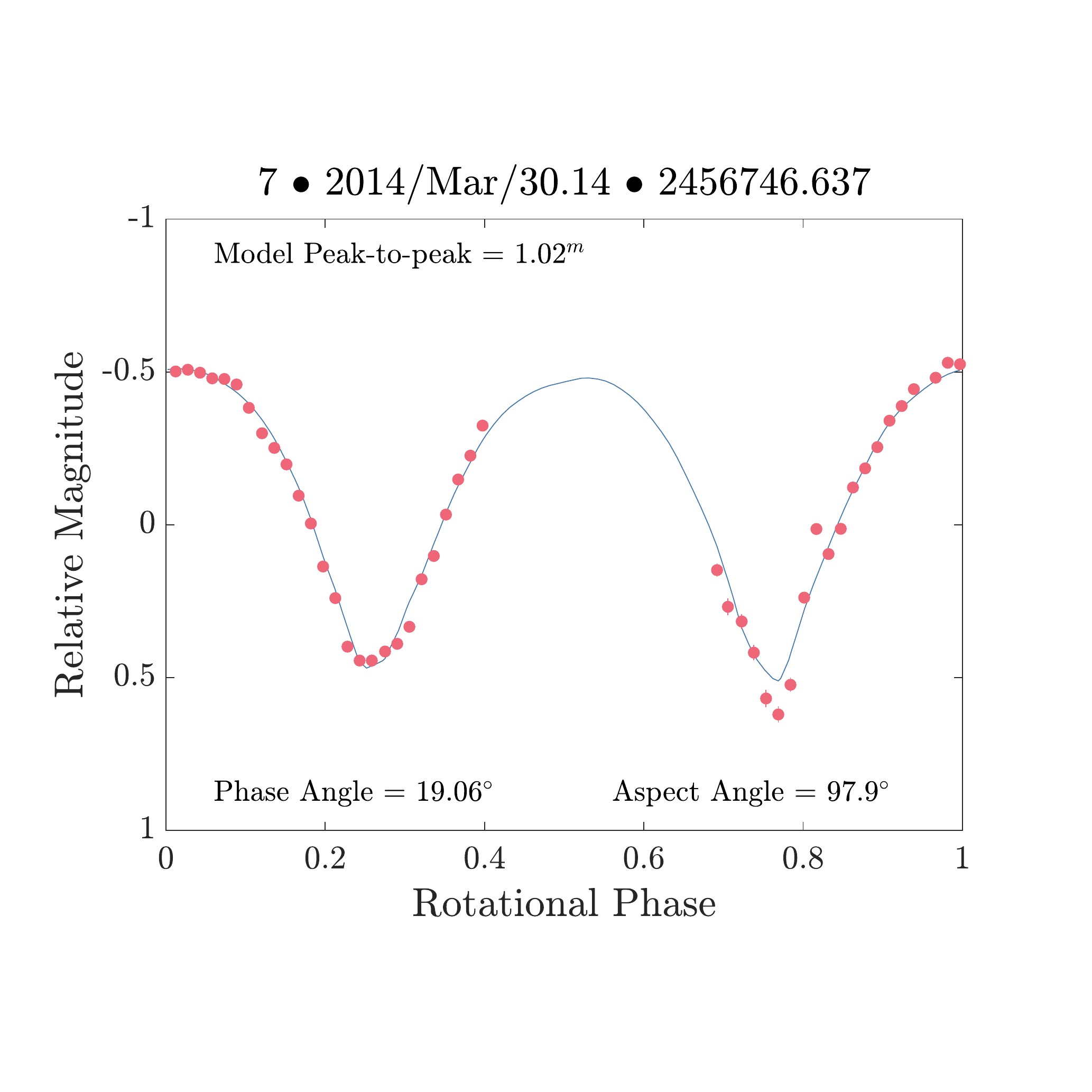}
			\includegraphics[width=.33\textwidth, trim=1cm 3.0cm 2.2cm 1.5cm, clip=true]{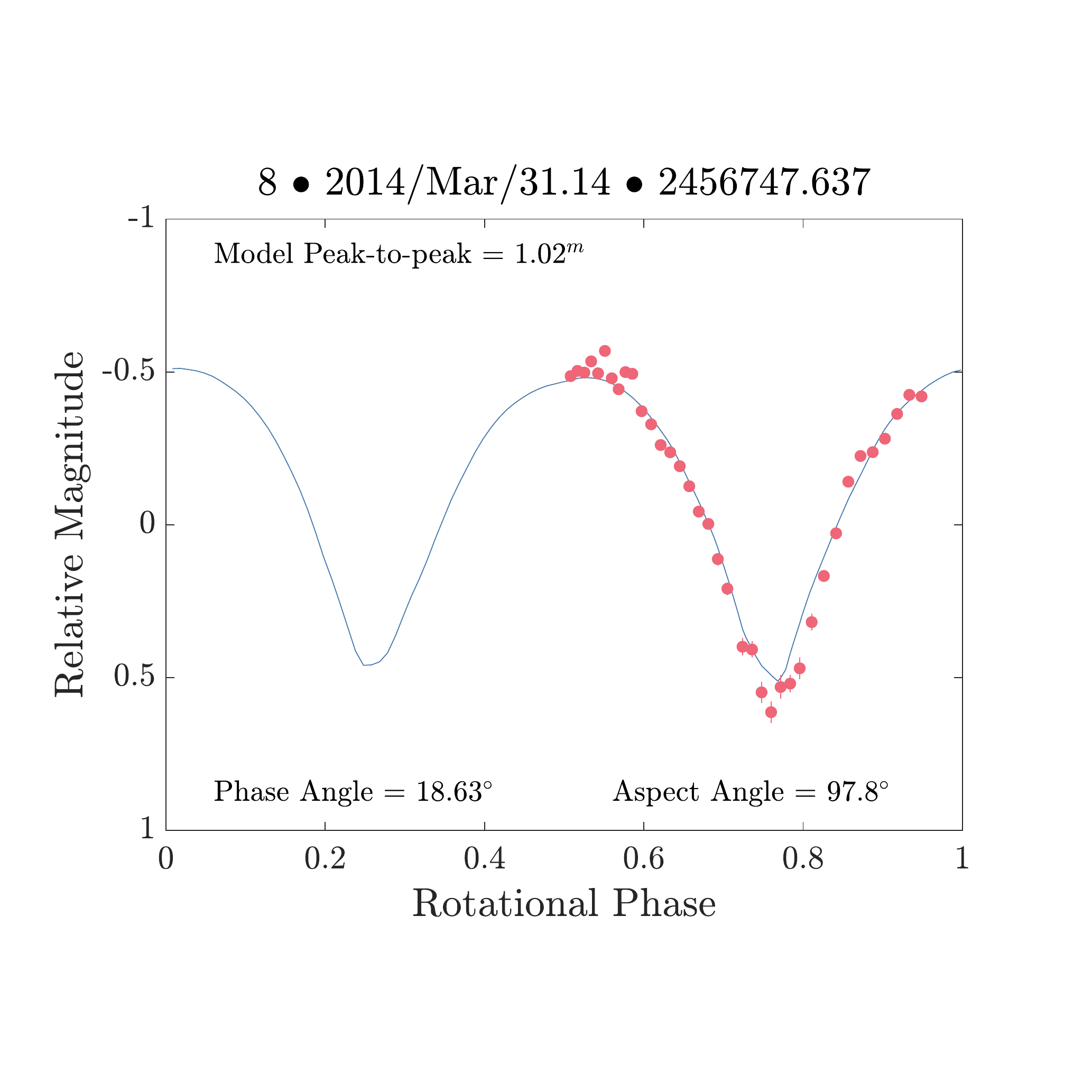}
			\includegraphics[width=.33\textwidth, trim=1cm 3.0cm 2.2cm 1.5cm, clip=true]{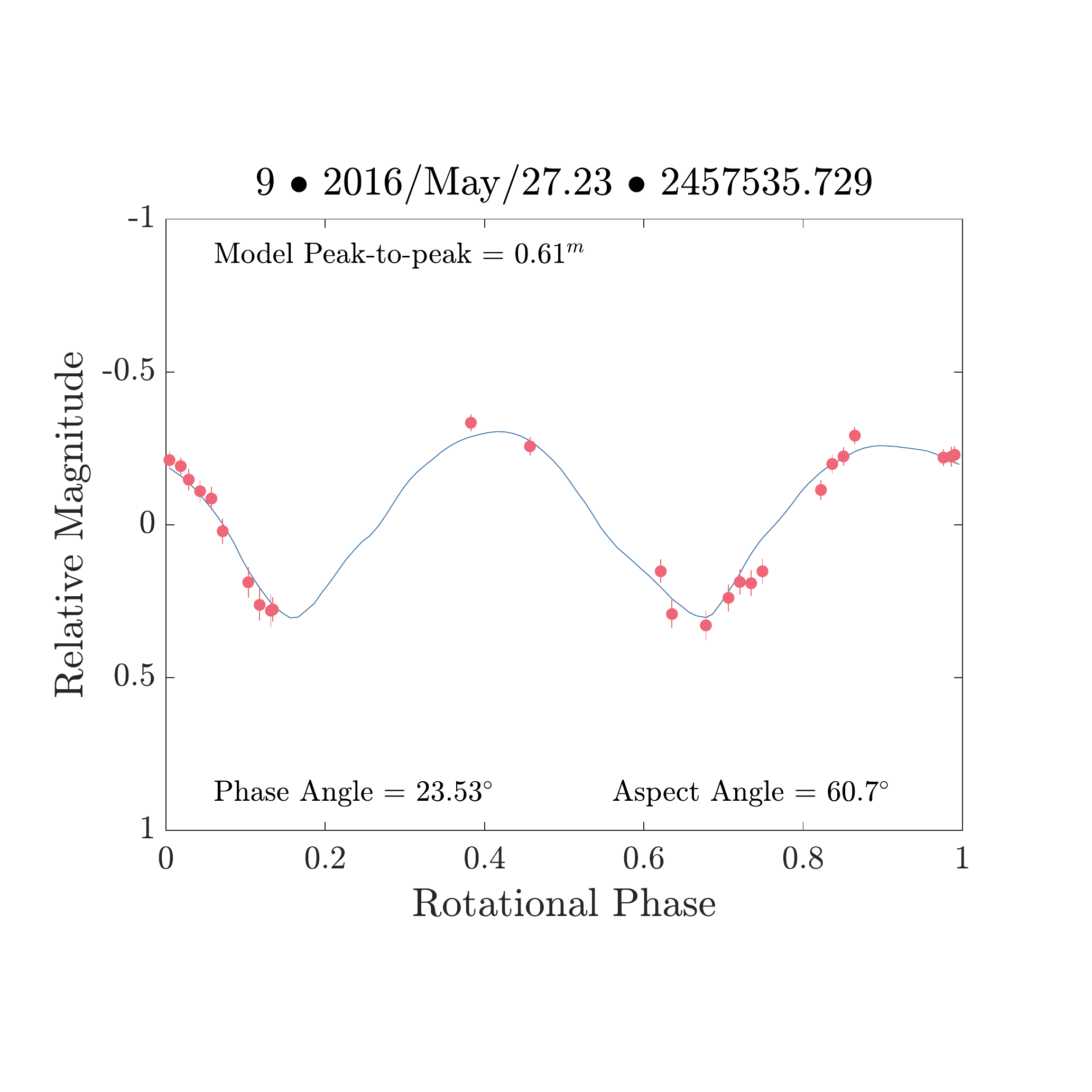}	
		}
		
		\resizebox{\hsize}{!}{	
			\includegraphics[width=.33\textwidth, trim=1cm 3.0cm 2.2cm 1.5cm, clip=true]{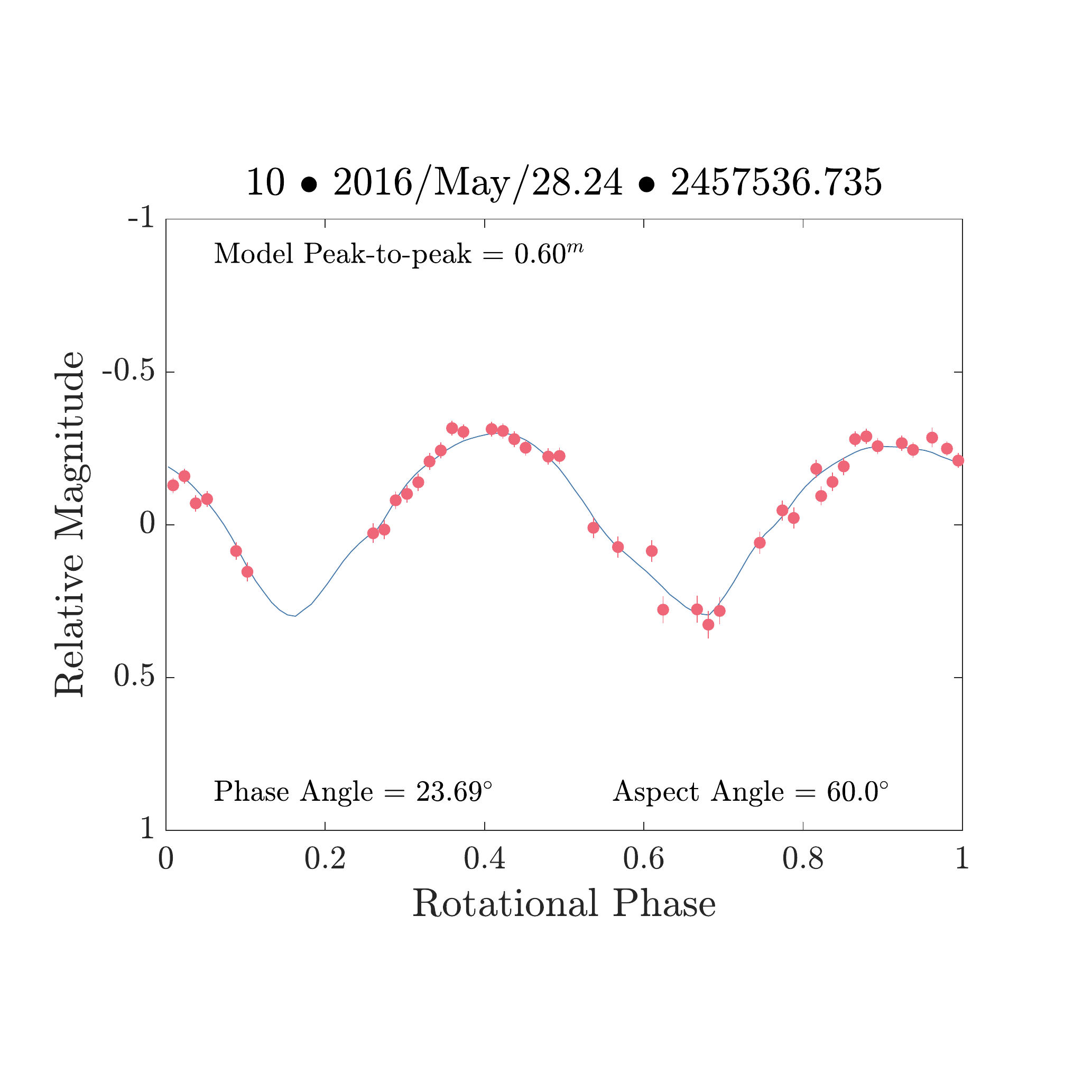}
			\includegraphics[width=.33\textwidth, trim=1cm 3.0cm 2.2cm 1.5cm, clip=true]{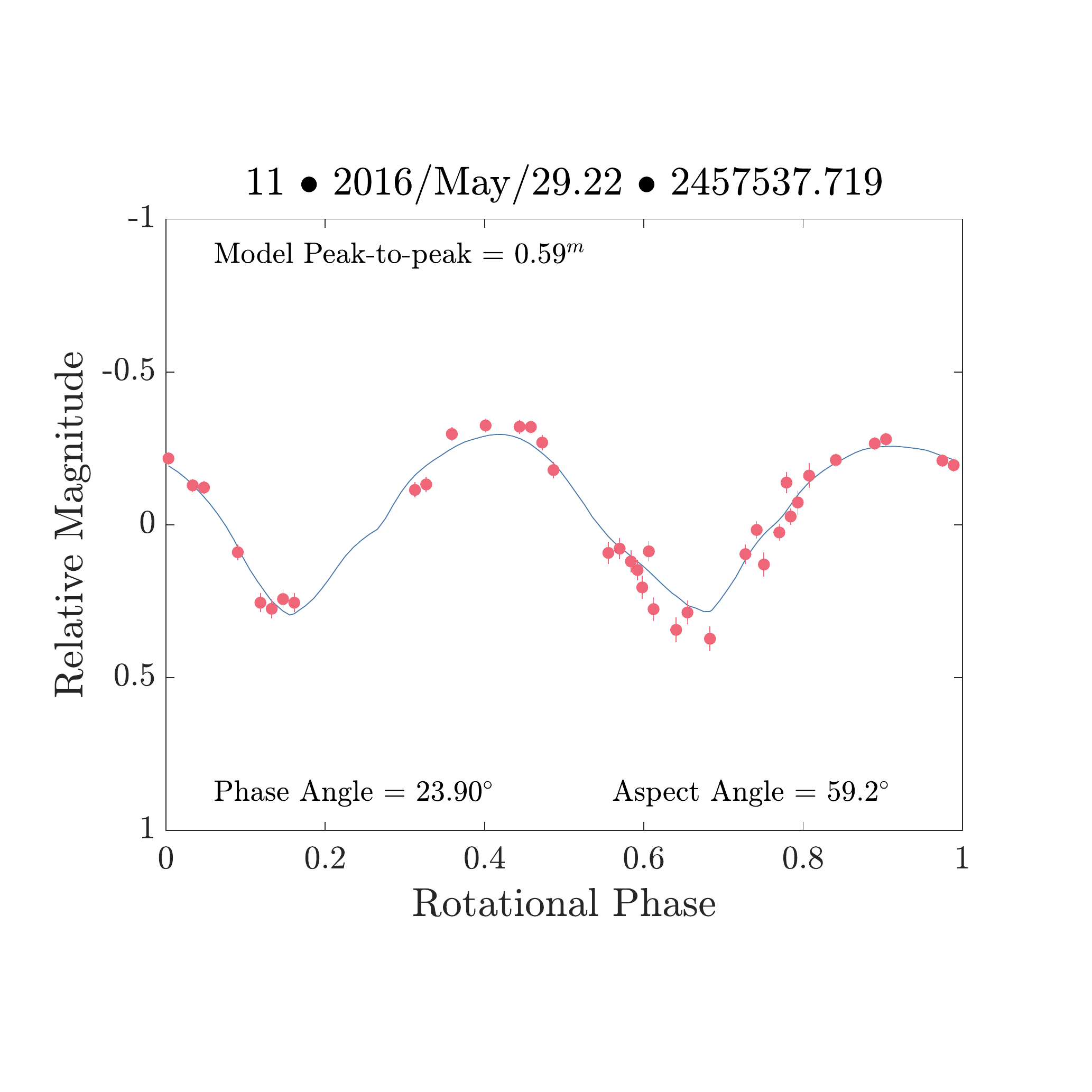}
			\includegraphics[width=.33\textwidth, trim=1cm 3.0cm 2.2cm 1.5cm, clip=true]{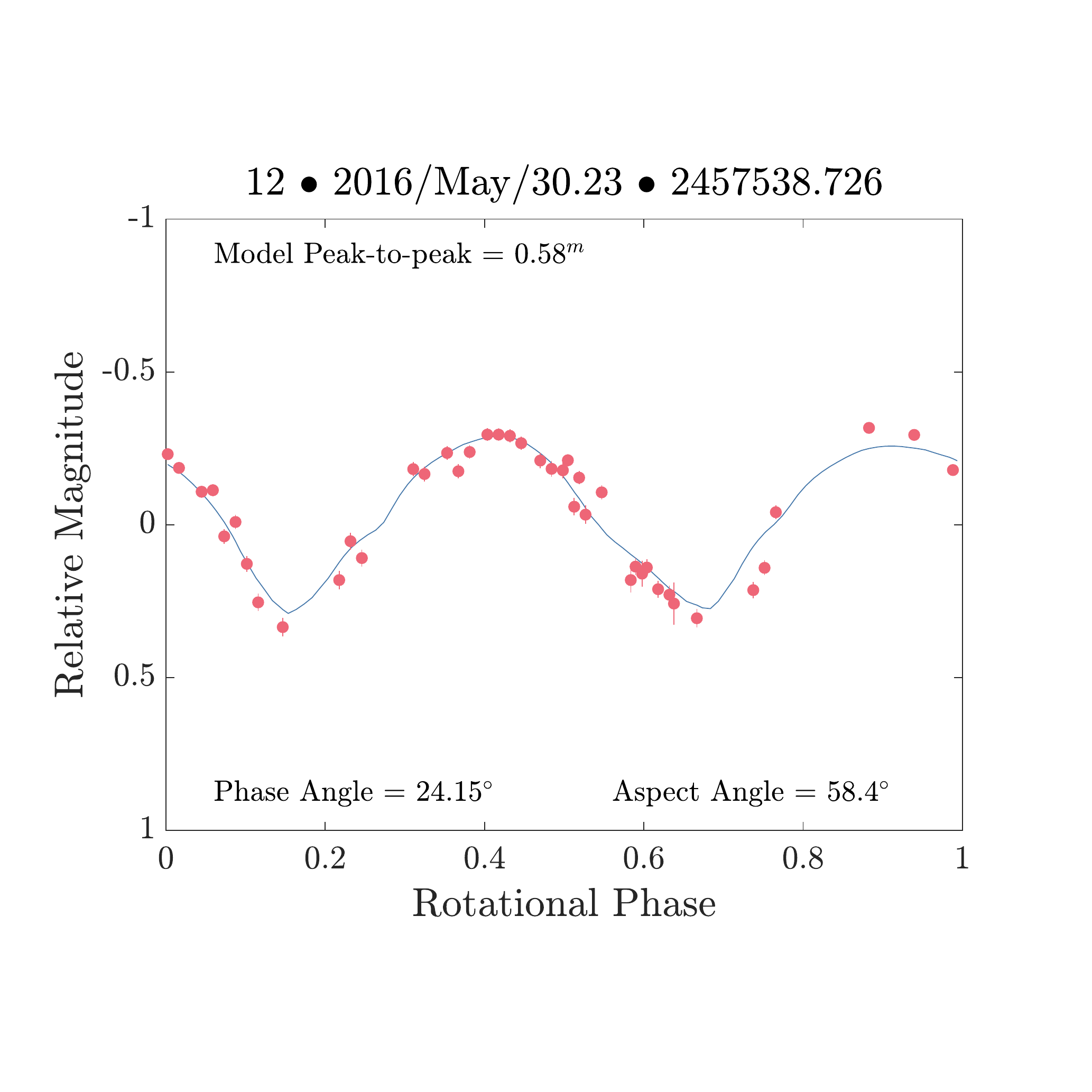}	
		}

		\caption{Synthetic light curves generated with the convex-inversion shape model of asteroid (68346) 2001 KZ66 (blue lines) plotted over all available light curve data (red dots). Light curve details can be found in Table~\ref{tab:obstable}. The model summary is given in Table~\ref{tab:models}.}
		\label{fig:conv-lightcurvefit1}
	\end{figure*}
	
	\setcounter{figure}{0}    
	\begin{figure*}
		
		\resizebox{\hsize}{!}{	
			\includegraphics[width=.33\textwidth, trim=1cm 3.0cm 2.2cm 1.5cm, clip=true]{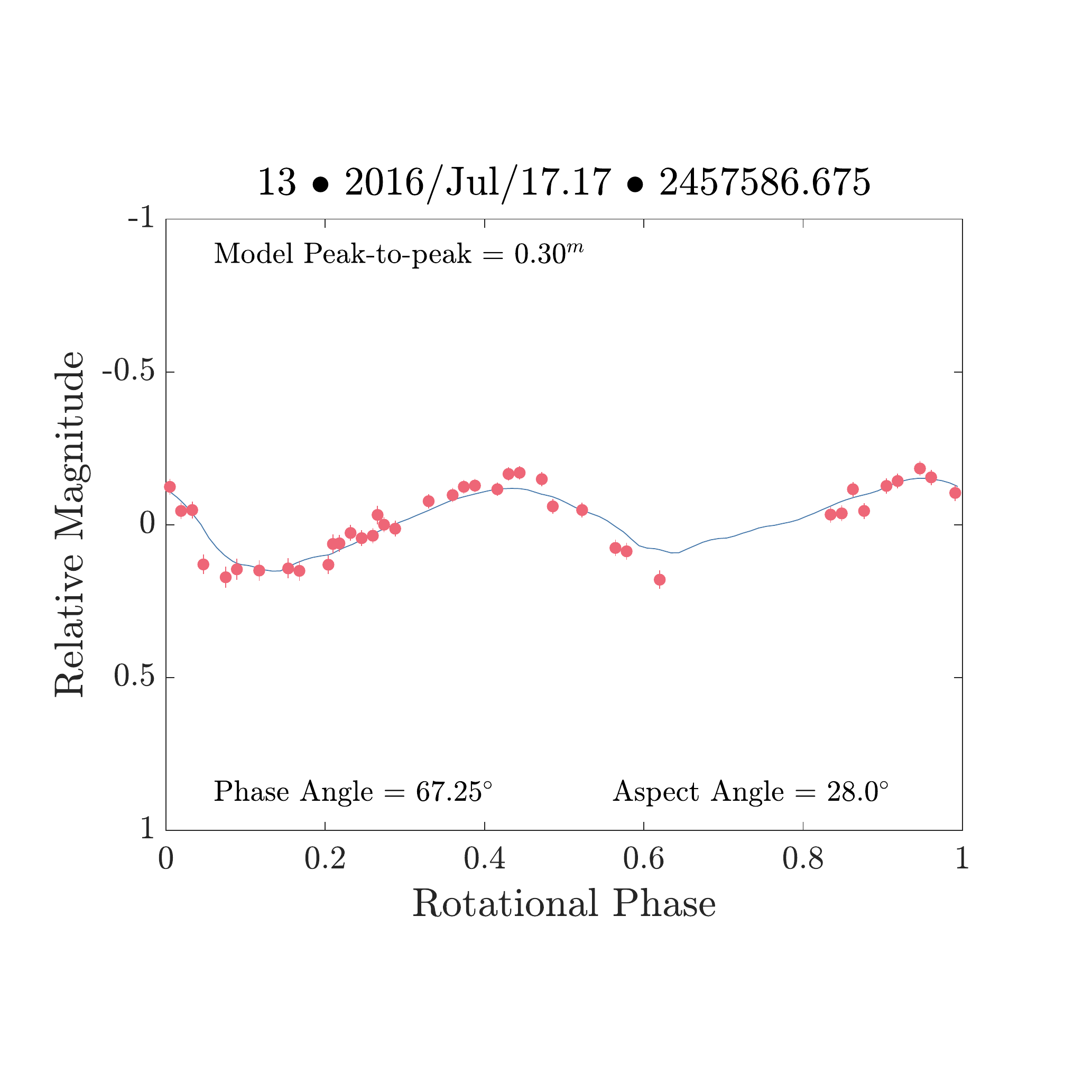}
			\includegraphics[width=.33\textwidth, trim=1cm 3.0cm 2.2cm 1.5cm, clip=true]{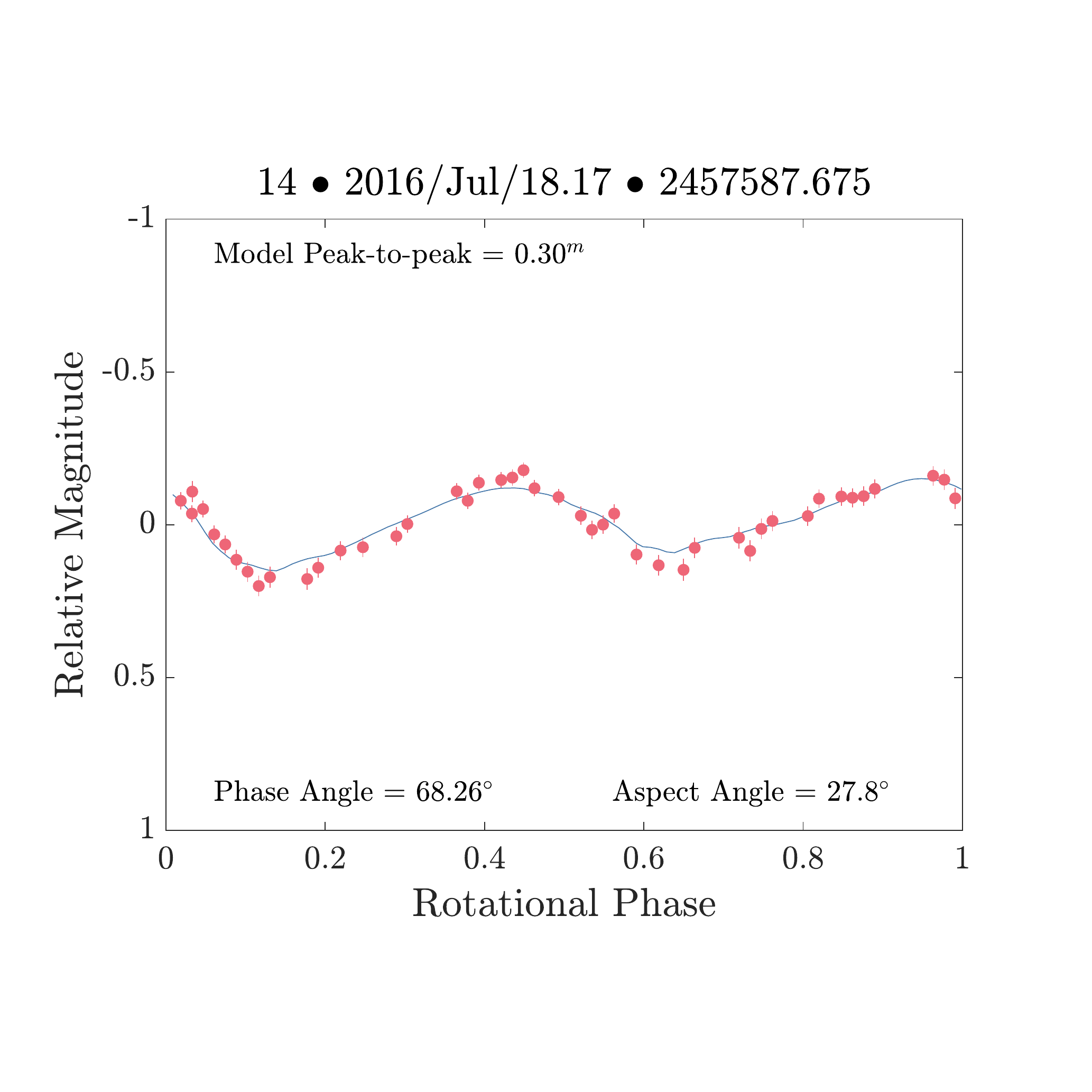}
			\includegraphics[width=.33\textwidth, trim=1cm 3.0cm 2.2cm 1.5cm, clip=true]{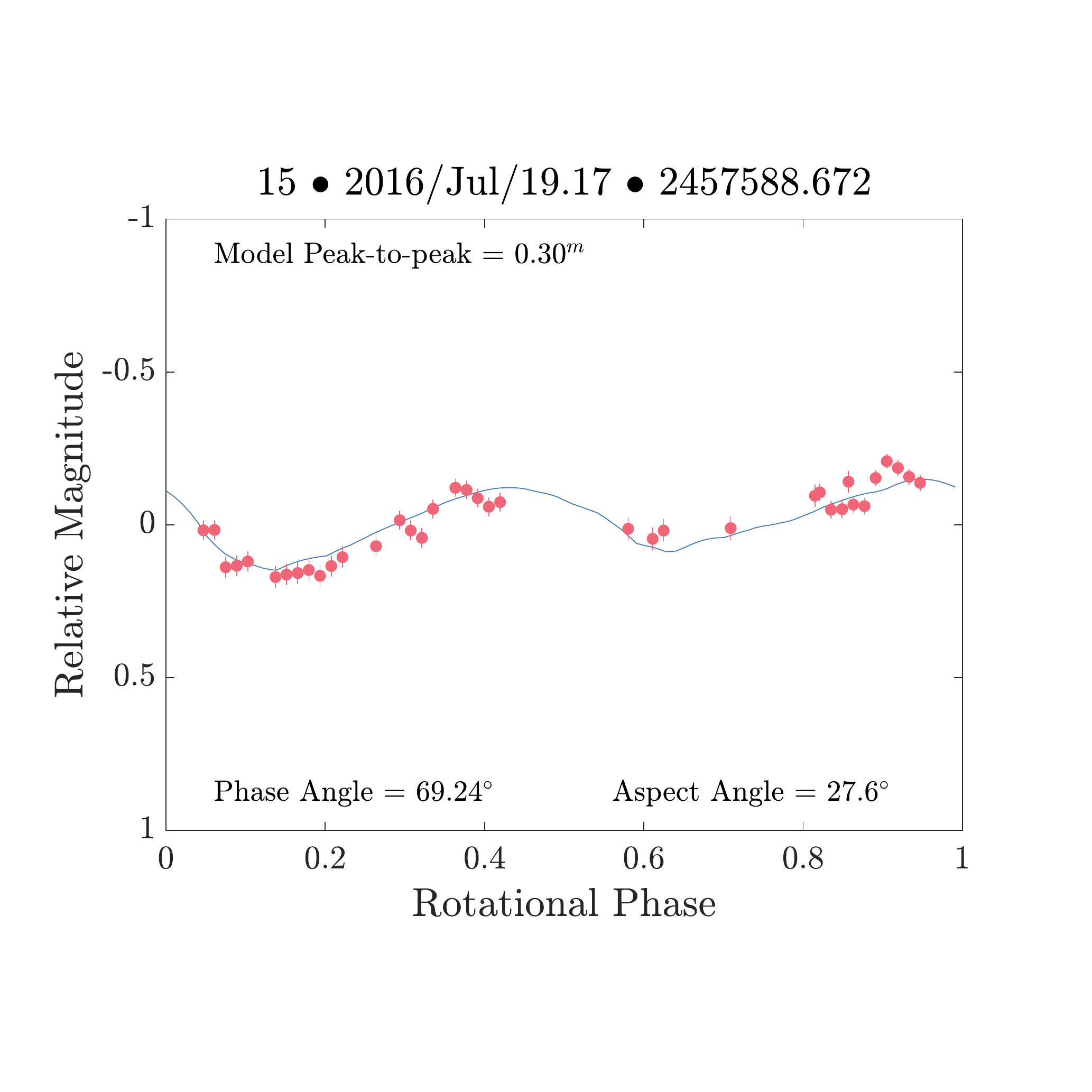}	
		}
		
		\resizebox{0.6666666\hsize}{!}{	
			\includegraphics[width=.33\textwidth, trim=1cm 3.0cm 2.2cm 1.5cm, clip=true]{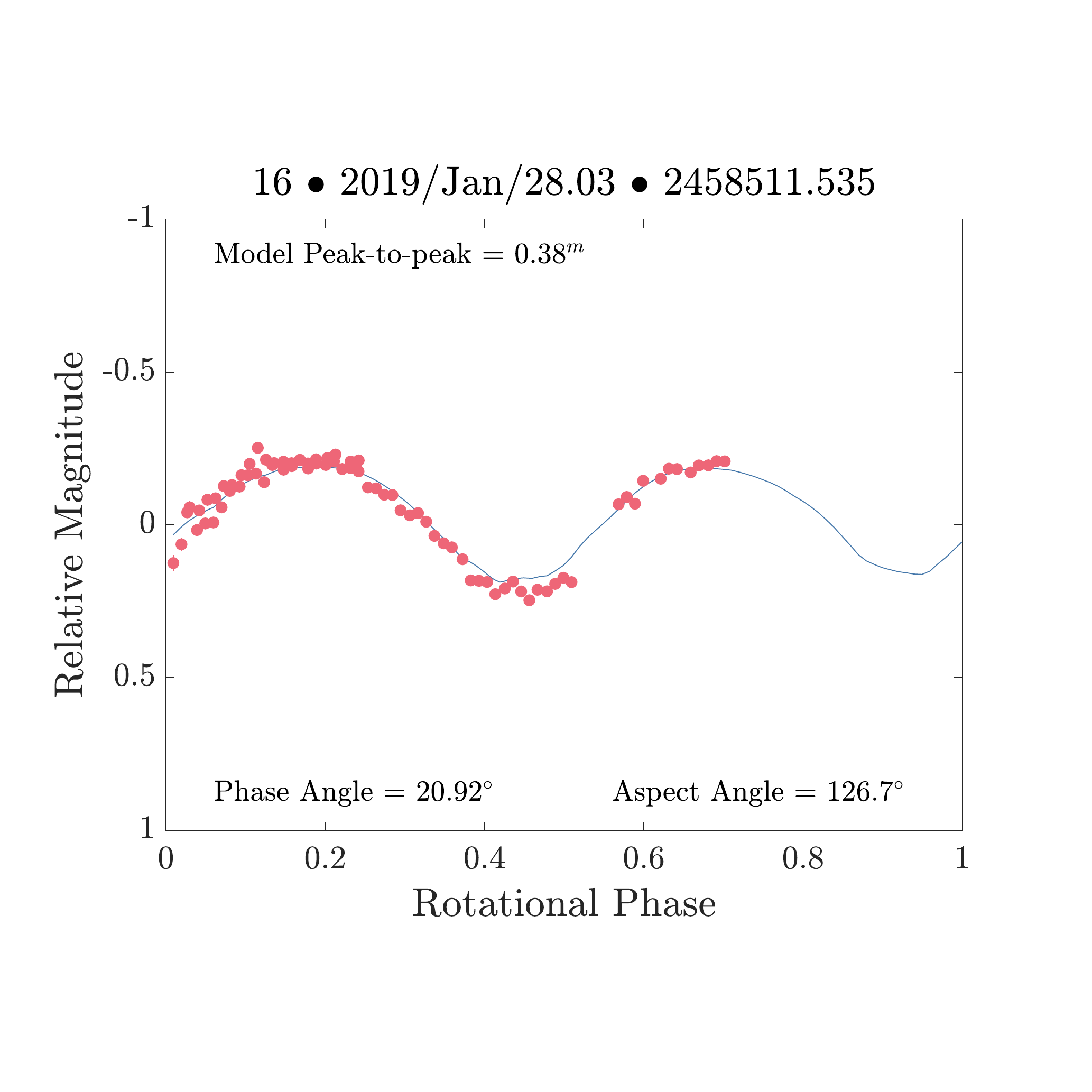}
			\includegraphics[width=.33\textwidth, trim=1cm 3.0cm 2.2cm 1.5cm, clip=true]{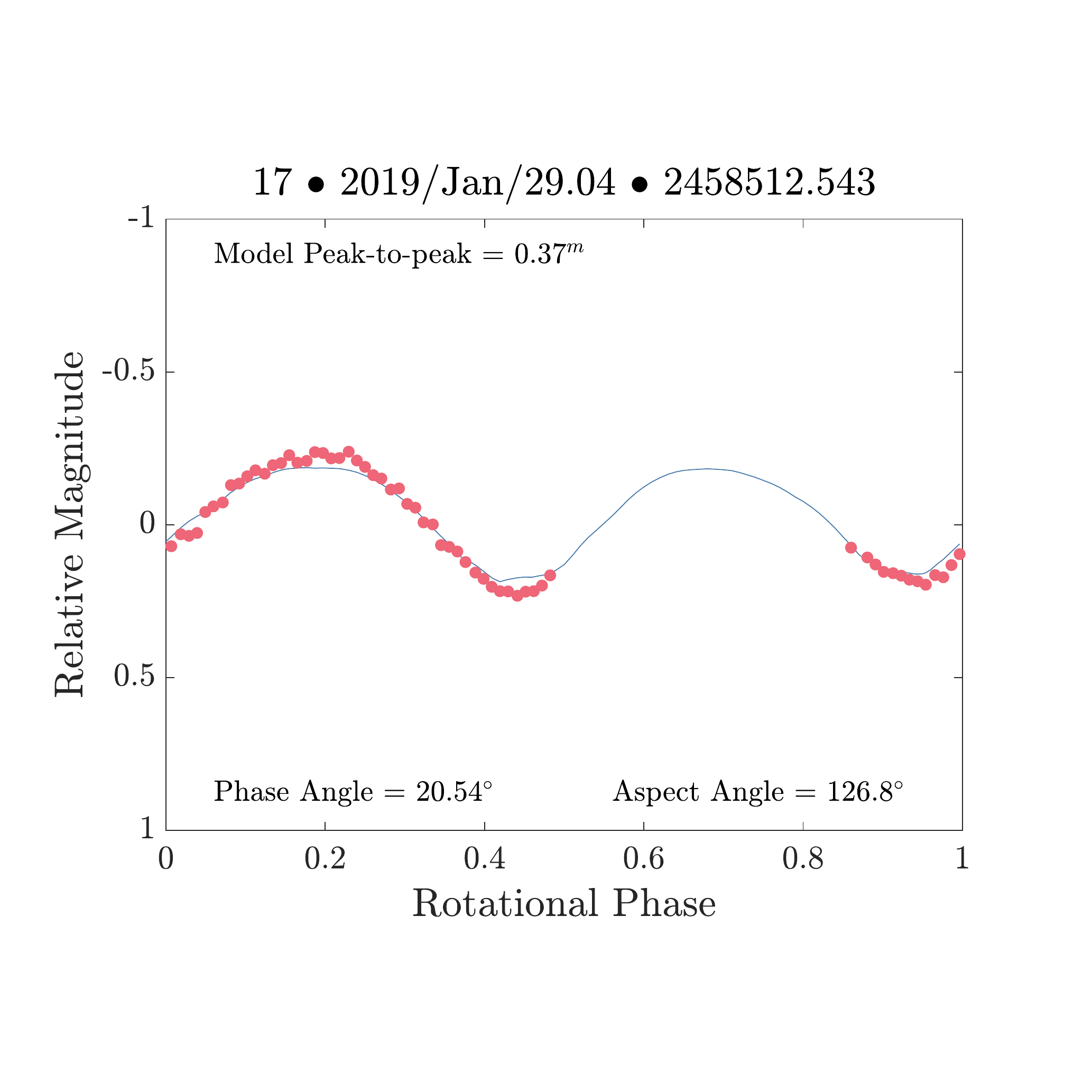}
		}
		
		\caption[]{(Continued.) }
		\label{fig:conv-lightcurvefit2}
	\end{figure*}

	\begin{figure*}
		
		\resizebox{\hsize}{!}{	
			\includegraphics[width=.33\textwidth, trim=1cm 3.0cm 2.2cm 1.5cm, clip=true]{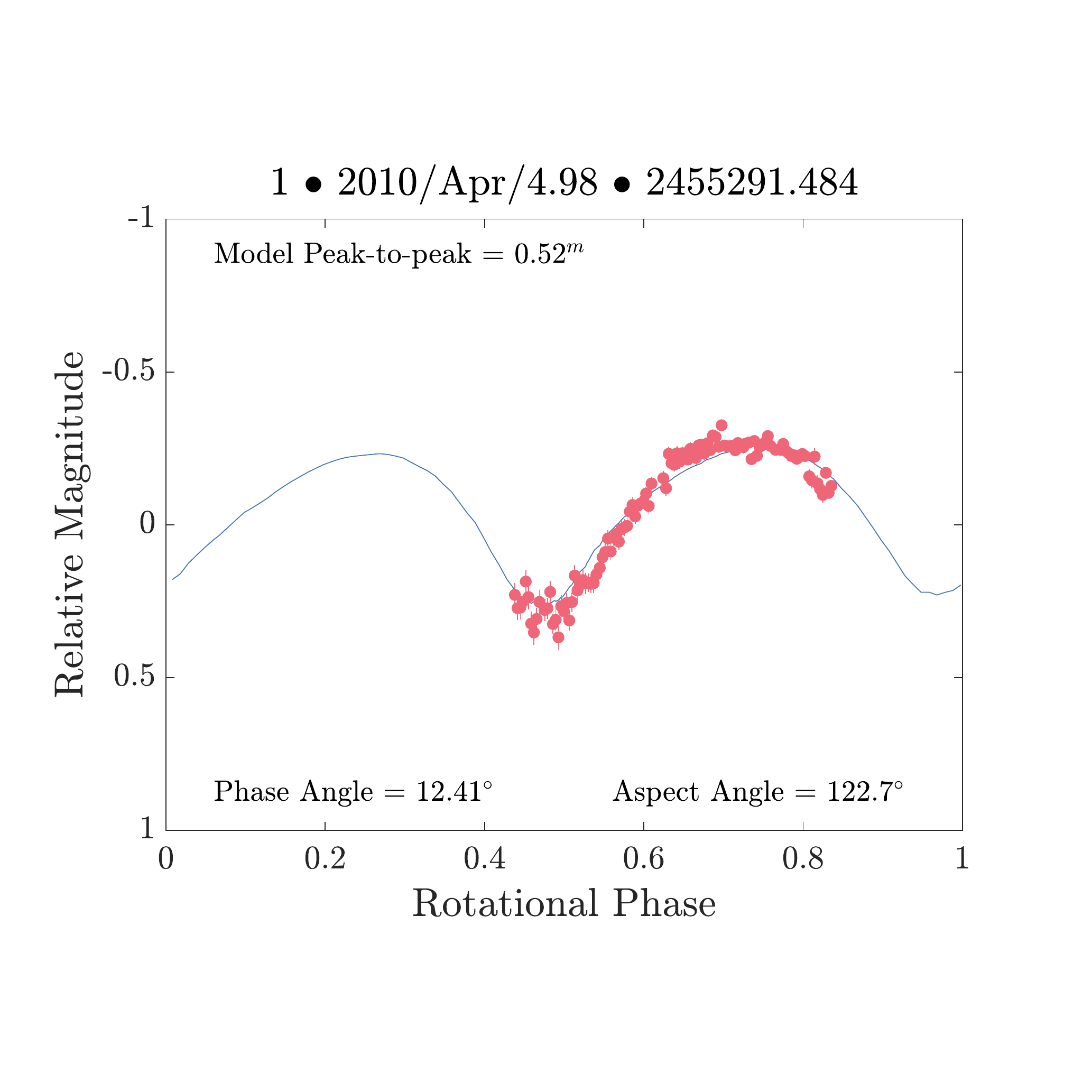}
			\includegraphics[width=.33\textwidth, trim=1cm 3.0cm 2.2cm 1.5cm, clip=true]{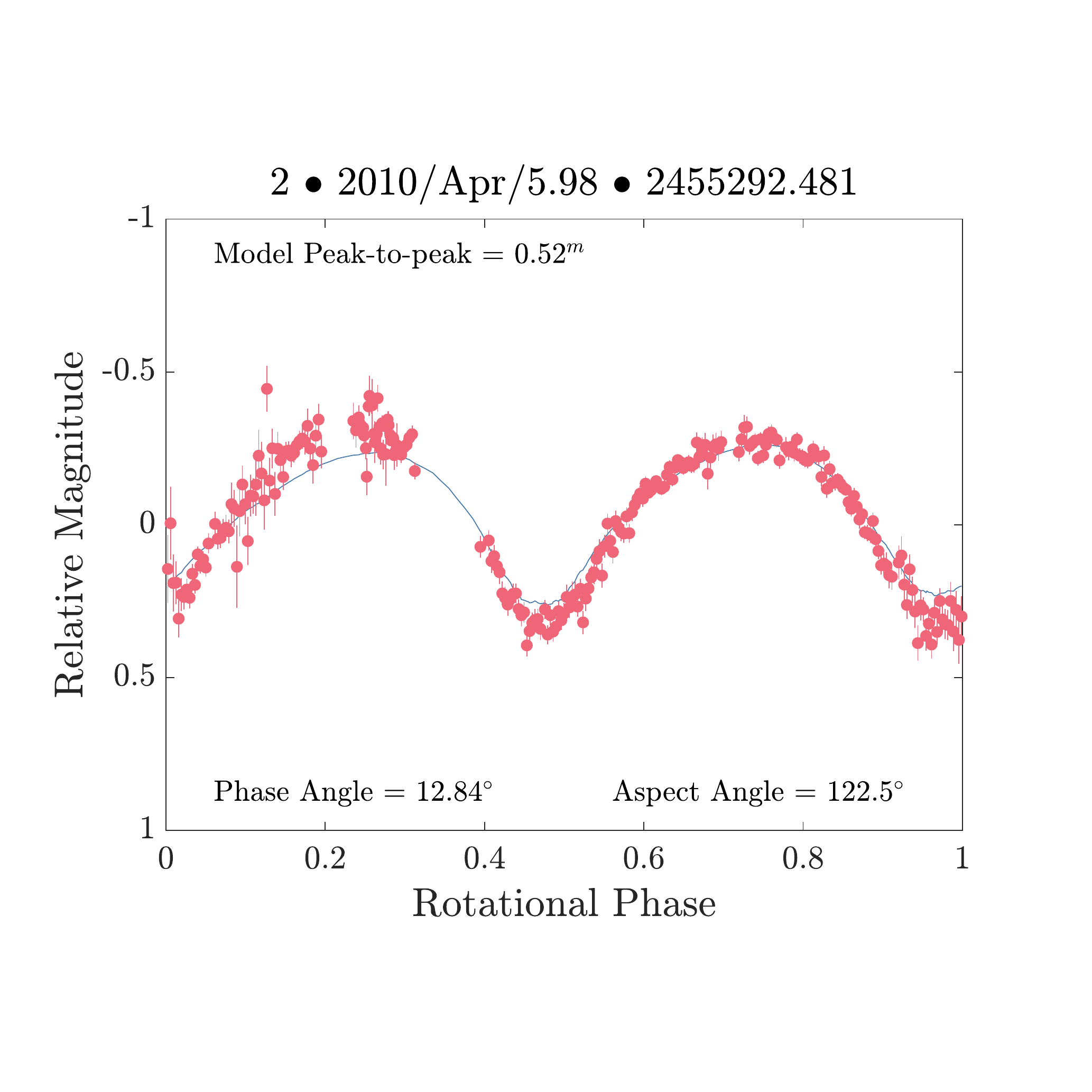}
			\includegraphics[width=.33\textwidth, trim=1cm 3.0cm 2.2cm 1.5cm, clip=true]{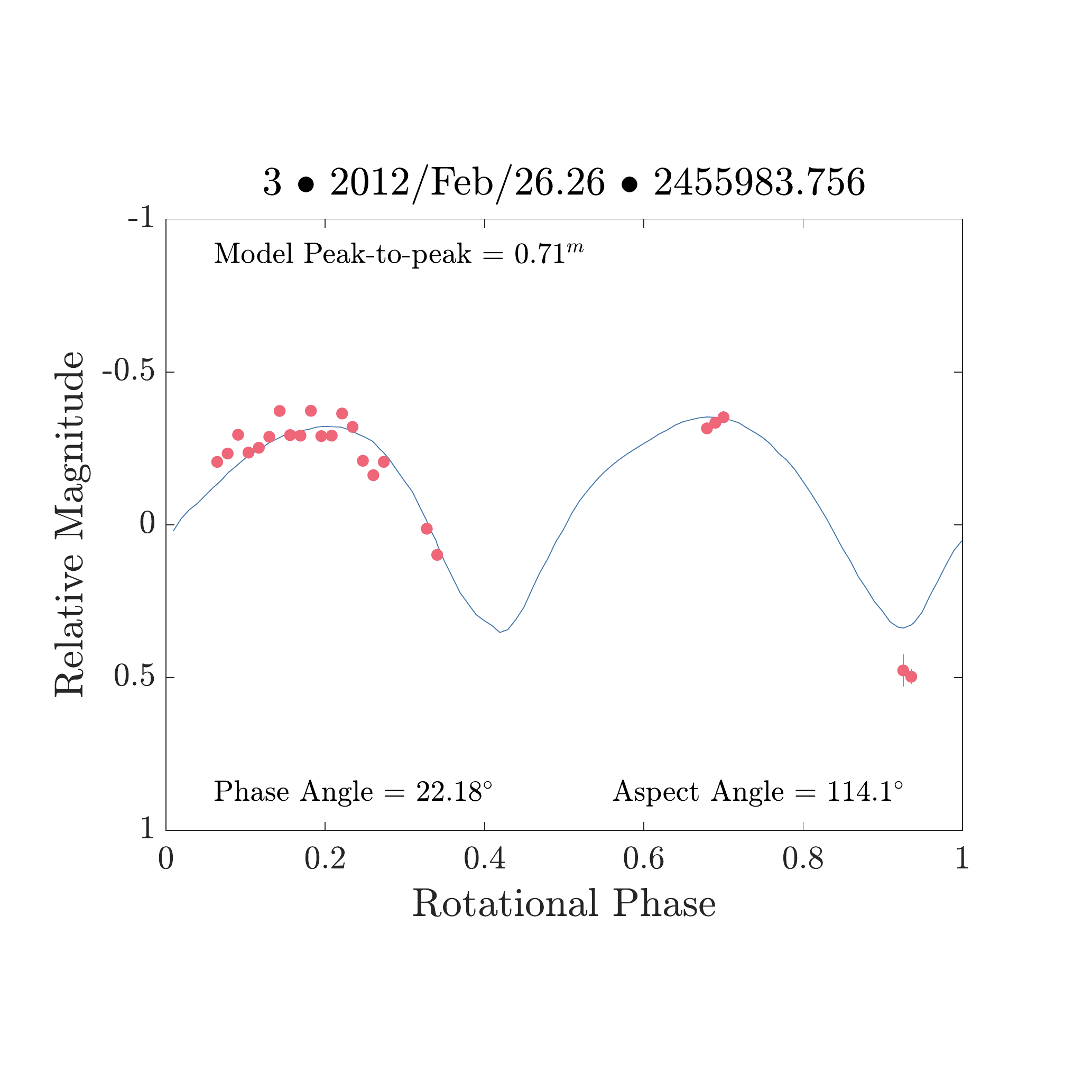}	
		}
		
		\resizebox{\hsize}{!}{	
			\includegraphics[width=.33\textwidth, trim=1cm 3.0cm 2.2cm 1.5cm, clip=true]{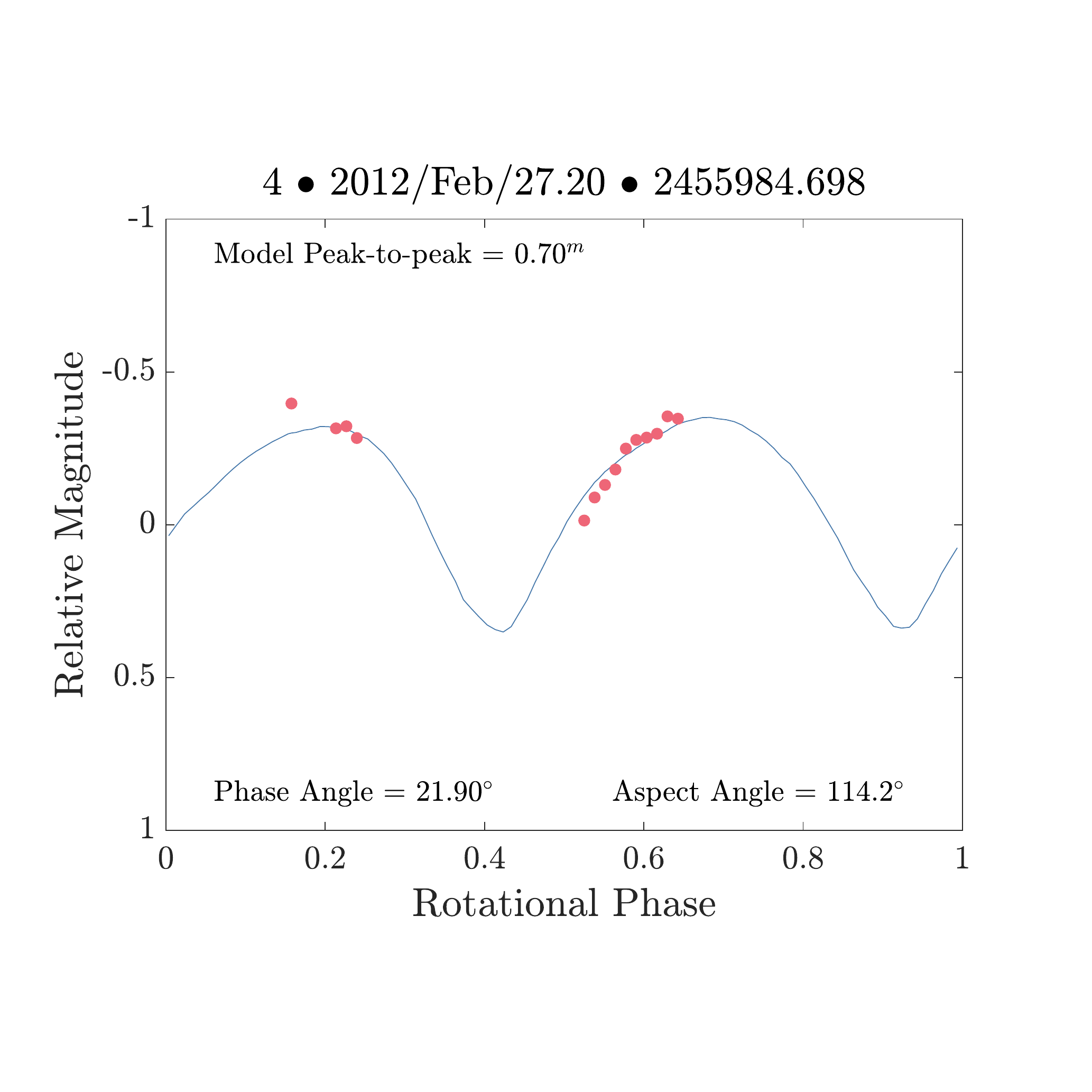}
			\includegraphics[width=.33\textwidth, trim=1cm 3.0cm 2.2cm 1.5cm, clip=true]{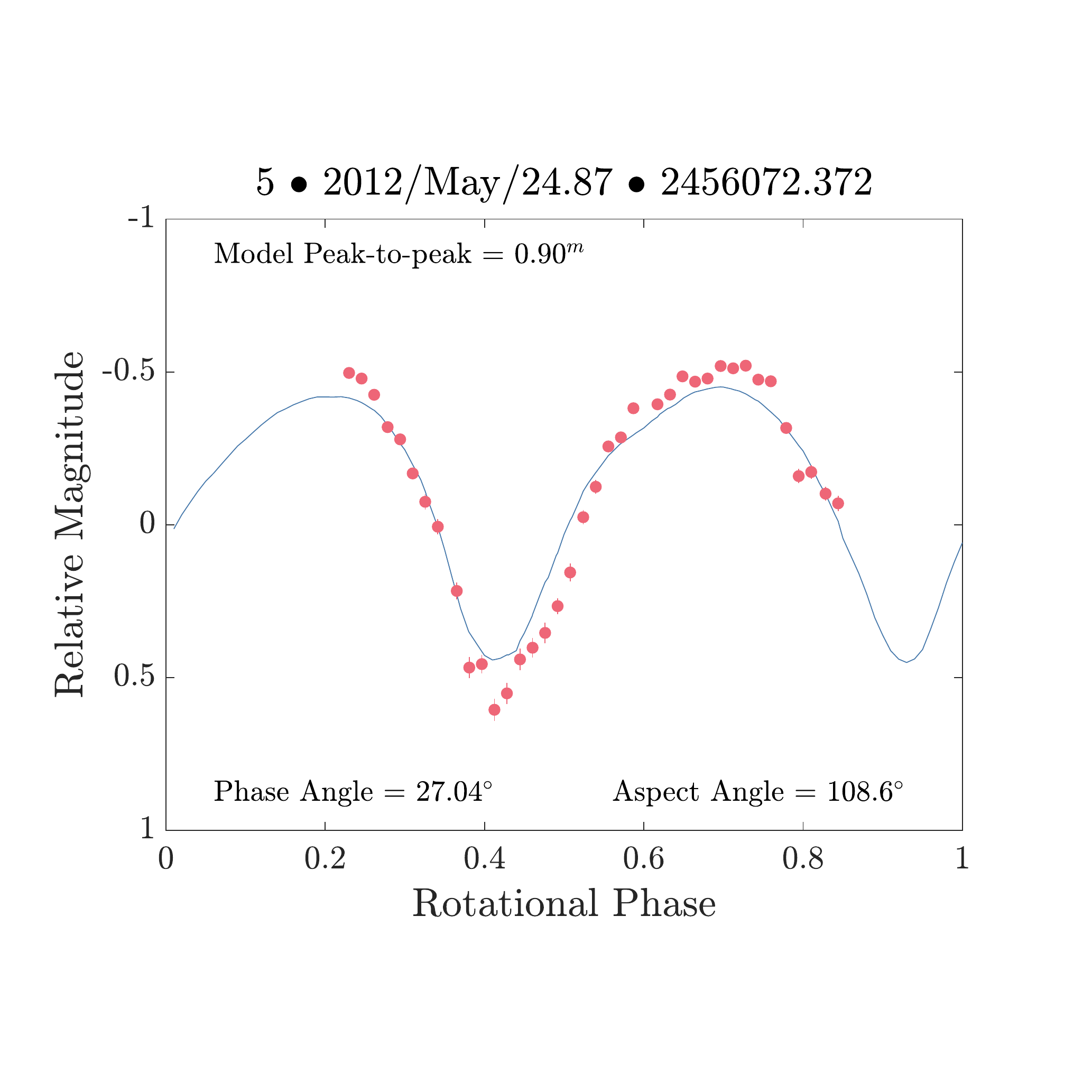}
			\includegraphics[width=.33\textwidth, trim=1cm 3.0cm 2.2cm 1.5cm, clip=true]{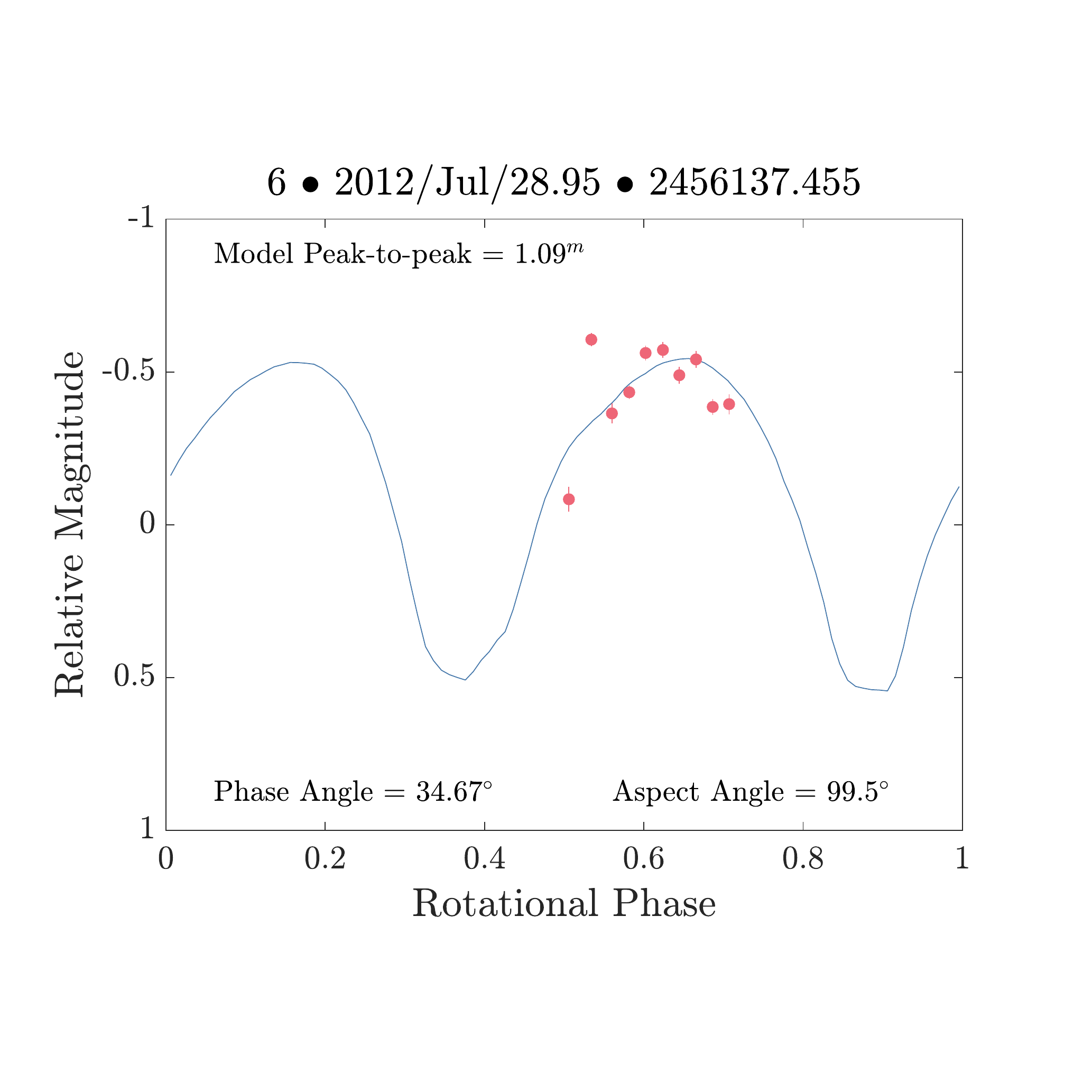}	
		}
		
		\resizebox{\hsize}{!}{	
			\includegraphics[width=.33\textwidth, trim=1cm 3.0cm 2.2cm 1.5cm, clip=true]{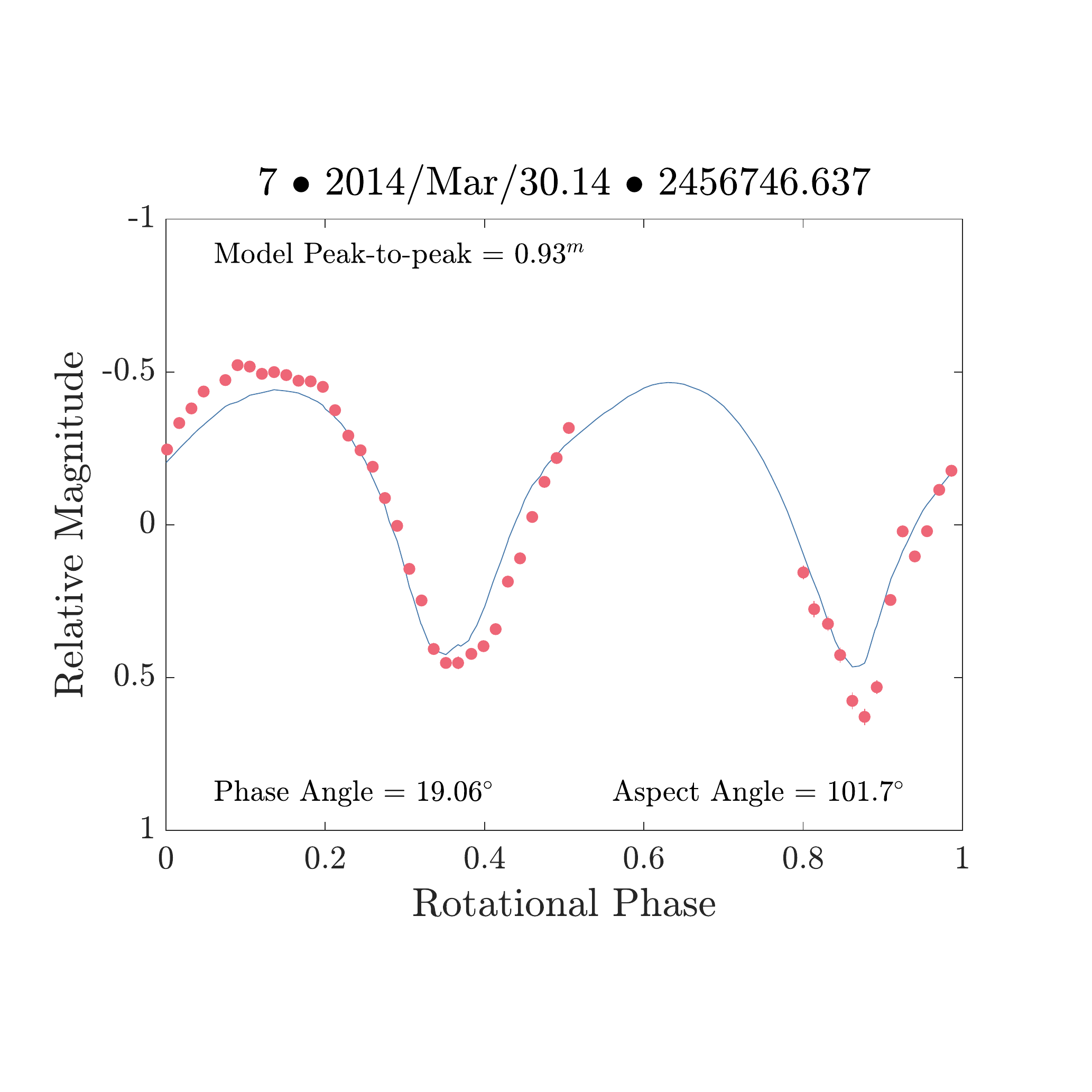}
			\includegraphics[width=.33\textwidth, trim=1cm 3.0cm 2.2cm 1.5cm, clip=true]{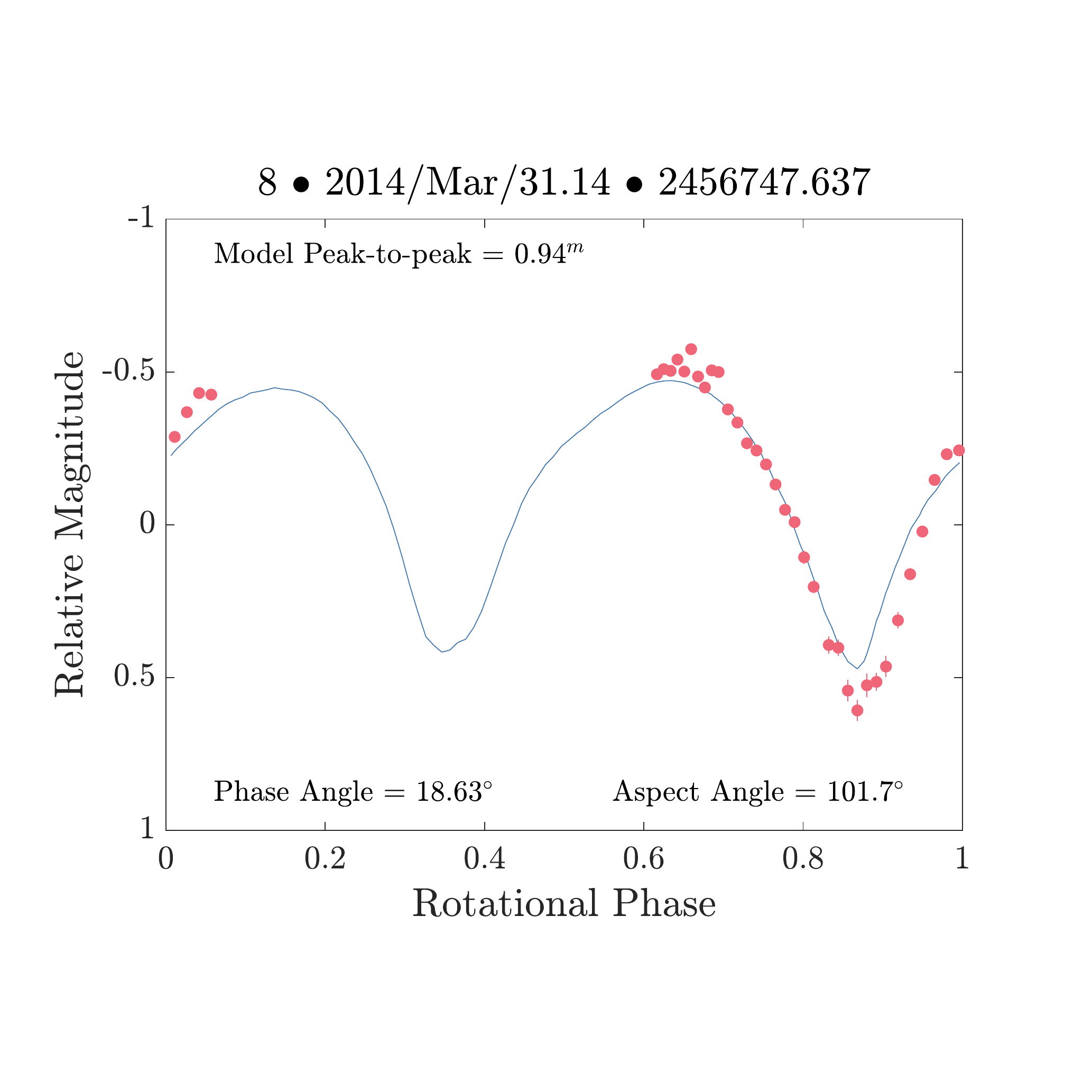}
			\includegraphics[width=.33\textwidth, trim=1cm 3.0cm 2.2cm 1.5cm, clip=true]{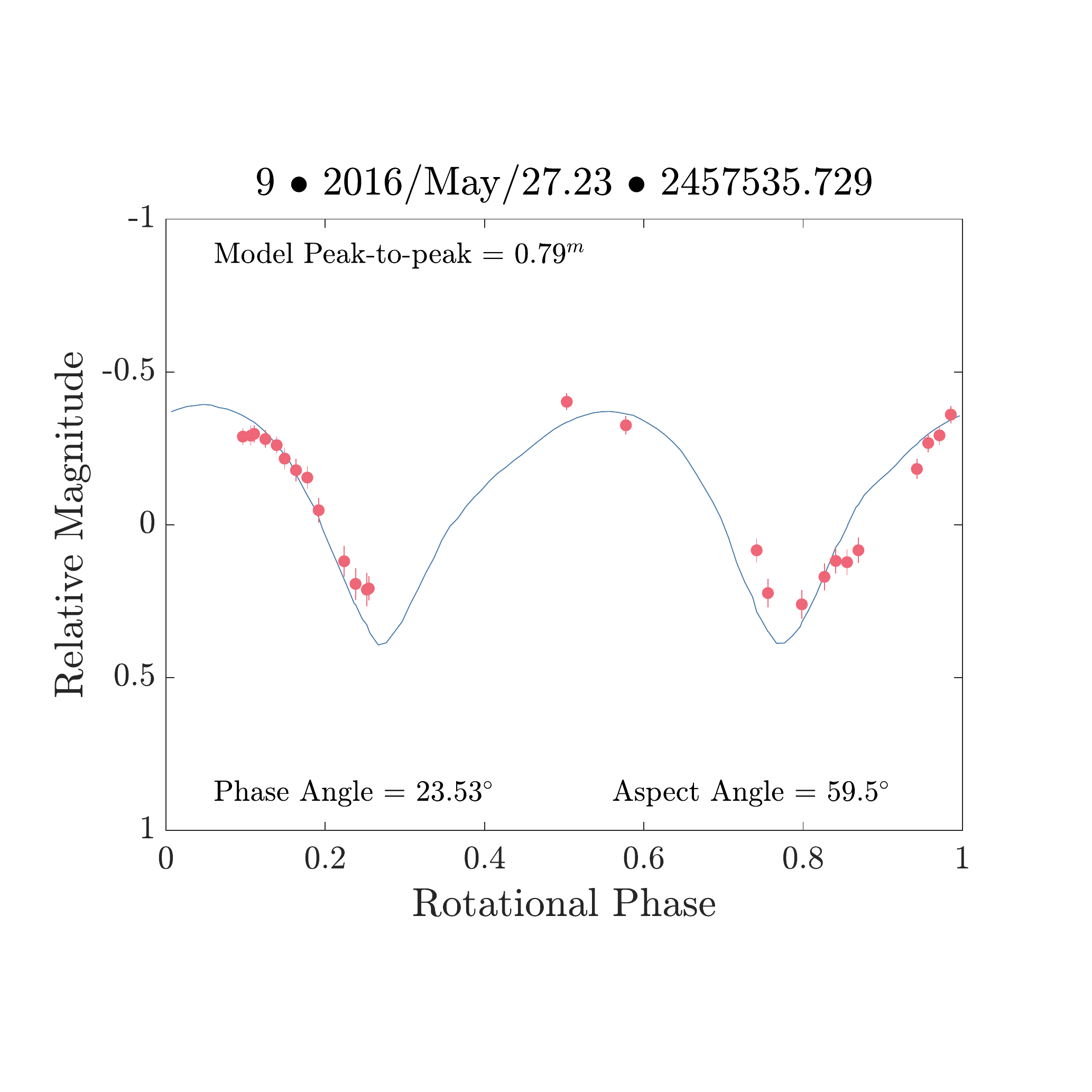}	
		}
		
		\resizebox{\hsize}{!}{	
			\includegraphics[width=.33\textwidth, trim=1cm 3.0cm 2.2cm 1.5cm, clip=true]{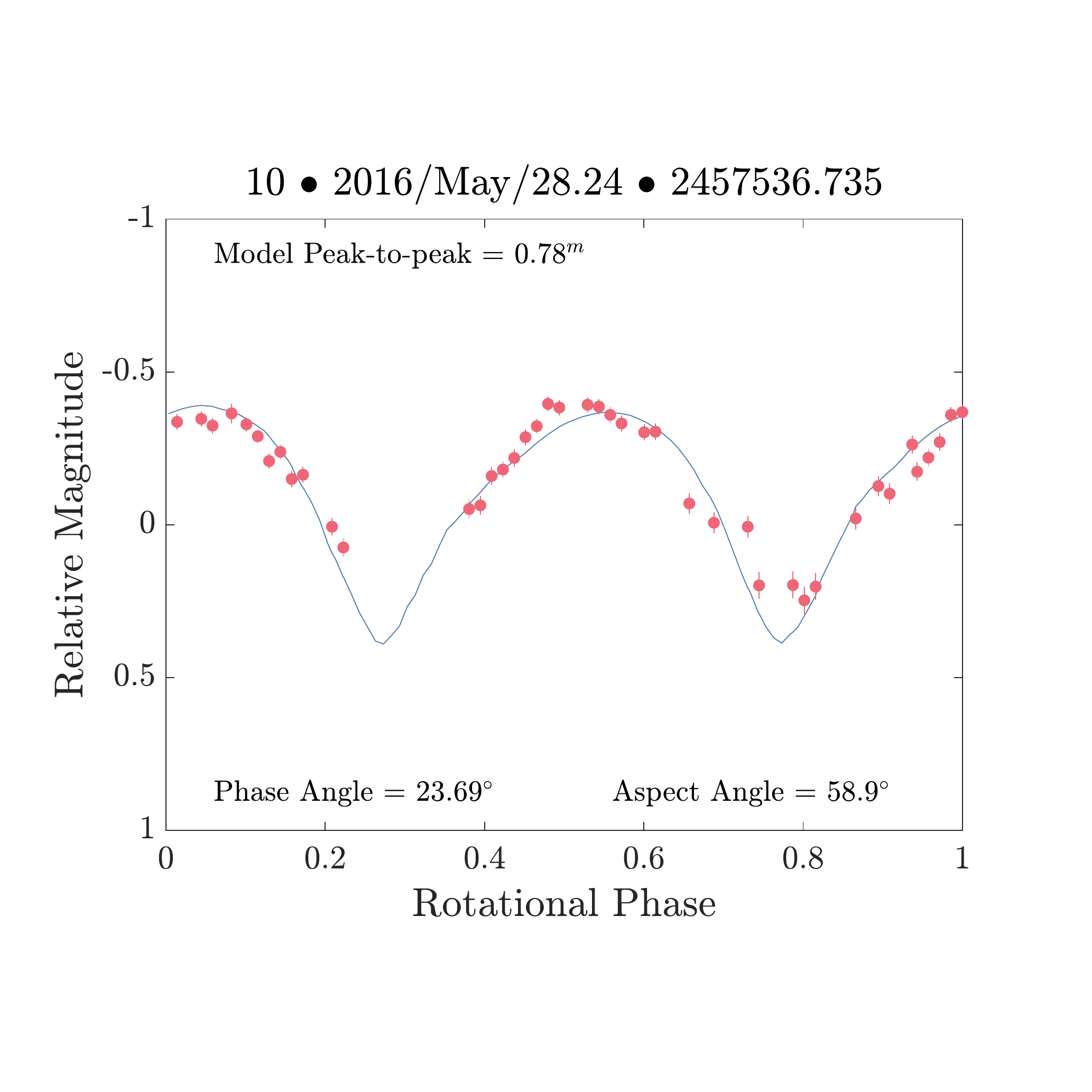}
			\includegraphics[width=.33\textwidth, trim=1cm 3.0cm 2.2cm 1.5cm, clip=true]{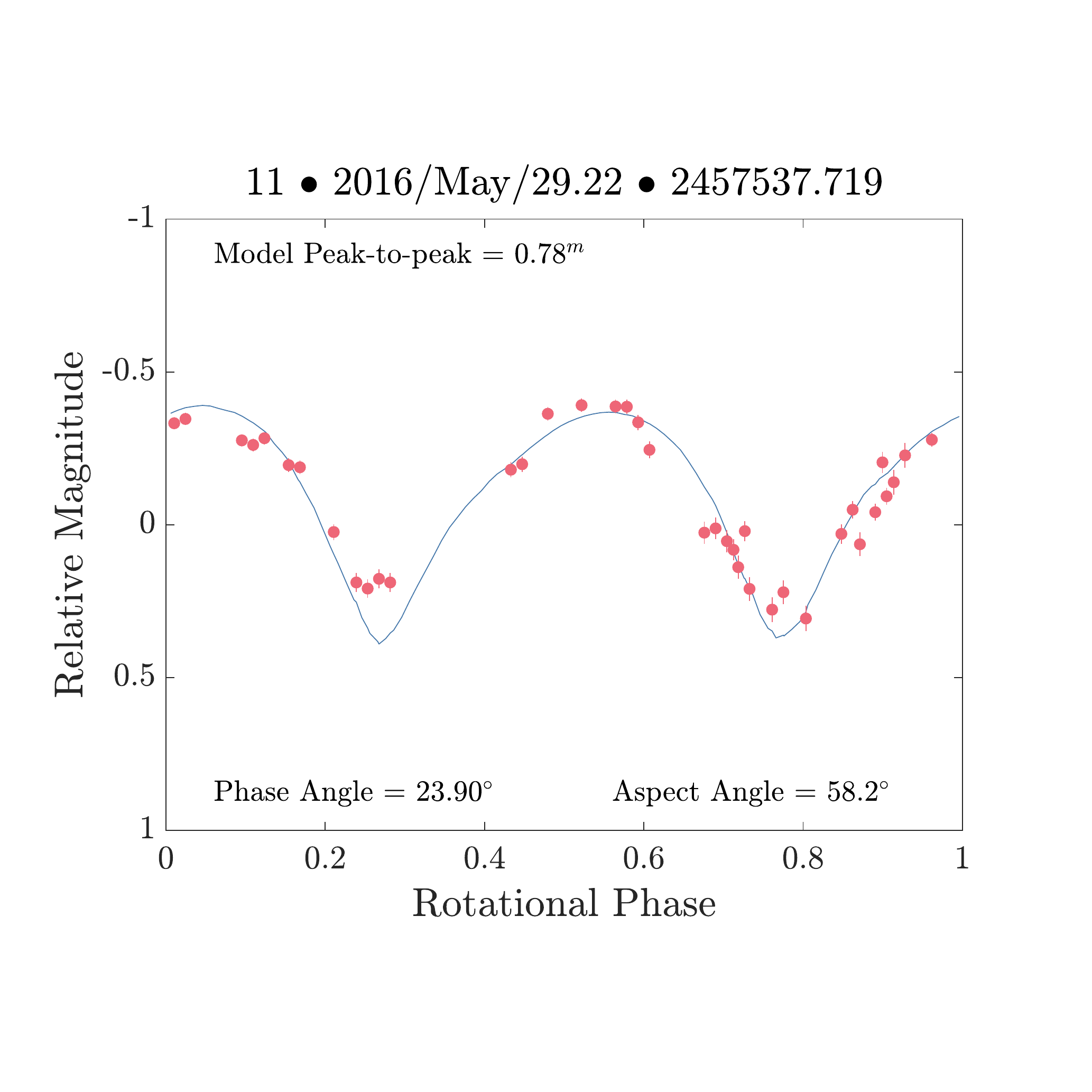}
			\includegraphics[width=.33\textwidth, trim=1cm 3.0cm 2.2cm 1.5cm, clip=true]{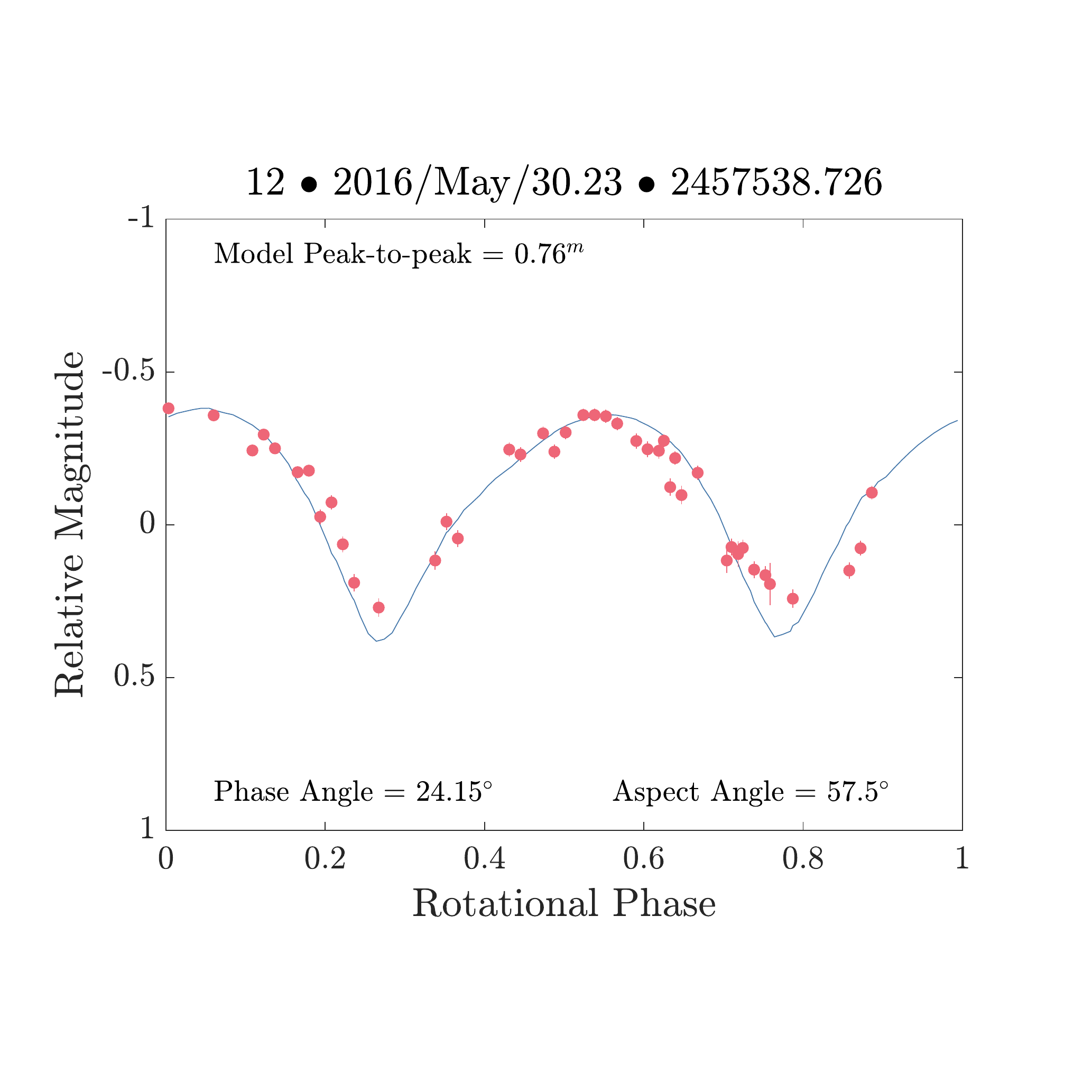}	
		}

		\caption{Same as Fig. \ref{fig:conv-lightcurvefit1}, but for the radar-derived shape model with a YORP acceleration of $8.43 \times 10^{-8} \rm ~rad ~day^{-2}$.}
		\label{fig:yorp-lightcurvefit1}
	\end{figure*}
	
	\setcounter{figure}{1}    
	\begin{figure*}
		
		\resizebox{\hsize}{!}{	
			\includegraphics[width=.33\textwidth, trim=1cm 3.0cm 2.2cm 1.5cm, clip=true]{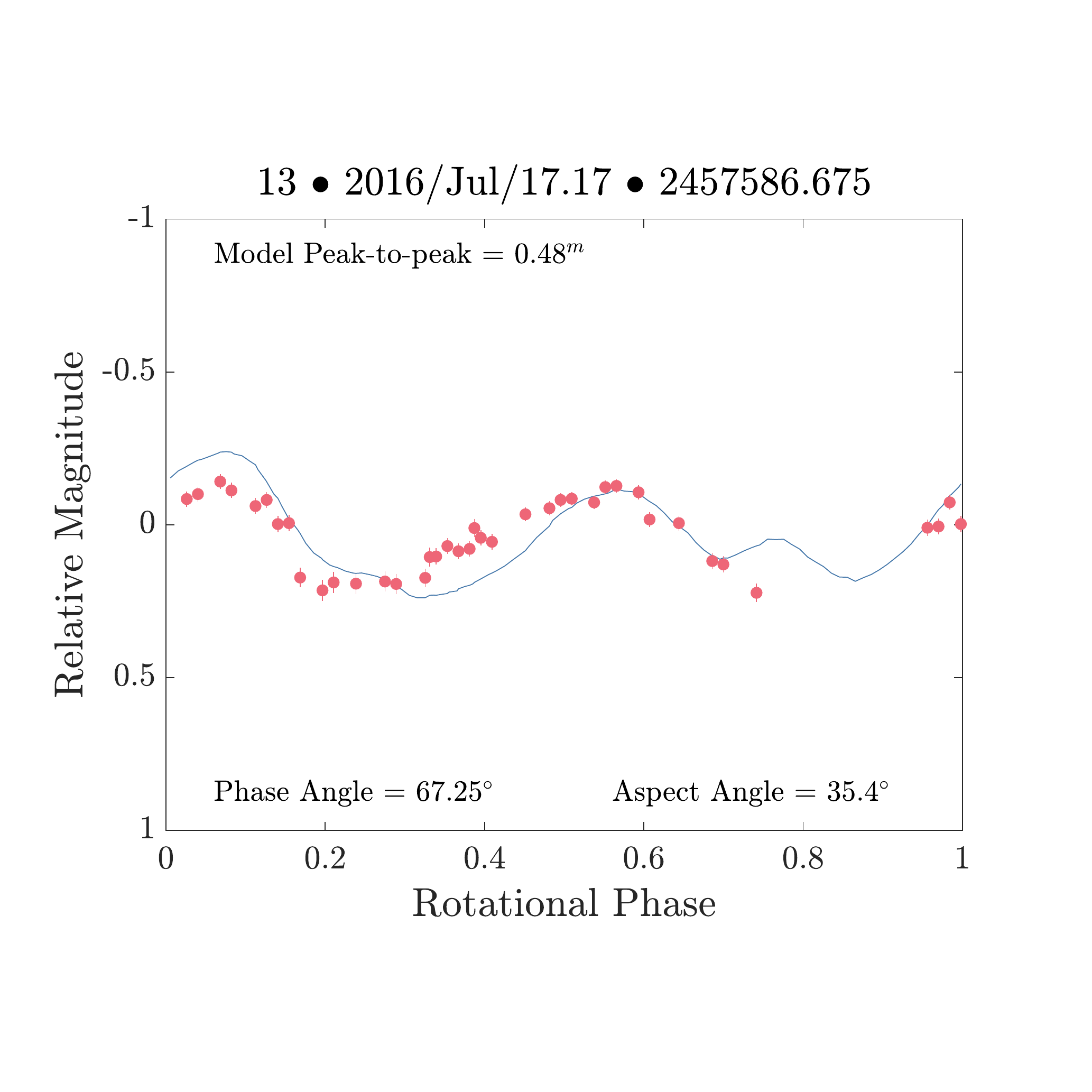}
			\includegraphics[width=.33\textwidth, trim=1cm 3.0cm 2.2cm 1.5cm, clip=true]{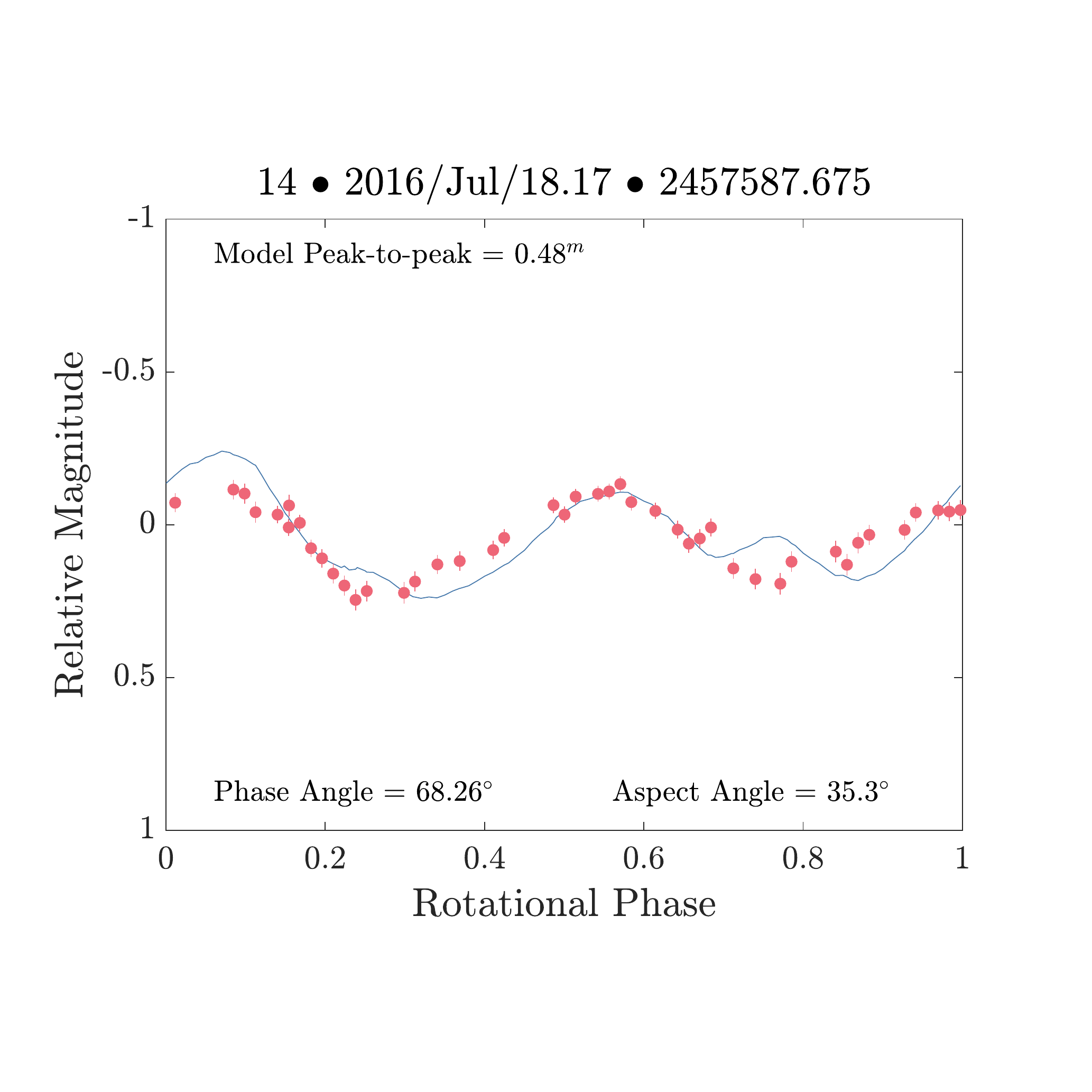}
			\includegraphics[width=.33\textwidth, trim=1cm 3.0cm 2.2cm 1.5cm, clip=true]{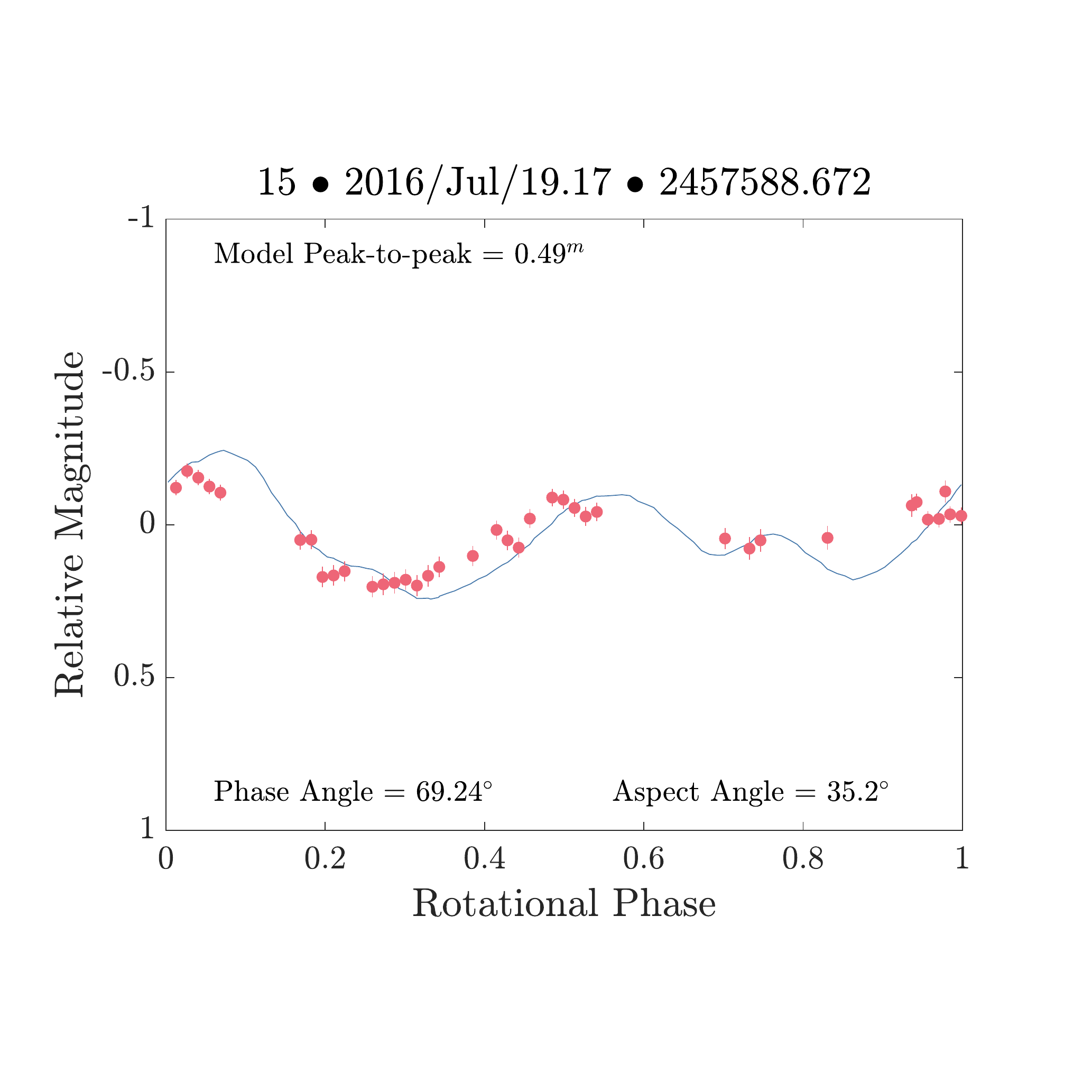}	
		}
		
		\resizebox{0.6666666\hsize}{!}{	
			\includegraphics[width=.33\textwidth, trim=1cm 3.0cm 2.2cm 1.5cm, clip=true]{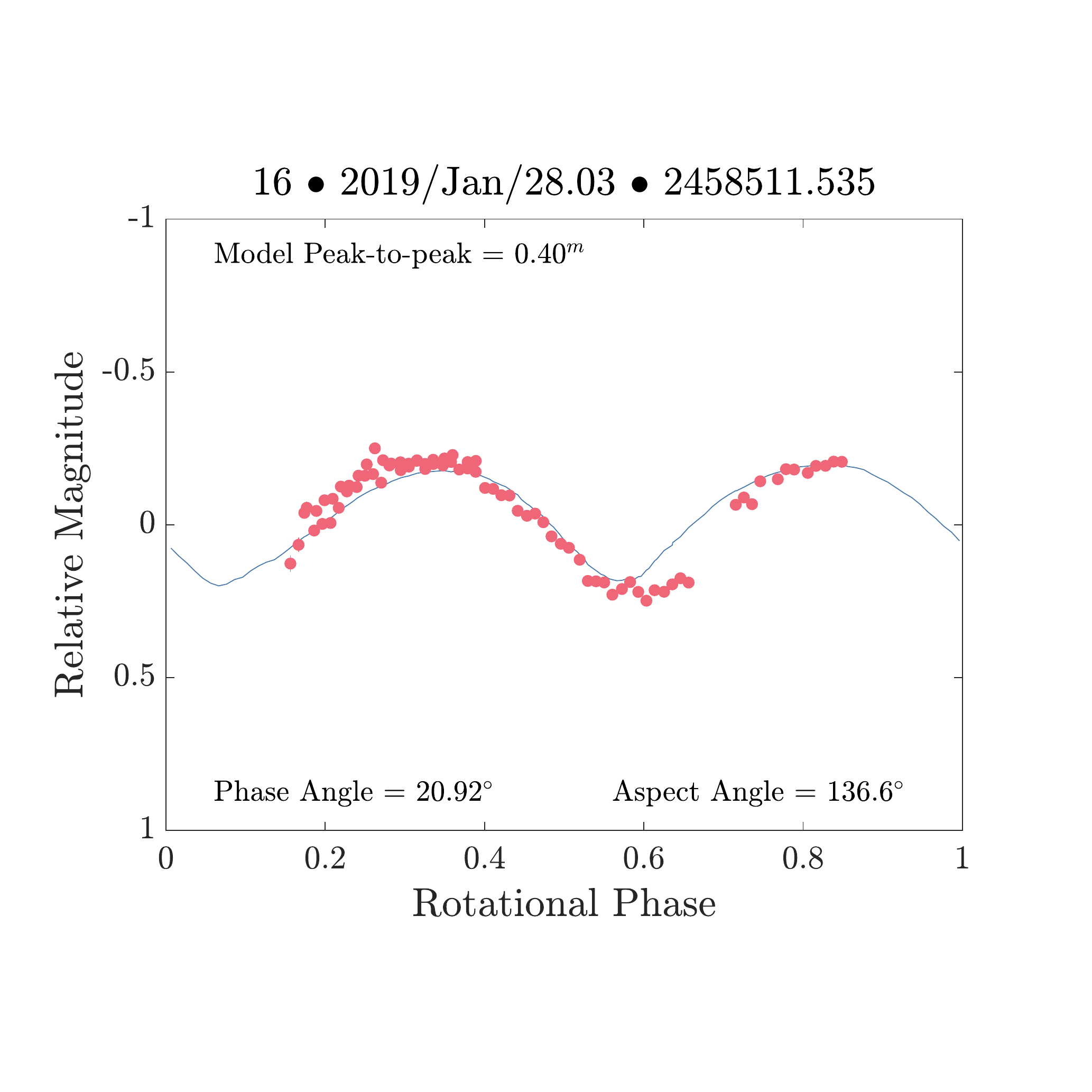}
			\includegraphics[width=.33\textwidth, trim=1cm 3.0cm 2.2cm 1.5cm, clip=true]{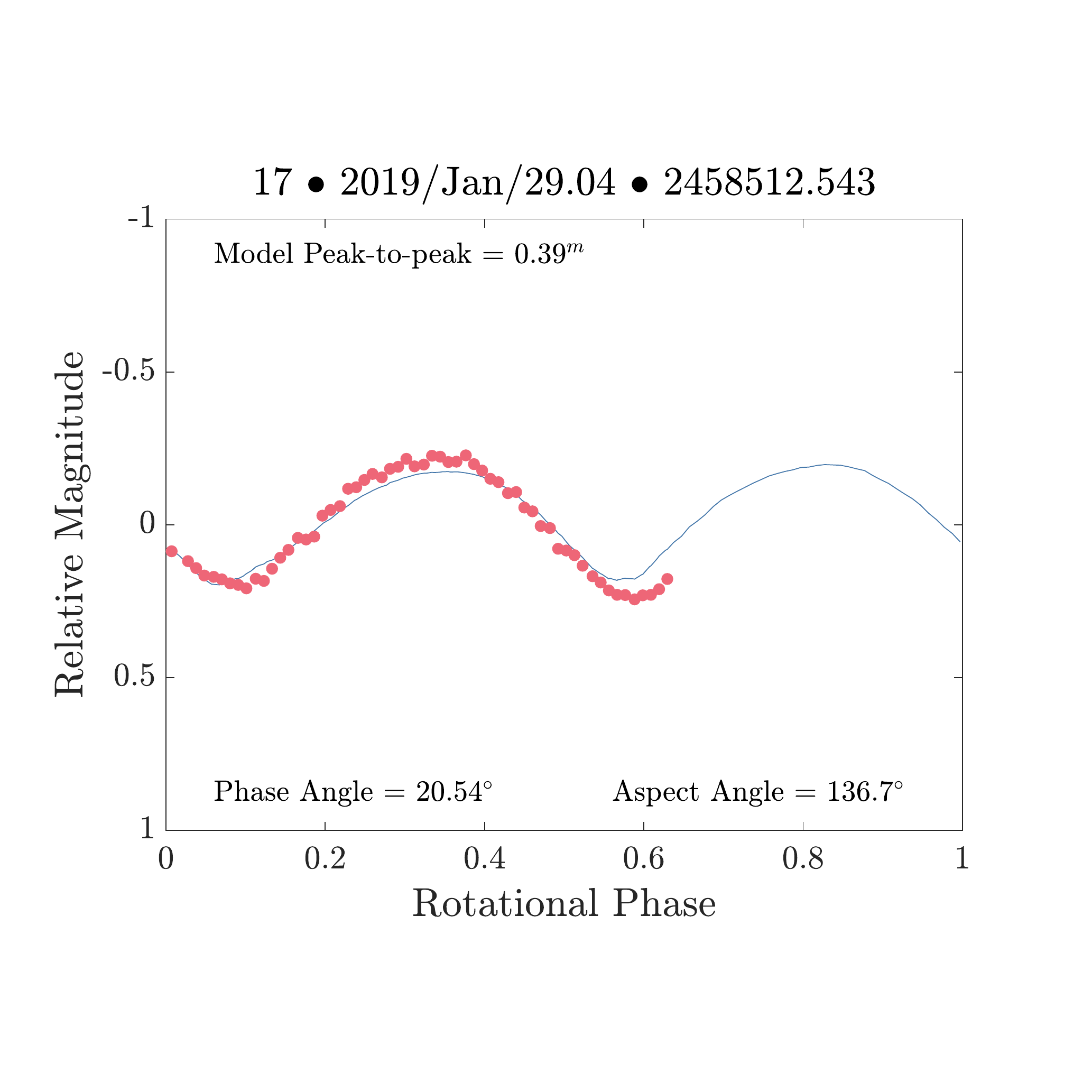}
		}
		\caption[]{(Continued.) }
		\label{fig:yorp-lightcurvefit2}
	\end{figure*}

	\begin{figure*}
		
		\resizebox{\hsize}{!}{	
			\includegraphics[width=.33\textwidth, trim=1cm 3.0cm 2.2cm 1.5cm, clip=true]{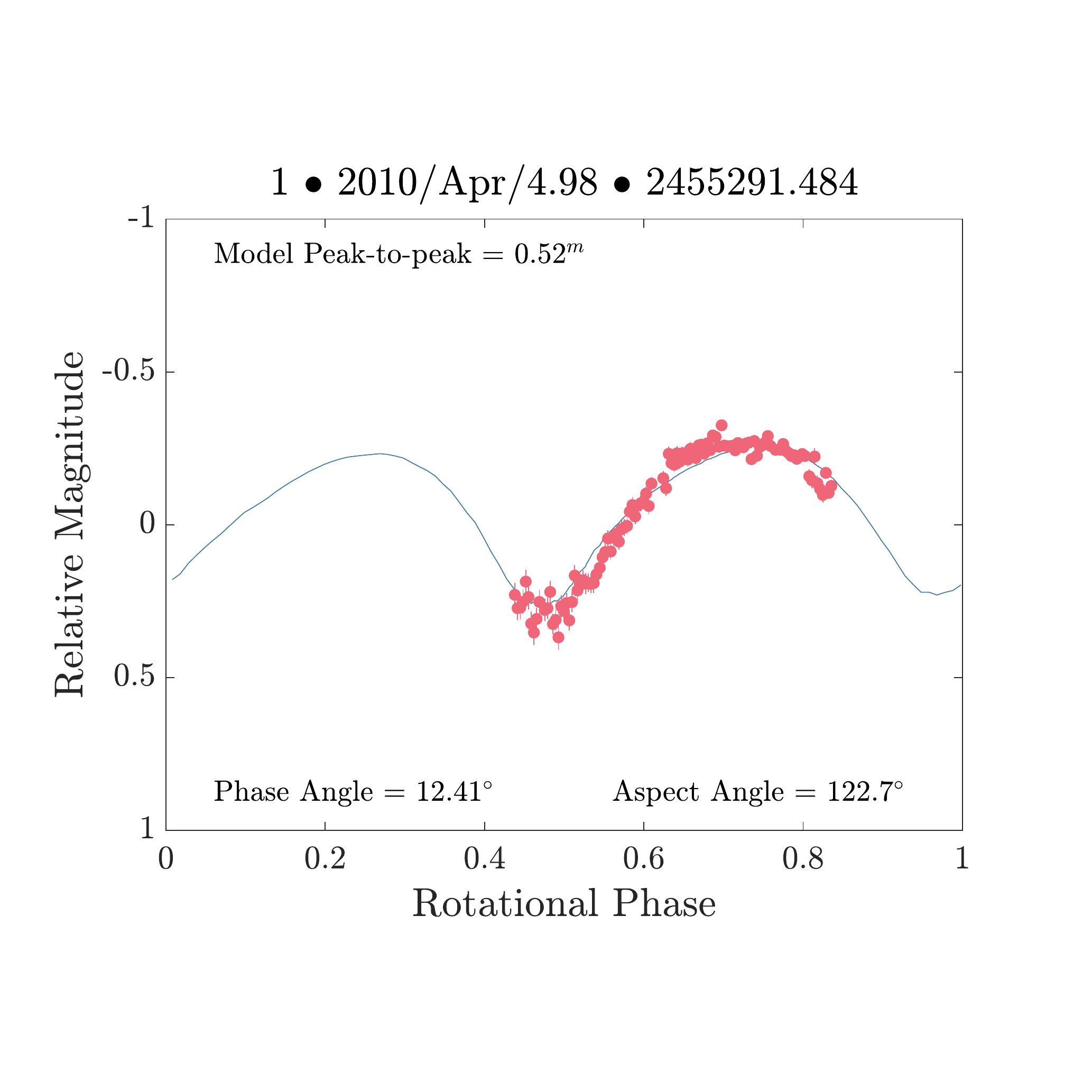}
			\includegraphics[width=.33\textwidth, trim=1cm 3.0cm 2.2cm 1.5cm, clip=true]{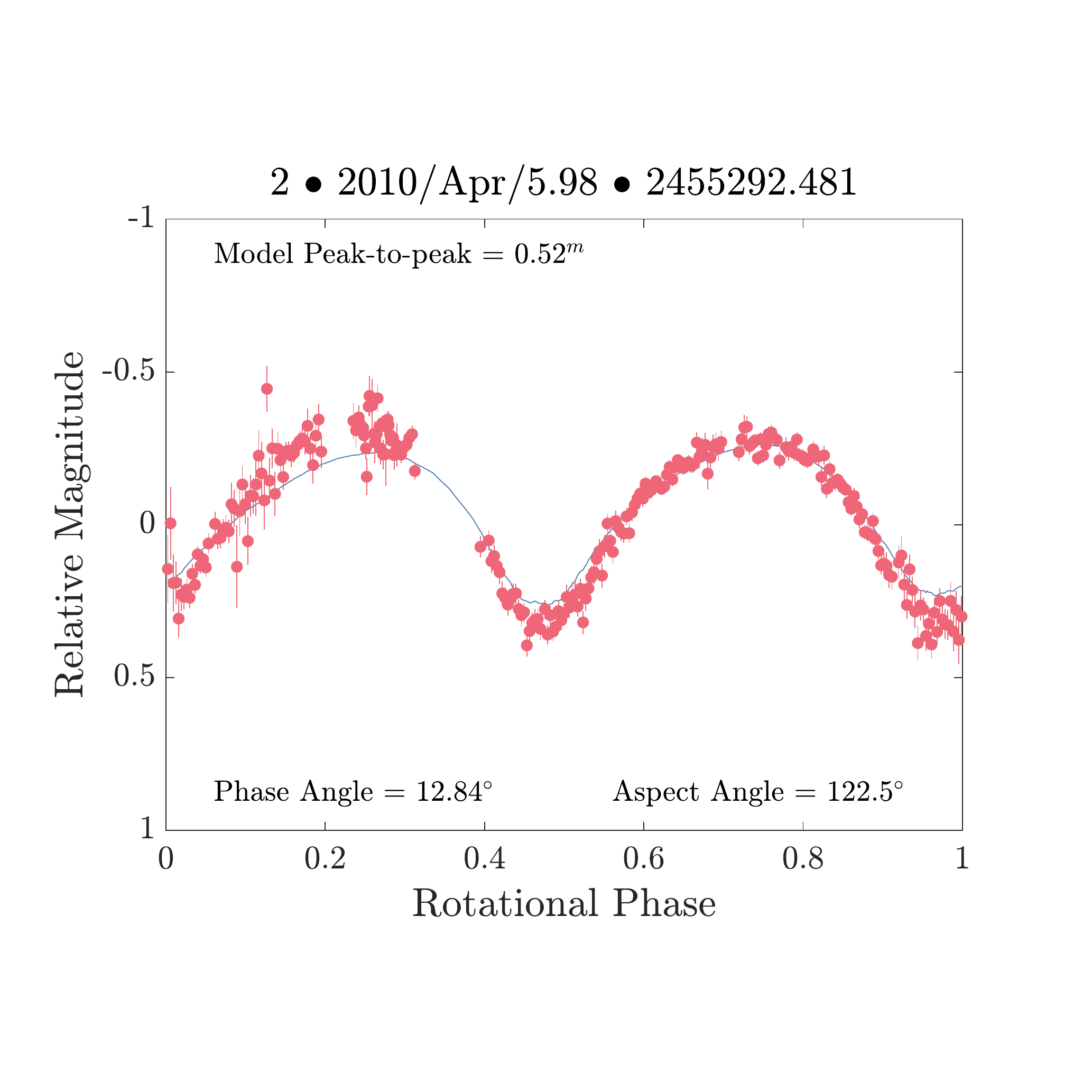}
			\includegraphics[width=.33\textwidth, trim=1cm 3.0cm 2.2cm 1.5cm, clip=true]{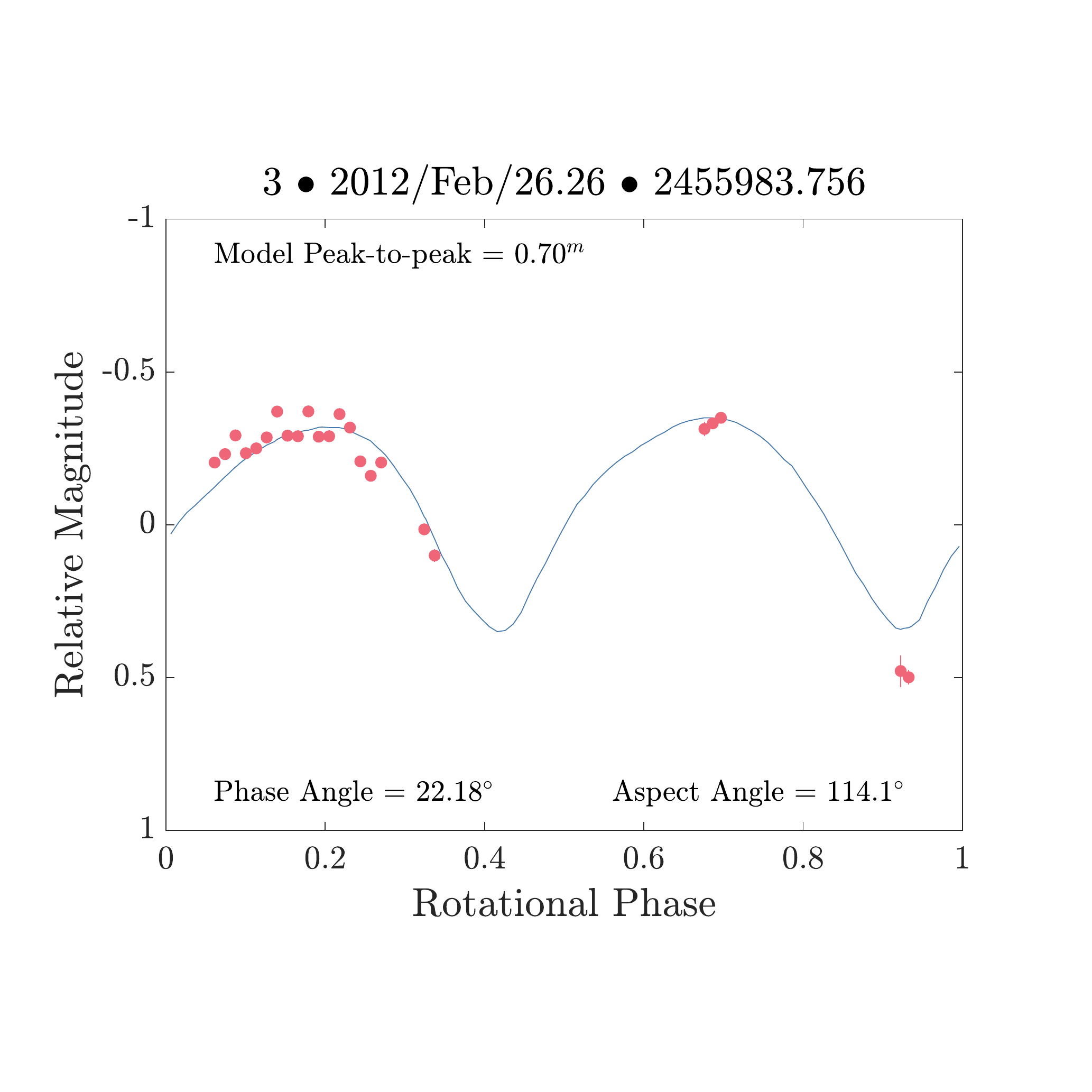}	
		}
		
		\resizebox{\hsize}{!}{	
			\includegraphics[width=.33\textwidth, trim=1cm 3.0cm 2.2cm 1.5cm, clip=true]{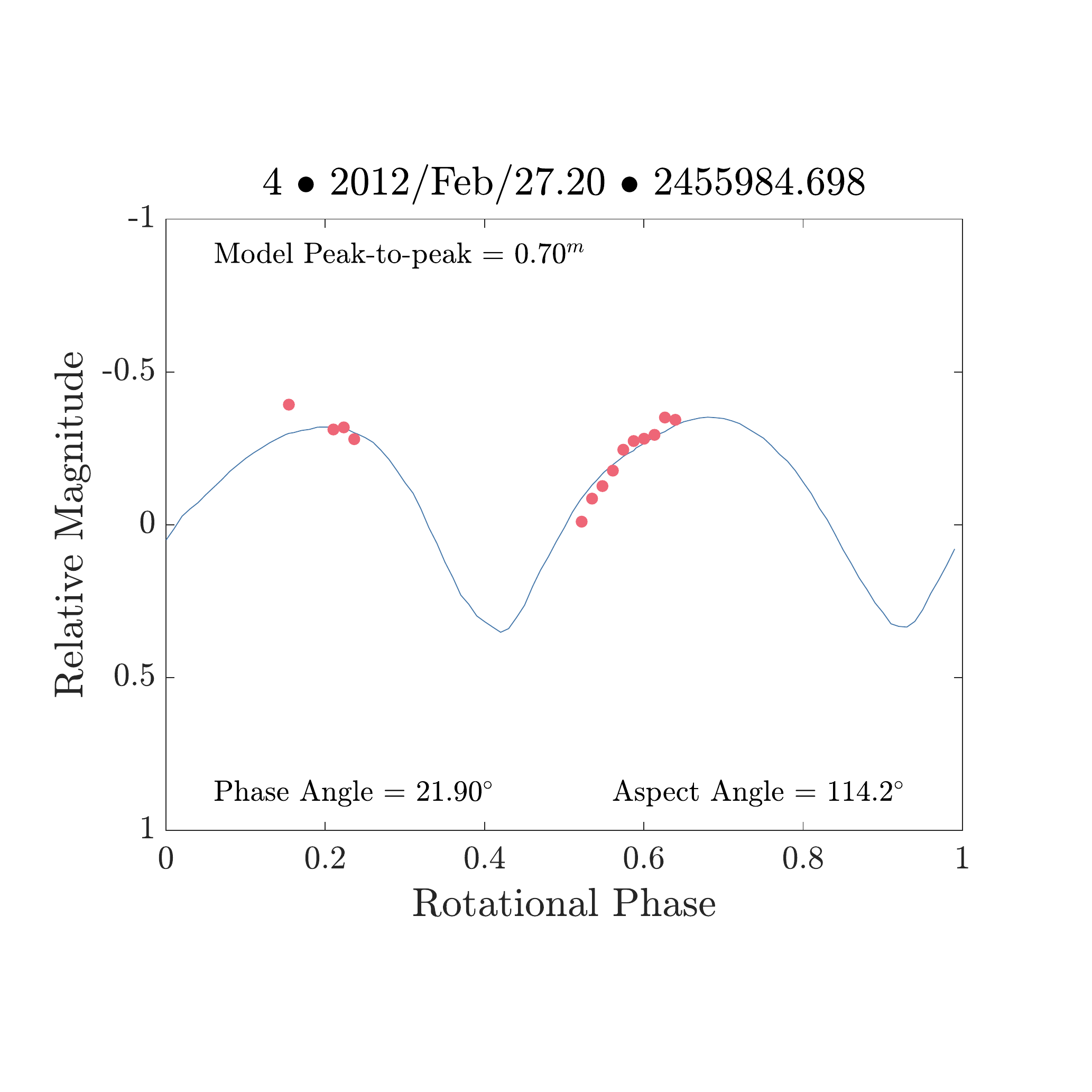}
			\includegraphics[width=.33\textwidth, trim=1cm 3.0cm 2.2cm 1.5cm, clip=true]{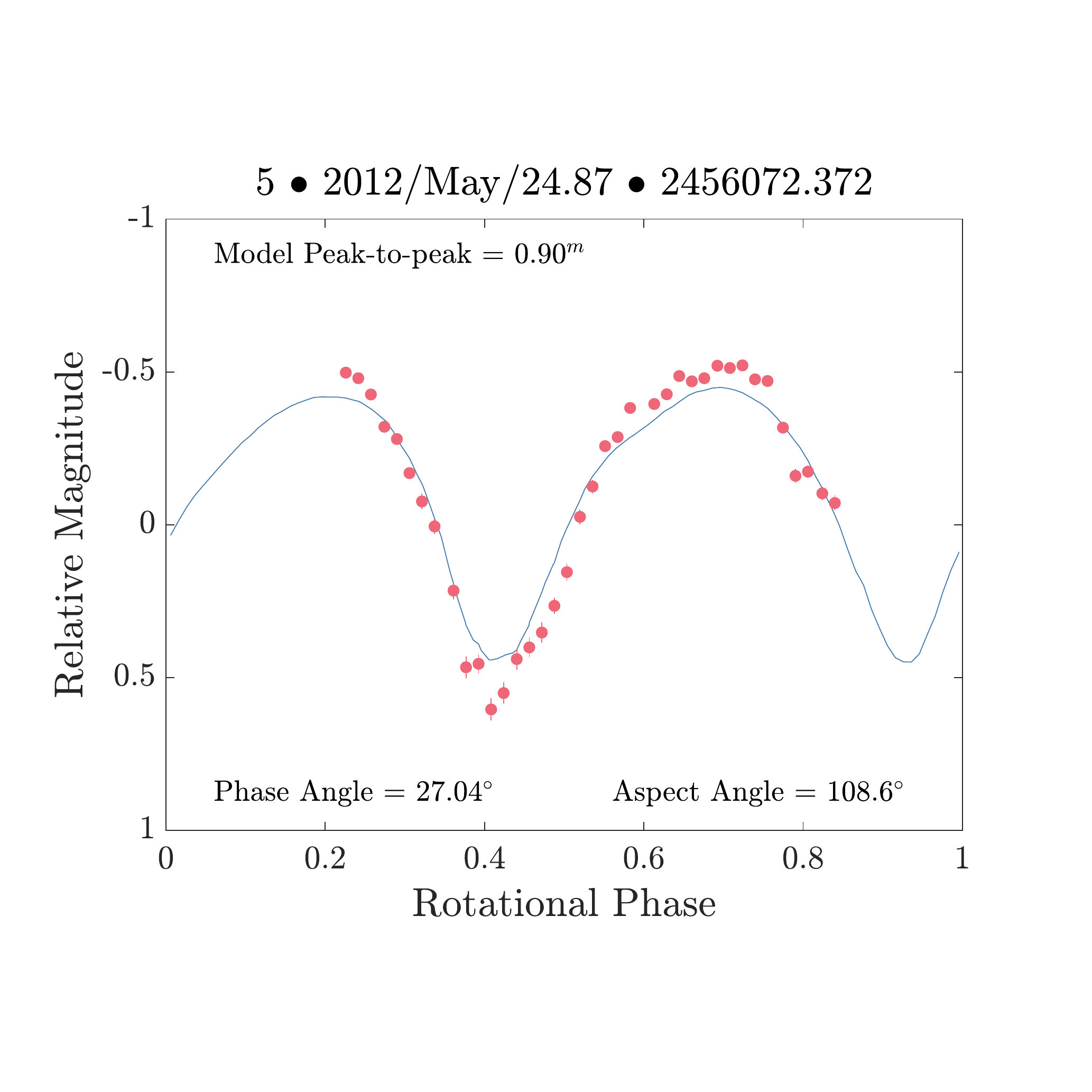}
			\includegraphics[width=.33\textwidth, trim=1cm 3.0cm 2.2cm 1.5cm, clip=true]{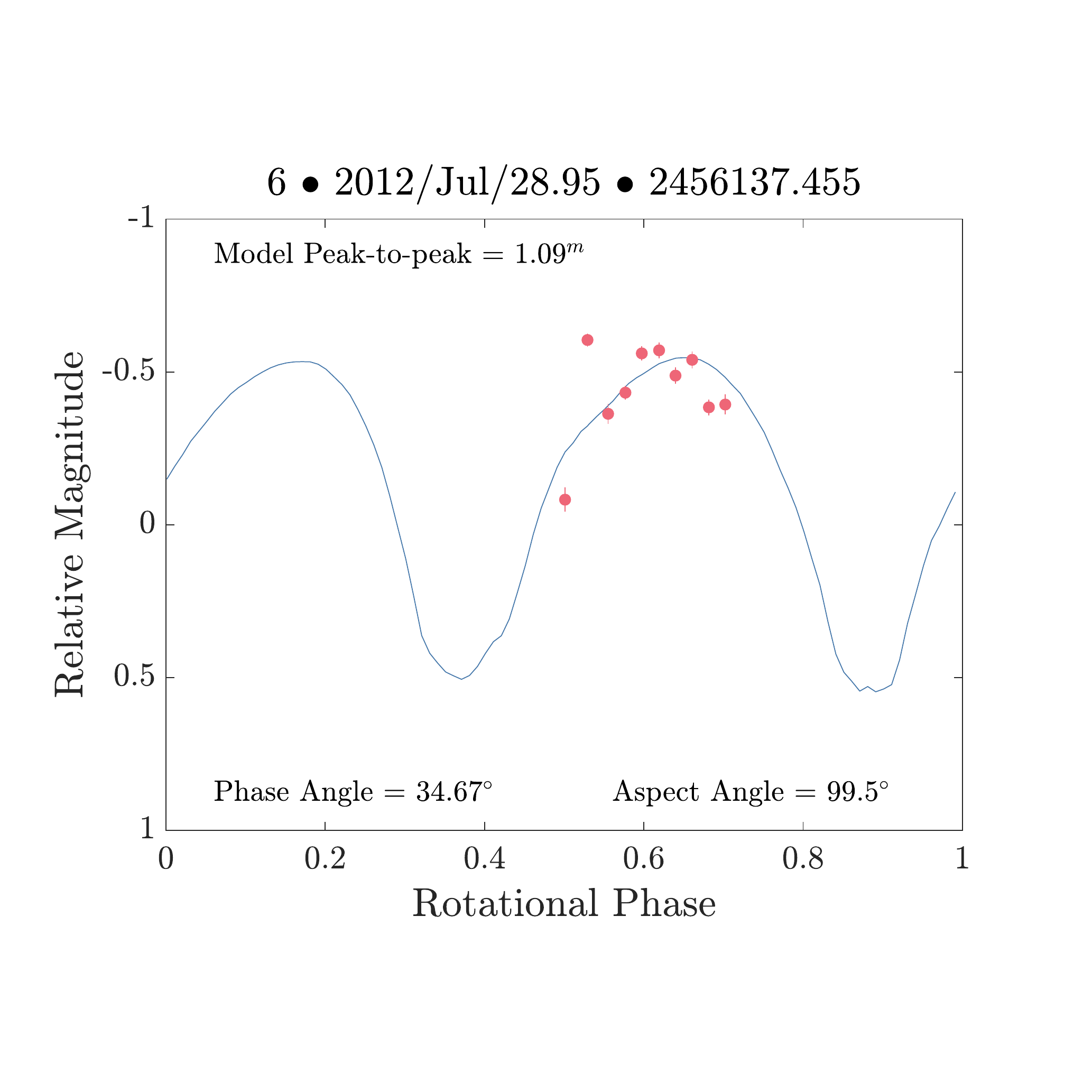}	
		}
		
		\resizebox{\hsize}{!}{	
			\includegraphics[width=.33\textwidth, trim=1cm 3.0cm 2.2cm 1.5cm, clip=true]{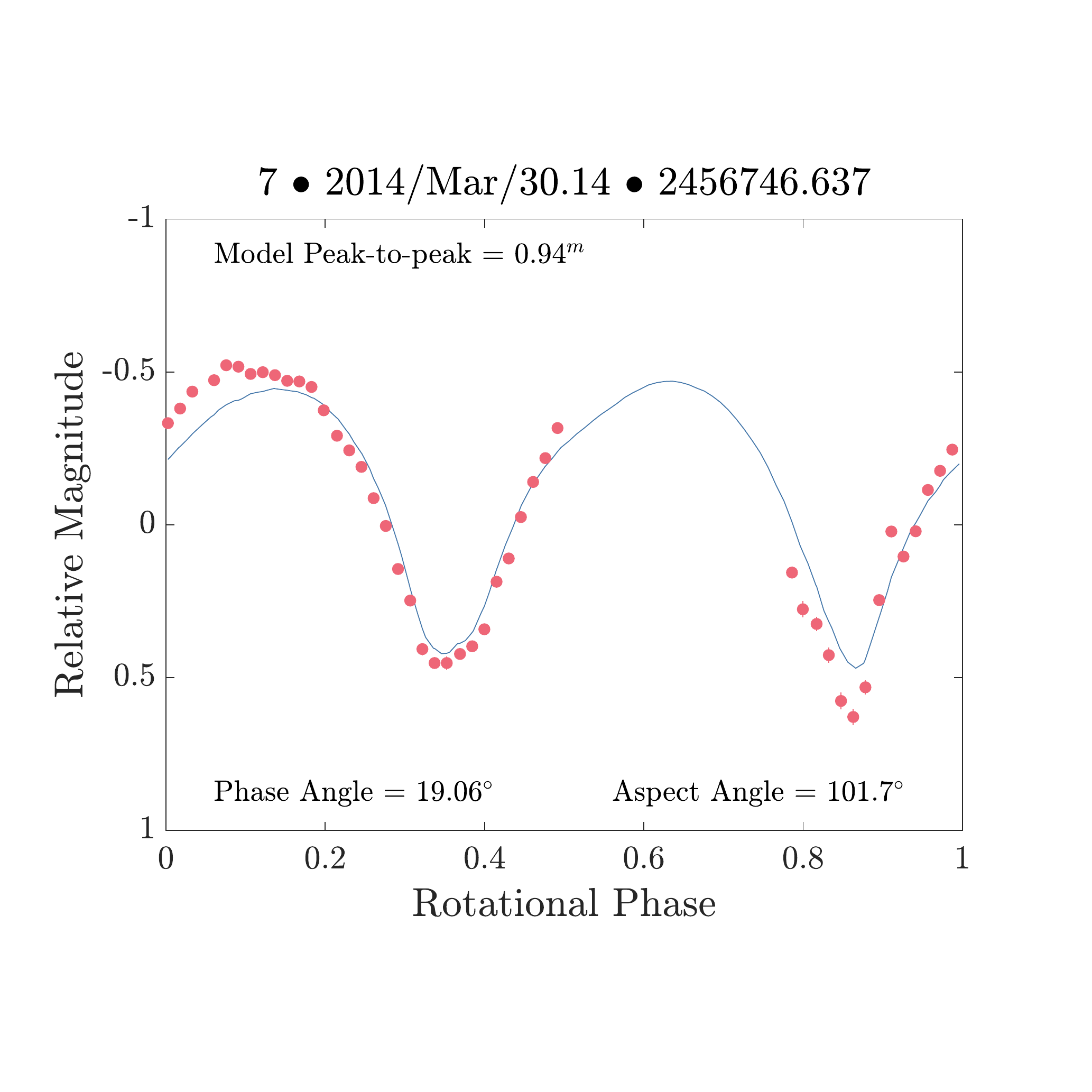}
			\includegraphics[width=.33\textwidth, trim=1cm 3.0cm 2.2cm 1.5cm, clip=true]{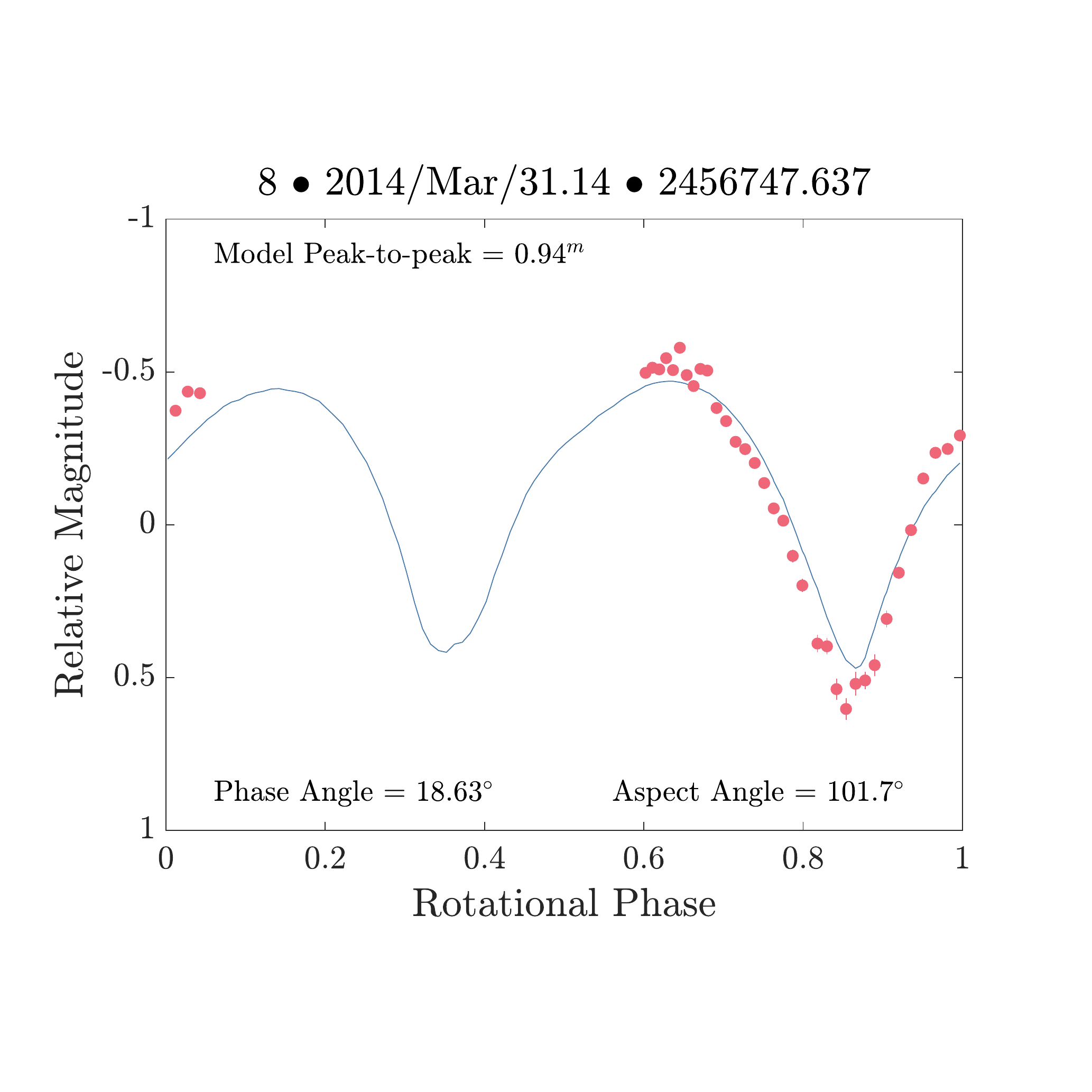}
			\includegraphics[width=.33\textwidth, trim=1cm 3.0cm 2.2cm 1.5cm, clip=true]{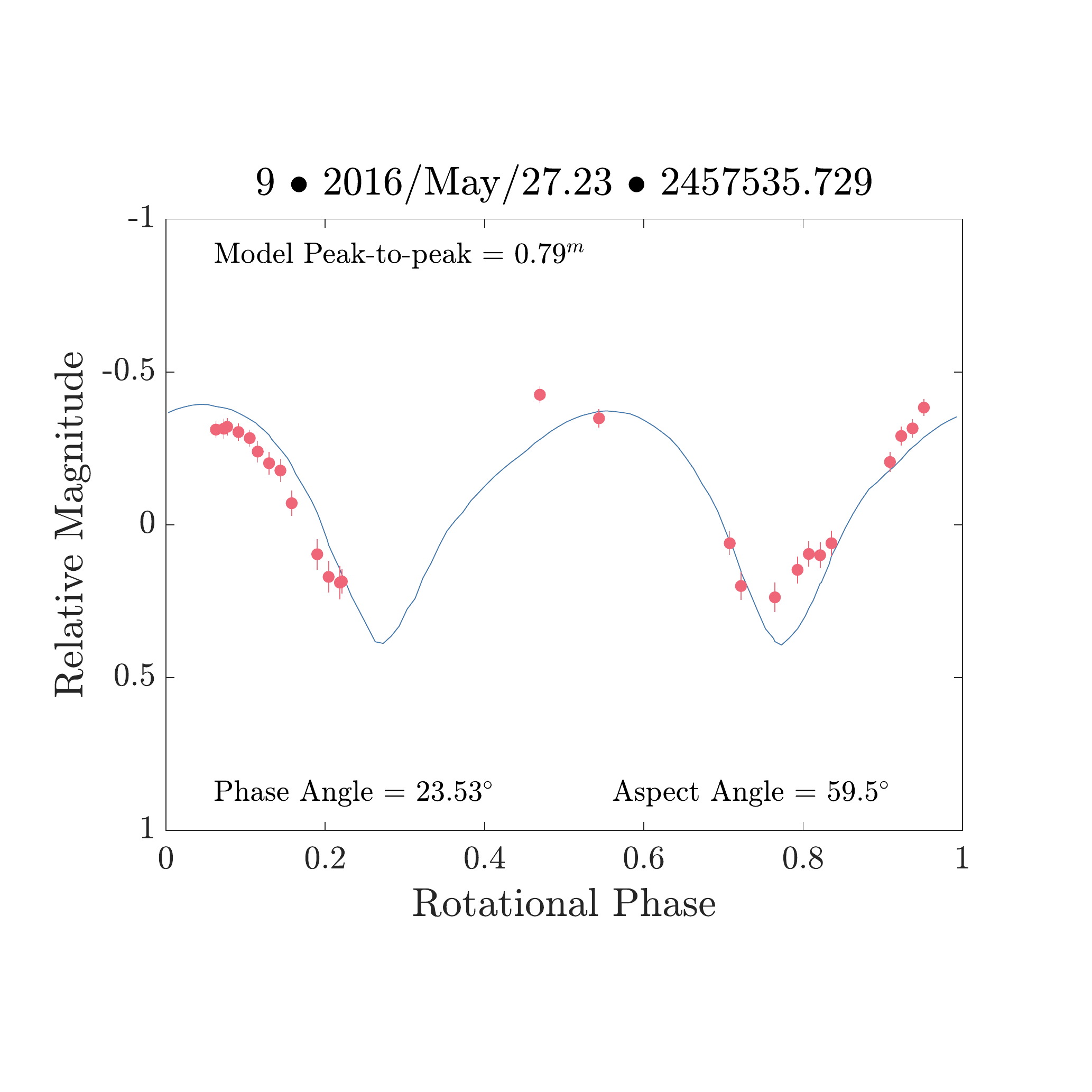}	
		}
		
		\resizebox{\hsize}{!}{	
			\includegraphics[width=.33\textwidth, trim=1cm 3.0cm 2.2cm 1.5cm, clip=true]{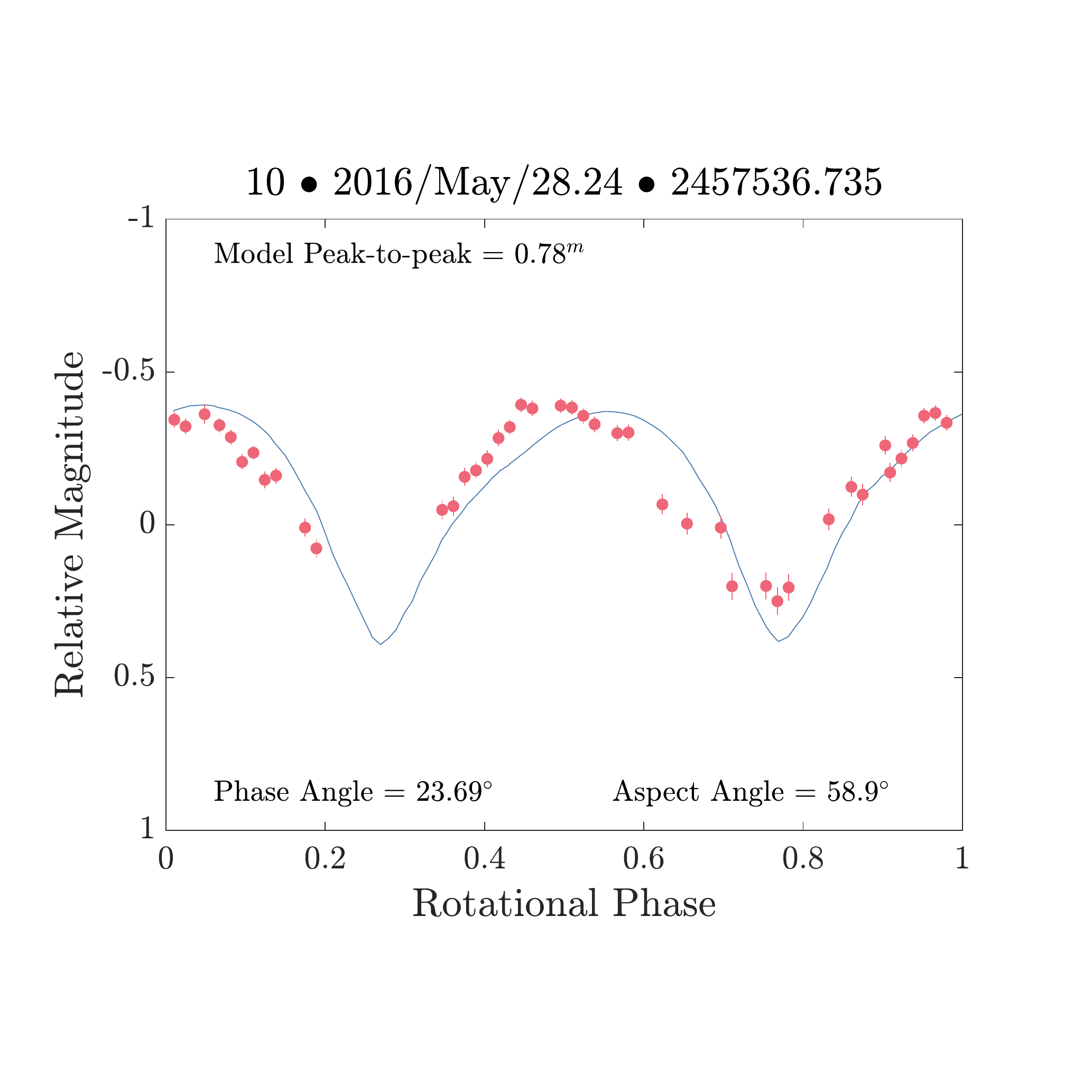}
			\includegraphics[width=.33\textwidth, trim=1cm 3.0cm 2.2cm 1.5cm, clip=true]{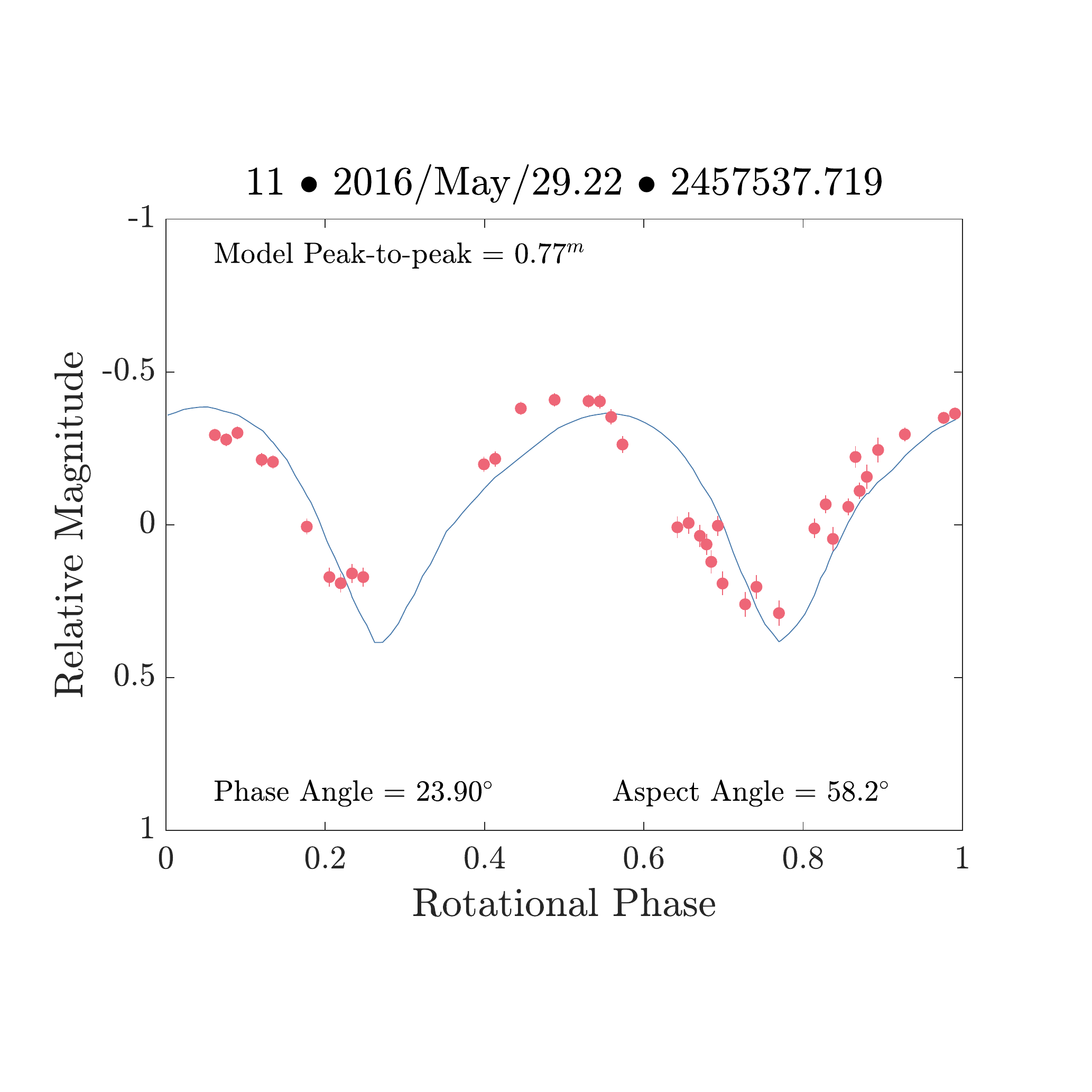}
			\includegraphics[width=.33\textwidth, trim=1cm 3.0cm 2.2cm 1.5cm, clip=true]{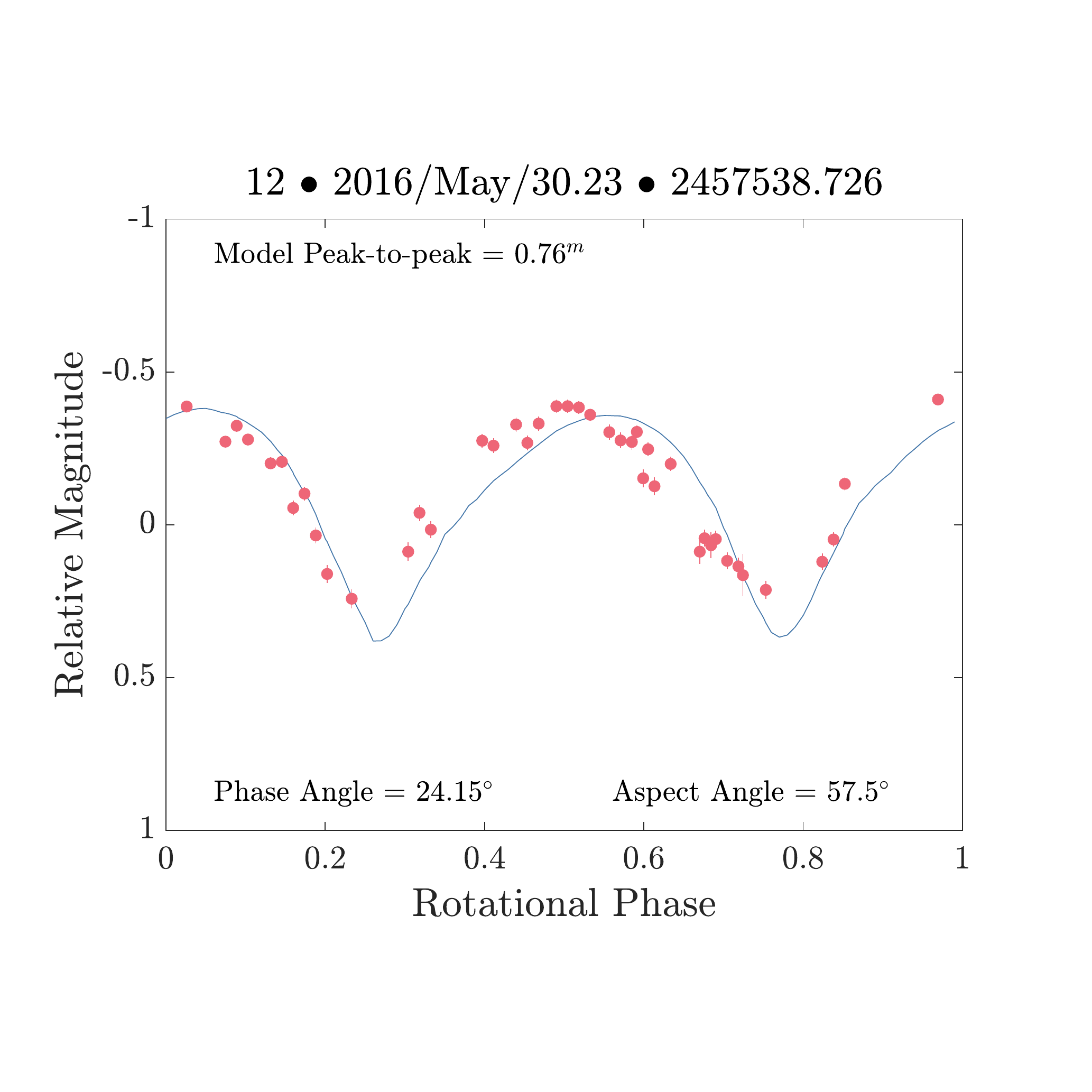}	
		}

		\caption{Same as Fig. \ref{fig:conv-lightcurvefit1}, but for the radar-derived shape model with a constant period.}
		\label{fig:noyorp-lightcurvefit1}
	\end{figure*}
	
	\setcounter{figure}{2}    
	\begin{figure*}
		
		\resizebox{\hsize}{!}{	
			\includegraphics[width=.33\textwidth, trim=1cm 3.0cm 2.2cm 1.5cm, clip=true]{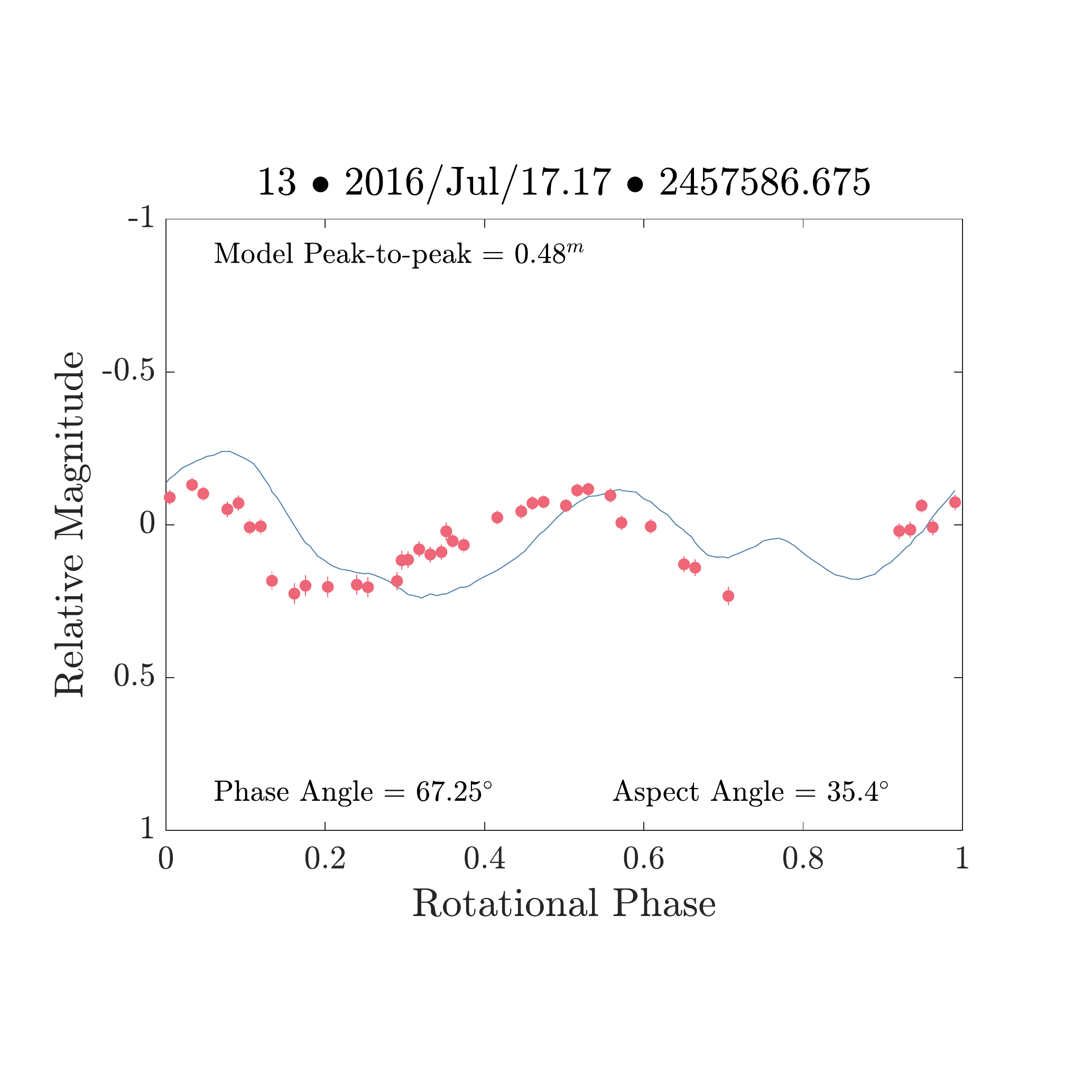}
			\includegraphics[width=.33\textwidth, trim=1cm 3.0cm 2.2cm 1.5cm, clip=true]{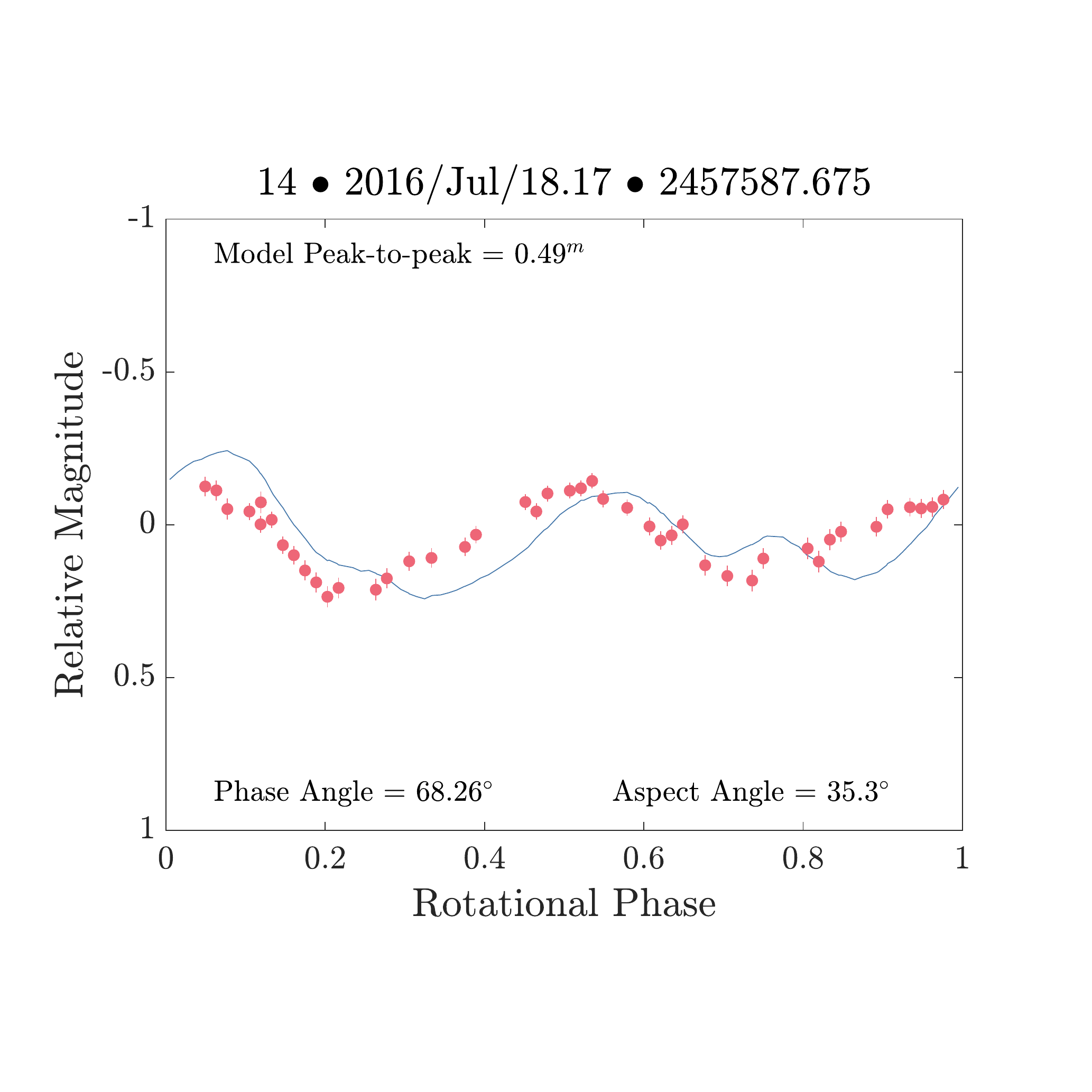}
			\includegraphics[width=.33\textwidth, trim=1cm 3.0cm 2.2cm 1.5cm, clip=true]{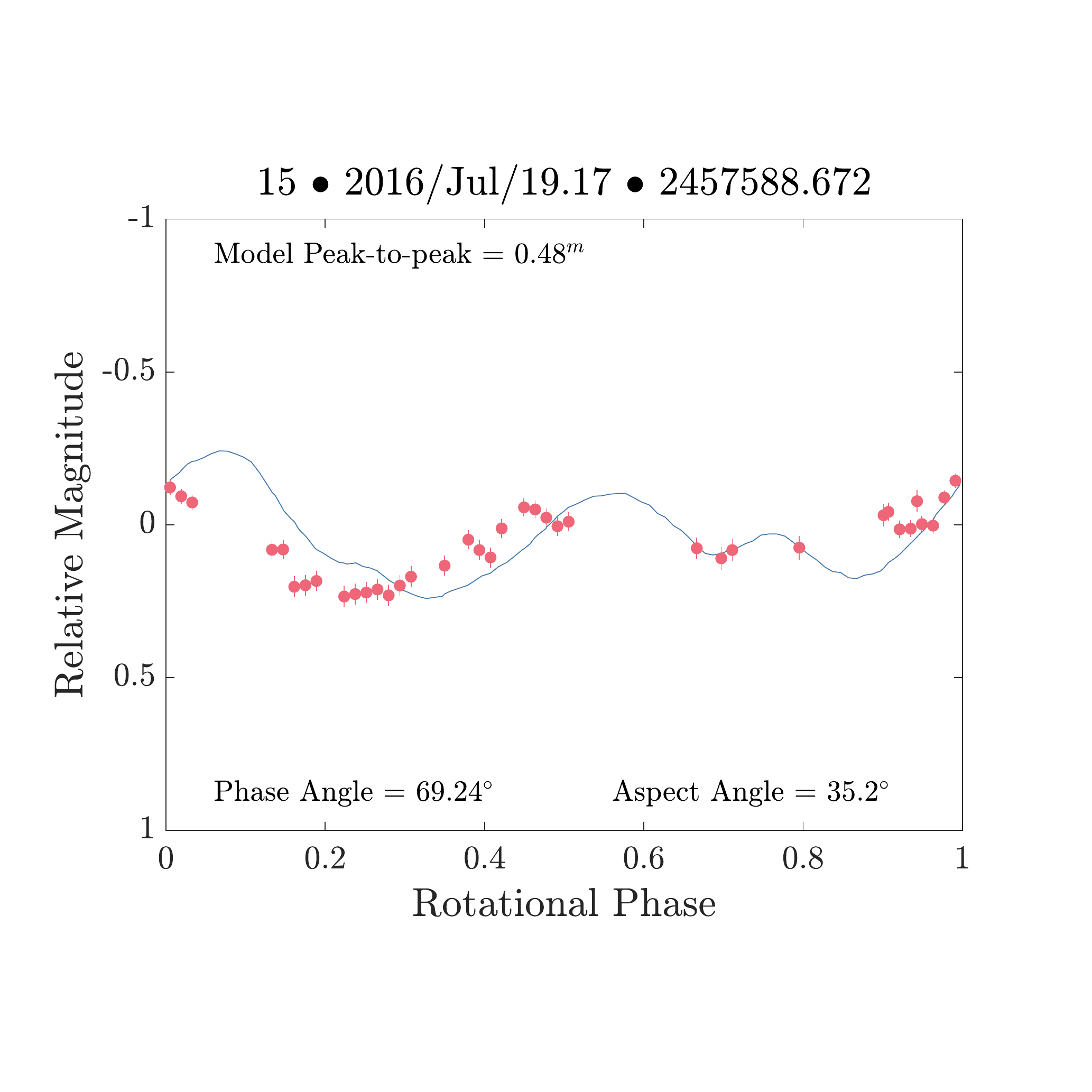}	
		}
		
		\resizebox{0.6666666\hsize}{!}{	
			\includegraphics[width=.33\textwidth, trim=1cm 3.0cm 2.2cm 1.5cm, clip=true]{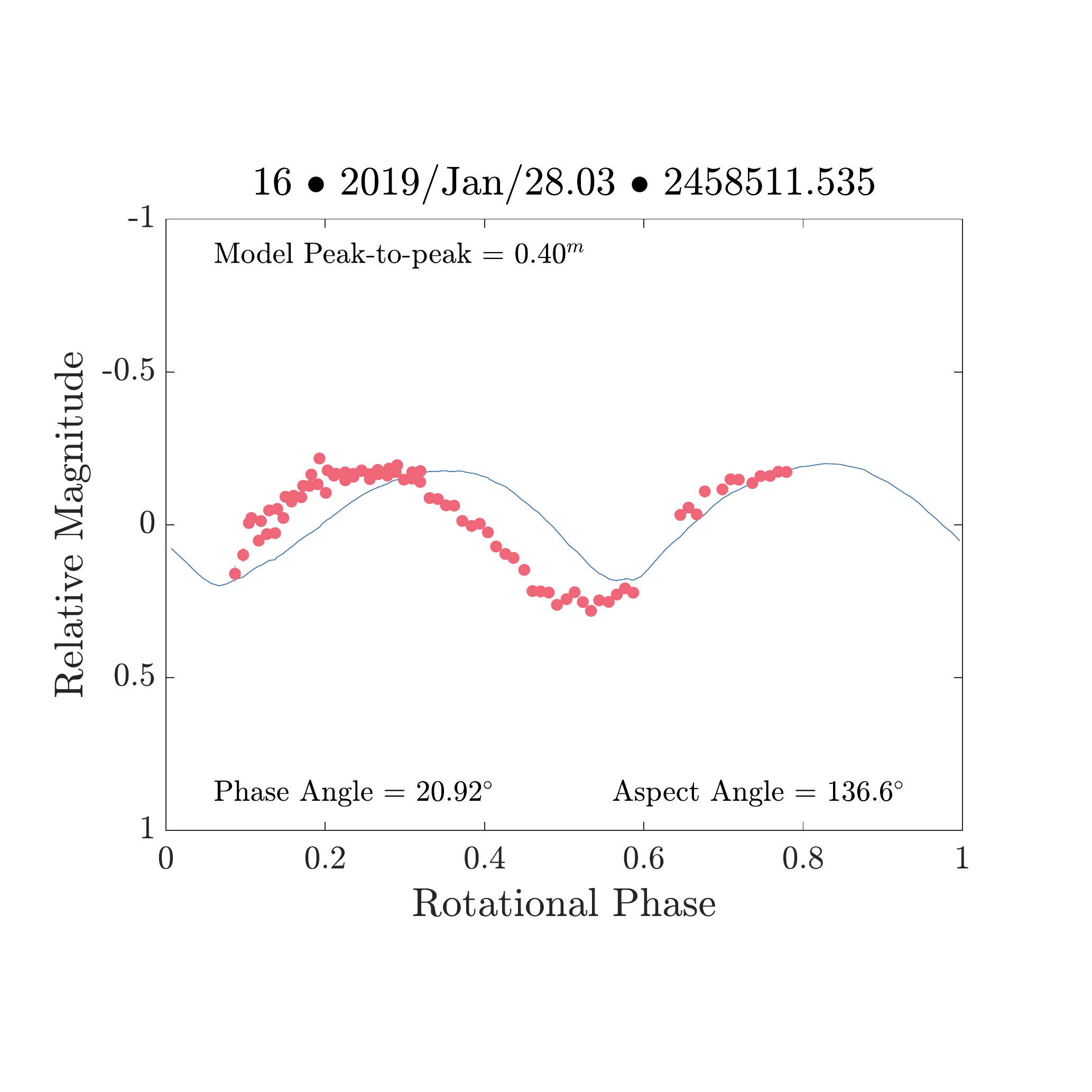}
			\includegraphics[width=.33\textwidth, trim=1cm 3.0cm 2.2cm 1.5cm, clip=true]{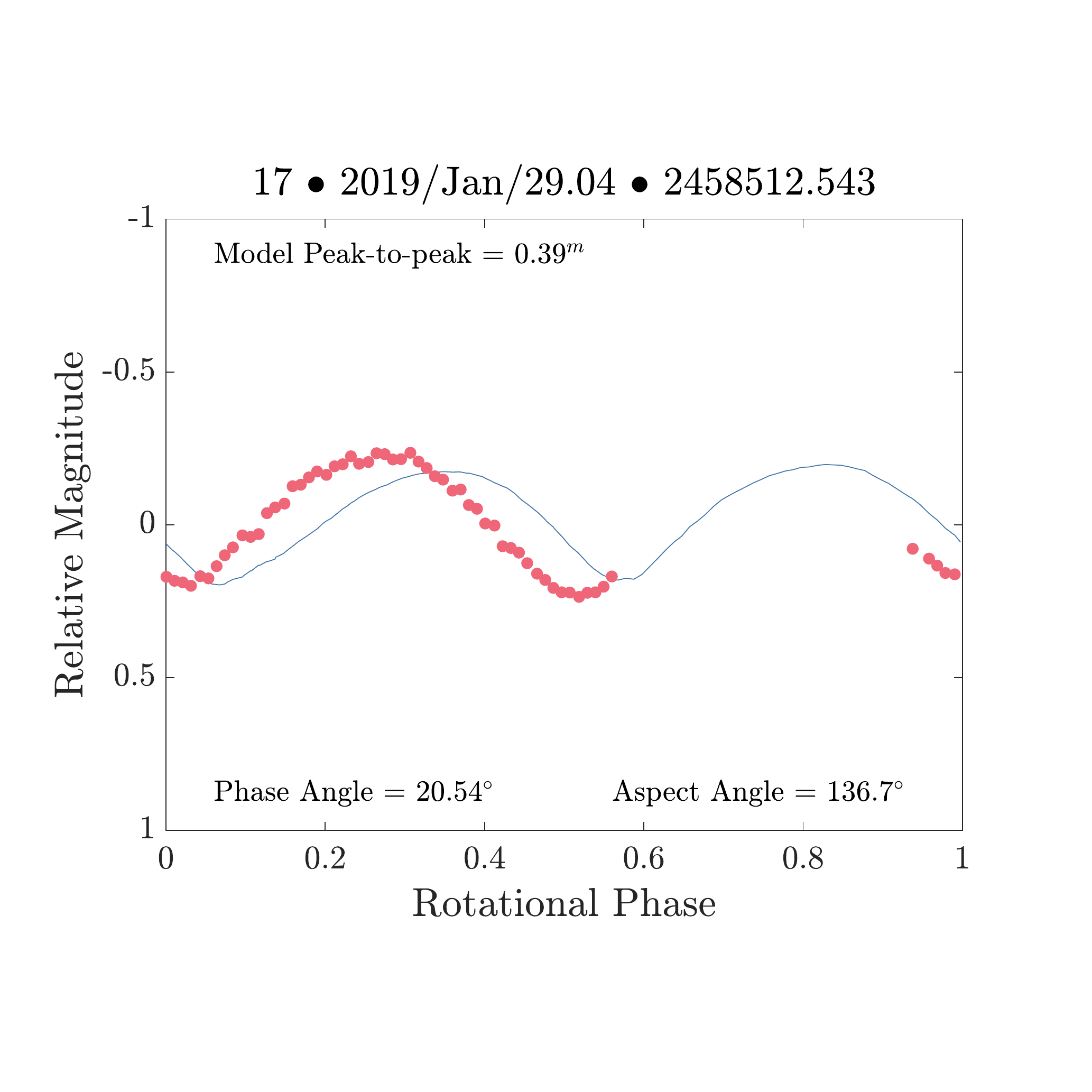}
		}
		\caption[]{(Continued.) }
		\label{fig:noyorp-lightcurvefit2}
	\end{figure*}

	\begin{table*}
		\center
		\begin{tabular}{lcccc}
			\hline \hline \noalign{\smallskip}
			\multicolumn{4}{l|}{Direction cosines of PAs with respect to body axes} & Ratio of moments of inertia  \\
			& x & y & \multicolumn{1}{c|}{z}				& to maximum moment of inertia\\
			\hline \noalign{\smallskip}
			\multicolumn{1}{l|}{PA1} &  0.996415 & -0.084589 & \multicolumn{1}{l|}{-0.001556} &  0.226285 \\
			\multicolumn{1}{l|}{PA2} &  0.084601 &  0.996352 & \multicolumn{1}{l|}{ 0.011208} &  1.000000 \\
			\multicolumn{1}{l|}{PA3} &  0.000602 & -0.011300 & \multicolumn{1}{l|}{ 0.999936} &  0.951949 \\
			\hline \noalign{\smallskip}
			\multicolumn{5}{c}{Angular offset between PA and body axes [deg]} \\
			\noalign{\smallskip}
			\multicolumn{2}{l}{PA1 \& x = 4.853186} & \multicolumn{2}{c}{PA2 \& y = 4.895571} & \multicolumn{1}{l}{PA3 \& z = 0.648350} \\
			\hline
		\end{tabular}
		\caption{A summary of the vertex shape model's moments of inertia and the alignment of the PAs to the model's body-centric axes. This table contains 
			a description of: 
			the direction cosines of each PA to each body-centric axis, this matrix transforms the body-centric axes to PAs 
			(if these were perfectly aligned: PA1 = 1, 0, 0; PA2 = 0, 1, 0; PA3 = 0, 0, 1); 
			the ratio of the moment of inertia for each axis to the axis with the maximum moment of inertia; 
			and the angular offset between each PA and its closest body-centric axes.
			An interpretation of the moments of inertia is discussed in Sect. \ref{sec:SHAPE}.}
		\label{table:Inertia&PAOrientation}
	\end{table*}

	\bsp	%
	\label{lastpage}
\end{document}